\documentclass[final,3p,times]{elsarticle}

\usepackage{amssymb}
\usepackage{amsthm}
\usepackage{amsmath,amsfonts,mathrsfs,color,longtable,bm}
\usepackage{hyperref}
\usepackage{graphicx}
\usepackage{array}
\usepackage{enumitem}
\usepackage[utf8]{inputenc}
\usepackage[normalem]{ulem}
\usepackage{cleveref}
\usepackage{xcolor}

\newcommand{\dd}{\mathrm{d}}

\newcommand{\ns}{n_{\rm s}}
\newcommand{\as}{\alpha_{\rm s}}
\newcommand{\bs}{\beta_{\rm s}}
\newcommand{\nt}{n_{\rm t}}
\newcommand{\at}{\alpha_{\rm t}}
\newcommand{\bt}{\beta_{\rm t}}
\newcommand{\Planck}{{\em Planck}}

\journal{Phys. Dark Universe}

\begin{document}

\begin{frontmatter}

\title{Third-order corrections to the slow-roll expansion: calculation and constraints with Planck, ACT, SPT, and BICEP/Keck}

\author[a,b,c]{Mario Ballardini}
\ead{mario.ballardini@unife.it}

\author{Alessandro Davoli}
\ead{alessandro.davoli91@gmail.com}

\author[a,b,d]{Salvatore Samuele Sirletti}
\ead{salvatoresamuele.sirletti@unife.it}

\affiliation[a]{organization={Dipartimento di Fisica e Scienze della Terra, Università degli Studi di Ferrara},
             addressline={via Giuseppe Saragat 1},
             city={Ferrara},
             postcode={44122},
             state={},
             country={Italy}}

 \affiliation[b]{organization={Istituto Nazionale di Fisica Nucleare, sezione di Ferrara},
             addressline={via Giuseppe Saragat 1},
             city={Ferrara},
             postcode={44122},
             state={},
             country={Italy}}

 \affiliation[c]{organization={INAF/OAS Bologna},
             addressline={via Piero Gobetti 101},
             city={Bologna},
             postcode={40129},
             state={},
             country={Italy}}

 \affiliation[d]{organization={Dipartimento di Fisica, Università di Trento},
             addressline={via Sommarive 14},
             city={Trento},
             postcode={38123},
             state={},
             country={Italy}}

\begin{abstract}
We investigate the primordial power spectra (PPS) of scalar and tensor perturbations, derived through the 
slow-roll approximation. By solving the Mukhanov-Sasaki equation and the tensor perturbation equation with Green's 
function techniques, we extend the PPS calculations to third-order corrections, providing a comprehensive expansion 
in terms of slow-roll parameters with an independent approach to the solution of the integrals compared to the one 
previously presented in the literature.
We investigate the accuracy of the analytic predictions starting from first-order corrections up to third-order ones 
with the numerical solutions of the perturbation equations for a selection of single-field slow-roll inflationary models. 
We derive the constraints on the Hubble flow functions $\epsilon_i$ from \Planck, ACT, SPT, and BICEP/Keck data. We 
find an upper bound $\epsilon_1 \lesssim 0.002$ at 95\% CL dominated by BICEP/Keck data and robust to all the different 
combination of datasets. 
We derive the constraint $\epsilon_2 \simeq 0.031 \pm 0.004$ at 68\% confidence level (CL) from the combination of 
\Planck\ data and late-time probes such as baryon acoustic oscillations, redshift space distortions, and supernovae data 
at first order in the slow-roll expansion. The uncertainty on $\epsilon_2$ gets larger including second- and third-order 
corrections, allowing for a non-vanishing running and running of the running respectively, leading to 
$\epsilon_2 \simeq 0.034 \pm 0.007$ at 68\% CL. We find $\epsilon_3 \simeq 0.1 \pm 0.4$ at 95\% CL both at second and 
at third order in the slow-roll expansion of the spectra. 
$\epsilon_4$ remains always unconstrained. The combination of \Planck\ and SPT data, compatible among each others, leads 
to slightly tighter constraints on $\epsilon_2$ and $\epsilon_3$. On the contrary, the combination of \Planck\ data with 
ACT measurements, which point to higher values of the scalar spectral index compared to \Planck\ findings, leads to 
shifts in the means and maximum likelihood values for $\epsilon_2$ and $\epsilon_3$. We discuss the results obtained 
for different dataset combinations and different multipole cuts.
\end{abstract}

\end{frontmatter}

\section{Introduction}
Cosmic inflation \cite{Starobinsky:1980te,Guth:1980zm,Linde:1981mu,Albrecht:1982wi,Hawking:1982ga,Linde:1983gd}, a period 
of accelerated expansion in the early universe, provides a compelling framework for understanding the initial conditions 
that led to the large-scale structure we observe today. During this epoch, quantum fluctuations in the {\em inflaton field}, 
a scalar field driving inflation, are stretched to macroscopic scales, seeding the primordial density perturbations and 
gravitational waves that later evolve into the cosmic microwave background (CMB) anisotropies and the inhomogeneous 
distribution of galaxies.

The dynamics of the scalar perturbations are encapsulated in the Mukhanov-Sasaki equation 
\cite{Mukhanov:1985rz,Sasaki:1986hm}, a second-order differential equation governing the evolution of perturbations in the 
inflating universe. An analogous equation can be defined for the tensor perturbations. Under the slow-roll approximation, 
where the inflaton field evolves slowly compared to the Hubble expansion rate, it is possible to derive approximate 
solutions through a perturbative expansion in terms of slow-roll parameters.

The primordial power spectra (PPS) are typically characterised by the scalar spectral index $\ns$ and the tensor-to-scalar 
ratio $r$, which are critical parameters for comparing theoretical predictions with observational data from CMB experiments 
\cite{Planck:2013jfk,Planck:2015sxf,Planck:2018jri}. Indeed, the standard phenomenological parameterisation of the 
dimensionless PPS of scalar and tensor perturbations corresponds to
\begin{equation} \label{eqn:spectra}
    {\cal P}_\zeta(k) = A_{\rm s}\left(\frac{k}{k_*}\right)^{n_{\rm s}-1} \,, \qquad
    {\cal P}_{\rm t}(k) = r A_{\rm s}\left(\frac{k}{k_*}\right)^{n_{\rm t}} \,,
\end{equation}
where $A_{\rm s}$ is the scalar amplitude, $n_{\rm s}$ ($n_{\rm t}$) is the scalar (tensor) spectral index, 
$r \equiv A_{\rm t}/A_{\rm s}$ is the tensor-to-scalar ratio, and $k_* = 0.05\,{\rm Mpc}^{-1}$ a reference pivot scale. 
Eqs. \eqref{eqn:spectra} can be improved by exploiting the analytic dependence of the slow-roll power spectra of primordial 
perturbations on the values of the Hubble parameter and the hierarchy of its time derivatives, known as the Hubble flow 
functions (HFFs) 
\begin{equation}\label{eqn:HFFs}
    \epsilon_{i+1}(N) = \frac{\dd \ln |\epsilon_i|}{\dd N} \,, \qquad \epsilon_1(N) = -\frac{\dd \ln H}{\dd N} \,,
\end{equation}
where $\dd N \equiv \dd \ln a$ is the number of $e$-folds.

Gong and Stewart used the Green's function technique in Ref. \cite{Gong:2001he} to solve the Mukhanov-Sasaki 
equation perturbatively; see 
Refs. \cite{Starobinsky:1979ty,Mukhanov:1985rz,Mukhanov:1988jd,Stewart:1993bc,Liddle:1994dx,Nakamura:1996da,Hoffman:2000ue} 
for earlier work based on the slow-roll perturbative expansion. This approach has provided valuable insights into the power 
spectra of scalar and tensor perturbations by incorporating higher-order corrections in the slow-roll expansion at second 
order or next-to-next-to-leading order (NNLO) in Refs. \cite{Gong:2001he,Schwarz:2001vv,Leach:2002ar} and at third order 
or next-to-next-to-next-to-leading order (N3LO) in Refs. \cite{Auclair:2022yxs,Bianchi:2024qyp}. 
In addition to the calculation based on the Green’s function method, derivations based on other approximation schemes are 
available in the literature, such as the uniform approximation \cite{Habib:2002yi}, the Wentzel-Kramers-Brillouin (WKB) 
approximation \cite{Martin:2002vn,Casadio:2005em}, or the method of comparison equations \cite{Casadio:2006wb}.

Such analytical methods allow one to accurately connect the expansion parameters $\epsilon_i$ to the physical parameters of 
specific single-field slow-roll inflationary models and to deal with a versatile framework to be applied for parameter 
inference \cite{Martin:2013tda,Martin:2024qnn,Ballardini:2024ado}. The accuracy of these analytical predictions is crucial, 
especially in the light of future cosmological surveys that will offer unprecedented precision in measuring the CMB 
anisotropies and the large-scale structure (LSS) of the Universe. 

Cosmological observations from future experiments dedicated to the measurements of CMB polarisation, such as Simons 
Observatory \cite{SimonsObservatory:2018koc} and CMB-S4 \cite{Abazajian:2019eic,CMB-S4:2020lpa}, from ground and from 
space, such as LiteBIRD \cite{LiteBIRD:2022cnt,Paoletti:2022anb}, will be able to reduce the uncertainties on the PPS 
of scalar and tensor fluctuations. 
The situation will be further improved by the complementarity of future galaxy survey experiments, such as from {\em Euclid} 
\cite{EUCLID:2011zbd,Amendola:2016saw,Euclid:2019clj,Euclid:2021qvm}, that will open for the opportunity of measuring 
ultra-large scales, thanks to its large observed volume, as well as small scales of matter distribution in the full 
nonlinear regime. 
Their combination will drastically improve our understanding of the early-Universe physics and of cosmic inflation 
reducing significantly the uncertainties on $\epsilon_1$, mostly from measurements of the B-mode of the CMB, and reducing the allowed 
$\epsilon_2$ - $\epsilon_3$ parameter space, mostly from small-scale measurements (and thanks to the increased the lever 
arm between the large and small angular scales).

In this paper, we calculate the PPS of scalar and tensor perturbations up to third-order corrections in the slow-roll 
expansion. While these results have been already presented by Auclair and Ringeval \cite{Auclair:2022yxs}, we obtain them 
with a different approach to the integrals. Moreover, we compare our final findings to the results obtained in 
Ref. \cite{Auclair:2022yxs}, and we systematically compare our results to the numerical solutions of the Mukhanov-Sasaki 
and tensor perturbation equations to validate the effectiveness and importance of the PPS solutions at third order.

Given the expected sensitivity of current CMB surveys, such as \Planck\ \cite{Planck:2018nkj}, ACT \cite{ACT:2020gnv}, 
SPT \cite{SPT-3G:2022hvq}, and BICEP/Keck \cite{BICEP:2021xfz}, we derive constraints on the HFF parameters considering 
different truncation of the PPS expansion at first order, second order, and third order, investigating the implications 
on the spectral indices, their runnings, and the running of the runnings. 
We also present forecasts for a futuristic cosmic-variance level CMB space mission.

We structure the paper as follows. After this introduction, we review the theoretical framework of the Mukhanov-Sasaki and 
tensor perturbation equations and their solution via Green's functions in the context of slow-roll inflation and present the 
derivation of the third-order equations of the PPS in \cref{sec:analytic}. In \cref{sec:numeric}, we compare the analytic 
results with numerical solutions and discuss the implications for a selection of two single-field inflationary models. 
Section \ref{sec:results} is dedicated to constraints on the slow-roll parameters. We conclude in \cref{sec:conclusions}.

\section{Third-Order Calculation} \label{sec:analytic}
Starting from the action for a single scalar field $\phi$ minimally coupled to gravity
\begin{equation}
    {\cal S} = \int \dd^4x \sqrt{-g} \left[\frac{M_{\rm Pl}^2}{2}R-\frac{1}{2}\partial_\mu\phi\partial^\mu\phi-V(\phi)\right]
\end{equation}
the Friedmann and Klein-Gordon equations for the cosmological background are, respectively 
\begin{equation}
    H^2 = \frac{1}{3 M_{\rm Pl}^2}\left(\frac{\dot{\phi}^2}{2} + V\right) \,,\qquad
    \ddot{\phi} + 3H\dot{\phi} + V_\phi = 0 \,,
    \label{eqn:dynamicalsystem}
\end{equation}
where $V_\phi \equiv \dd V / \dd \phi$ and the background metric is chosen to be the flat 
Friedmann-Lema{\^i}tre-Robertson-Walker one given by 
\begin{equation}
    \dd s^2 = -\dd t^2 + a^2(t)\, \dd x^2 \,,
\end{equation}
where $t$ is the proper time, $a(t)$ is the scale factor, and $x^i$ are the three-dimensional comoving spatial coordinates. 
The equation of motion for the gauge-invariant quantity $v$ at leading order in perturbations, also known as Mukhanov-Sasaki 
equation for the scalar perturbations, is 
\begin{equation} \label{eqn:mukhanov_1}
    v_k^{\rm (s,t)}{\,''}(\tau) + \left[k^2 - U^{\rm (s,t)}(\tau)\right]v_k^{\rm (s,t)}(\tau) = 0 \,,
\end{equation}
where $v_k$ is the mode function in Fourier space, and $'$ means derivative with respect to the conformal time $\tau$, 
defined as $\dd t \equiv a(\tau)\dd \tau$. For the scalar perturbations, the mode function corresponds to the Mukhanov 
variable $v^{\rm (s)}_k(\tau)\equiv a(\tau)\sqrt{2\epsilon_1(\tau)}\zeta_k(\tau)$ \cite{Mukhanov:1981xt} and the potential 
corresponds to $U^{\rm (s)}(\tau) \equiv z''/z$ where $z(\tau) \equiv a(\tau)\sqrt{2\epsilon_1(\tau)}$. For the tensor 
perturbations, the mode function corresponds to $v^{\rm (t)}_k(\tau)\equiv a(\tau) h_k(\tau)$ and the potential to 
$U^{\rm (t)}(\tau) \equiv a''/a$. 
Following the procedure introduced in Ref.~\cite{Gong:2001he},~\cref{eqn:mukhanov_1} can be rewritten as
\begin{equation} \label{eqn:mukhanov_2}
    \frac{\dd^2y^{\rm (s,t)}}{\dd x^2} + \left[1 - \frac{U^{\rm (s,t)}(\tau)}{k^2}\right]y^{\rm (s,t)} = 0 \,,
\end{equation}
where $y^{\rm (s,t)} \equiv \sqrt{2k}\,v^{\rm (s,t)}_k$ and $x\equiv -k\tau$; both $y(x)$ satisfy the asymptotic flat-space vacuum 
condition, i.e. Bunch-Davies initial conditions
\begin{equation}\label{eqn:bunchdavis}
    \lim_{x\to\infty}y(x) = e^{ix} \,.
\end{equation}

Let us then introduce the function $g^{\rm (s,t)}(\ln x) \equiv \tau^2 U^{\rm (s,t)}(x) - 2$. In terms of these functions, 
we can rewrite~\cref{eqn:mukhanov_2} as 
\begin{equation} \label{eqn:mukhanov_3}
    \frac{\dd^2y^{\rm (s,t)}}{\dd x^2} + \left(1-\frac{2}{x^2}\right)y^{\rm (s,t)} = \frac{g^{\rm (s,t)}(\ln x)}{x^2}\,y^{\rm (s,t)} \,.
\end{equation}
The solution to~\cref{eqn:mukhanov_3} for both scalar and tensor perturbations, using Green’s function method, developed in 
Ref.~\cite{Gong:2001he}, can be written in an integral form as
\begin{equation} \label{eqn:solution_y}
    y(x) = y_0(x) + 
    \frac{i}{2}\int_x^\infty\,\frac{\dd u}{u^2}\,g\left(\ln u\right)\,y(u)\left[y_0^*(u)y_0(x)-y_0^*(x)y_0(u)\right] \,,
\end{equation}
where $y_0(x)$ is the solution to the corresponding homogeneous equation, i.e.
\begin{equation} \label{eqn:y_0}
    y_0(x) = \left(1+\frac{i}{x}\right)e^{ix} \,.
\end{equation}
The idea is to perform a series expansion for $g$ for both scalar and tensor perturbations as 
\begin{equation} \label{eqn:expansion_g}
    g(\ln x) \equiv \sum_{n=0}^\infty\frac{g_{n+1}}{n!} \left(\ln x\right)^n \,,
\end{equation}
where quantity $x$ can then be expressed in terms of the slow-roll parameters up to third order as
\begin{equation} \label{eqn:x_expansion}
    x = -k\int\frac{\dd t}{a} \simeq 
    \frac{k}{aH} \left[1 + \epsilon_{1} + \epsilon_{1}^2 + \epsilon_{1}^3 + \epsilon_{1} \epsilon_{2} + 3 \epsilon_{1}^2 \epsilon_{2} + \epsilon_{1} \epsilon_{2}^2 + \epsilon_{1} \epsilon_{2} \epsilon_{3} + {\cal O}(\epsilon^4)\right] \,.
\end{equation}
Finally, in order to find a perturbatively solution to~\cref{eqn:solution_y}, we can expand $y$ as 
\begin{equation} \label{eqn:expansion_y}
    y(x) \equiv \sum_{n=0}^\infty y_n(x) \,,
\end{equation}
where $y_0(x)$ is the homogeneous solution~\cref{eqn:y_0} and $y_n(x)$ is of order $n$ in the slow-roll expansion.

\subsection{Primordial Scalar Perturbations}
Following Ref.~\cite{Gong:2001he}, it is convenient to introduce the function
\begin{equation} \label{eqn:definition_fs}
    f^{\rm (s)}(\ln x) \equiv zx \,,
\end{equation}
where
\begin{equation}
    g^{\rm (s)}(\ln x) = \frac{1}{f^{\rm (s)}} \left[\frac{\dd^2 f^{\rm (s)}}{{\left(\dd\ln x\right)}^2} - 3\frac{\dd f^{\rm (s)}}{\dd\ln x}\right] \,.
\end{equation}
We can express the $g_n$ coefficients in~\cref{eqn:expansion_g}, for both scalar and tensor perturbations, up to third order as
\begin{subequations} \label{eqn:g_from_f}
\begin{align}    
    g_1 &\simeq -3\frac{f_1}{f_0}+\frac{f_2}{f_0} \\
    g_2 &\simeq 3\frac{f_1^2}{f_0^2}-3\frac{f_2}{f_0}-\frac{f_1 f_2}{f_0^2}+\frac{f_3}{f_0} \\
    g_3 &\simeq -3\frac{f_1^3}{f_0^3}+\frac{9}{2}\frac{f_1 f_2}{f_0^2}-\frac{3}{2}\frac{f_3}{f_0} \,,
\end{align}
\end{subequations}
where, in general, the ratio $f_n/f_0$ is of order $n$ in the slow-roll parameters. We perform a series expansion also for 
$f$ as 
\begin{equation} \label{eqn:expansion_f}
    f(\ln x) \equiv \sum_{n=0}^\infty\frac{f_n}{n!} \left(\ln x\right)^n \,,
\end{equation}
where the coefficients $f_n$ can be computed from
\begin{equation}
    f_n = \left.\frac{\dd^n f}{{\left(\dd\ln x\right)}^n}\right|_{x=1} \,.
\end{equation}
We Taylor expand around $x = 1$ using \cref{eqn:x_expansion}. This leads to different coefficient compared to the expansion 
performed in Ref.~\cite{Gong:2001he} around horizon crossing $aH = k$ as explained in Ref.~\cite{Martin:2024qnn}.
Keeping terms up to third order in the slow-roll parameters, we have
\begin{subequations}
\begin{align}
    f^{\rm (s)}_0 \simeq &\left. \frac{a\dot\phi}{H} \right|_{x=1} \,,\\
    f^{\rm (s)}_1 \simeq & \frac{a\dot\phi}{H} \left[1 - \frac{a H}{k} \left(1 + \frac{\epsilon_2}{2} \right)\right]_{x=1} \,,\\
    f^{\rm (s)}_2 \simeq & \frac{a\dot\phi}{H} \left[1 
    - 3\frac{a H}{k}\left(1 + \frac{\epsilon_2}{2} \right)
    + \left(\frac{a H}{k}\right)^2 \left(2 -\epsilon_1 +\frac{3 \epsilon_2}{2} -\frac{\epsilon_1 \epsilon_2}{2}+\frac{\epsilon_2 \epsilon_3}{2}+\frac{\epsilon_2^2}{4} \right) \right]_{x=1} \,,\\
    f^{\rm (s)}_3 \simeq & \frac{a\dot\phi}{H} \left[1 
    -7\frac{a H}{k}\left(1+\frac{\epsilon_2}{2}\right) 
    + \left(\frac{a H}{k}\right)^2 \left(12 - 6\epsilon_1 + 9\epsilon_2 - 3 \epsilon_1 \epsilon_2 + \frac{3 \epsilon_2^2}{2} 
    + 3\epsilon_2\epsilon_3 \right) \right. \notag\\    
    &\left. - \left(\frac{a H}{k}\right)^3 \left(6 - 7 \epsilon_1 + \frac{11 \epsilon_2}{2} + 2 \epsilon_1^2 + \frac{3\epsilon_2^2}{2} + 3 \epsilon_2 \epsilon_3 - 6 \epsilon_1 \epsilon_2 \right.\right. \notag\\
    &\left.\left. +\epsilon_1^2 \epsilon_2 - \frac{5 \epsilon_1 \epsilon_2^2}{4} 
    -\frac{3 \epsilon_1 \epsilon_2 \epsilon_3}{2}   
    + \frac{3 \epsilon_2^2 \epsilon_3}{4} + \frac{\epsilon_2^3}{8} + \frac{\epsilon_2 \epsilon_3^2}{2}  
    + \frac{\epsilon_2 \epsilon_3 \epsilon_4}{2} \right) \right]_{x=1} 
\end{align}
\end{subequations}
and then using~\cref{eqn:g_from_f}, we find for the primordial scalar perturbations
\begin{subequations} \label{eqn:g_s}
\begin{align}
    g^{\rm (s)}_1 &\simeq 3\epsilon_1 + \frac{3}{2}\epsilon_2 + 4\epsilon_1^2 + \frac{13}{2}\epsilon_1\epsilon_2 + \frac{1}{4}\epsilon_2^2 + 5\epsilon_1^3 + \frac{35}{2}\epsilon_1^2\epsilon_2 + \frac{15}{2}\epsilon_1\epsilon_2^2 + \frac{1}{2}\epsilon_2\epsilon_3 + 5\epsilon_1\epsilon_2\epsilon_3 \,,\\
    g^{\rm (s)}_2 &\simeq -3\epsilon_1\epsilon_2 - 11\epsilon_1^2\epsilon_2 - \frac{13}{2}\epsilon_1\epsilon_2^2 - \frac{3}{2}\epsilon_2\epsilon_3 - 8\epsilon_1\epsilon_2\epsilon_3 - \frac{1}{2}\epsilon_2^2\epsilon_3 - \frac{1}{2}\epsilon_2\epsilon_3^2 - \frac{1}{2}\epsilon_2\epsilon_3\epsilon_4 \,,\\
    g^{\rm (s)}_3 &\simeq \frac{3}{2}\epsilon_1\epsilon_2^2 + \frac{3}{2}\epsilon_1\epsilon_2\epsilon_3 + \frac{3}{4}\epsilon_2\epsilon_3^2 + \frac{3}{4}\epsilon_2\epsilon_3\epsilon_4 \,.
\end{align}
\end{subequations}

\subsection{Primordial Tensor Perturbations}
We can repeat the same procedure applied to the scalar perturbations on the quantity
\begin{equation} \label{eqn:definition_ft}
    f^{\rm (t)}(\ln x) \equiv ax \,,
\end{equation}
where
\begin{equation}
    g^{\rm (t)}(\ln x) = \frac{1}{f^{\rm (t)}} \left[\frac{\dd^2 f^{\rm (t)}}{{\left(\dd\ln x\right)}^2} - 3\frac{\dd f^{\rm (t)}}{\dd\ln x}\right] \,.
\end{equation}
For tensor perturbations, $f$ and $g$ coefficients up to third order correspond to 
\begin{subequations}
\begin{align}
f^{\rm (t)}_0 \simeq &\left.a\right|_{x=1} \,,\\
f^{\rm (t)}_1 \simeq\, &a\left(1 - \frac{aH}{k}\right)_{x=1} \,,\\
f^{\rm (t)}_2 \simeq\, &a\left[1 - 3\frac{aH}{k} + \left(\frac{aH}{k}\right)^2\left(2 - \epsilon_1\right) \right]_{x=1} \,,\\
f^{\rm (t)}_3 \simeq\, &a\left[1 - 7\frac{aH}{k} + \left(\frac{aH}{k}\right)^2\left(12 - 6\epsilon_1\right) 
- \left(\frac{aH}{k}\right)^3\left(6 - 7\epsilon_1 + 2\epsilon_1^2 - \epsilon_1\epsilon_2 \right) \right]_{x=1} 
\end{align}
\end{subequations}
and 
\begin{subequations} \label{eqn:g_t}
    \begin{align}
    g^{\rm (t)}_1 &\simeq 3\epsilon_1 + 4\epsilon_1^2 + 4\epsilon_1\epsilon_2 + 5\epsilon_1^3 + 14\epsilon_1^2\epsilon_2 + 4\epsilon_1\epsilon_2^2 + 4\epsilon_1\epsilon_2\epsilon_3 \,,\\
    g^{\rm (t)}_2 &\simeq -3\epsilon_1\epsilon_2 - 11\epsilon_1^2\epsilon_2 - 4\epsilon_1\epsilon_2^2 - 4\epsilon_1\epsilon_2\epsilon_3 \,,\\
    g^{\rm (t)}_3 &\simeq \frac{3}{2}\epsilon_1\epsilon_2^2 + \frac{3}{2}\epsilon_1\epsilon_2\epsilon_3 \,.
    \end{align}
\end{subequations}

\subsection{The Primordial Power Spectra Including Third-Order Corrections}
Now that we have the $g_n$ for both scalar and tensor perturbations up to third-order in slow-roll parameters, in order to 
solve~\cref{eqn:mukhanov_3} we need to calculate the recursive solutions of~\cref{eqn:solution_y} up to third-order 
corrections, which corresponds to
\begin{equation} \label{eqn:y}
    y(x) \simeq y_0(x) + y_1(x) + y_{2}(x) + y_{3}(x) \,.
\end{equation}
In this case there are unique solutions for $y_0$, $y_1$, $y_2$, and $y_3$ since the dependence on the slow-roll parameters, 
thus the differences between scalar and tensor perturbations, are only encoded in the $g_n$ functions.

Since the right hand side of~\cref{eqn:mukhanov_3} is at least of order one in the slow-roll expansion, the lowest order 
solution to that equation is simply the homogeneous solution $y_0(x)$,~\cref{eqn:y_0}.

The first-order correction to~\cref{eqn:y_0} is obtained by substituting $g = g_1$ and $y = y_0$ into the right hand side 
of~\cref{eqn:solution_y}, we can easily find 
\begin{align}
    y_1(x) &= \frac{i g_1}{2}\int_x^\infty\,\frac{\dd u}{u^2}\,\,y_0(u)\left[y_0^*(u)y_0(x)-y_0^*(x)y_0(u)\right] \notag\\
    &= \frac{i\,g_1}{3}\left[\frac{2}{x}e^{ix} + i\,y_0^*(x)\int_x^\infty\frac{\dd u}{u}\,e^{2iu}\right] \,.
\label{eqn:y_1}
\end{align}

The second-order correction is made of two pieces, namely
\begin{equation}
    y_2(x) = y_{21}(x) + y_{22}(x) \,,
\end{equation}
where we substitute $g = g_1 + g_2 \ln x$ and $y = y_0 + y_1$ into~\cref{eqn:solution_y} finding
\begin{subequations}
\begin{align}
    y_{21}(x) &= \frac{i g_1}{2}\int_x^\infty\frac{\dd u}{u^2}\,y_1(u)\left[y_0^*(u)y_0(x)-y_0^*(x)y_0(u)\right] \notag\\
    &= -\frac{i g_1^2}{9}\left[\frac{2}{3x}e^{ix} - \left(\frac{5}{3x}-\frac{i}{3}\right)e^{-ix}\int_x^\infty\frac{\dd u}{u}\,e^{2iu} + i\,y_0(x)\int_x^\infty\frac{\dd u}{u}\,e^{-2iu}\int_u^\infty\frac{\dd t}{t}\,e^{2it}\right] \label{eqn:y_21}\\
    y_{22}(x) &= \frac{i g_2}{2}\int_x^\infty\frac{\dd u}{u^2}\,y_0(u)\ln u\left[y_0^*(u)y_0(x)-y_0^*(x)y_0(u)\right] \notag\\
    &= \frac{i g_2}{3}\left[\frac{8}{3x}e^{ix} + \frac{2}{x}\ln x \  e^{ix} + \frac{7i}{3}\,y_0^*(x)\int_x^\infty\frac{\dd u}{u}\,e^{2iu}+i\,y_0^*(x)\int_x^\infty\frac{\dd u}{u}\,e^{2iu}\ln u\right] \label{eqn:y_22} \,.
\end{align}
\end{subequations}

The third-order correction is made of three pieces, namely
\begin{equation}
    y_3(x) = y_{31}(x) + y_{32}(x) + y_{33}(x) \,.
\end{equation}
Here, we substitute $g = g_1 + g_2 \ln x + g_3 \ln^2 x$ and $y = y_0 + y_1 + y_2$ into~\cref{eqn:solution_y} finding
\begin{subequations}
\begin{align}
    y_{31}(x) &\equiv \frac{i g_1}{2}\int_x^\infty\frac{\dd u}{u^2}\,y_2(u)\left[y_0^*(u)y_0(x)-y_0^*(x)y_0(u)\right] \\
    y_{32}(x) &\equiv \frac{i g_2}{2}\int_x^\infty\frac{\dd u}{u^2}\,y_1(u)\ln u\left[y_0^*(u)y_0(x)-y_0^*(x)y_0(u)\right] \\
    y_{33}(x) &\equiv \frac{i g_3}{4}\int_x^\infty\frac{\dd u}{u^2}\,y_0(u)\ln^2u\left[y_0^*(u)y_0(x)-y_0^*(x)y_0(u)\right] \,.
\end{align}
\end{subequations}

Because of the large number of terms, it is worth considering these three contributions separately. Splitting the first 
contribution to the third-order correction, taking into account~\cref{eqn:y_21,eqn:y_22}, we have
\begin{subequations} \label{eqn:y_31}
\begin{align}\notag
y_{31}(x)&=\frac{g_1^3}{18}\int_x^\infty\frac{\dd u}{u^2}\left[\frac{2e^{iu}}{3u}-\left(\frac{5}{3u}-\frac{i}{3}\right)e^{-iu}\int_u^\infty\frac{\dd t}{t}\,e^{2it} + i y_0(u)\int_u^\infty\frac{\dd t}{t}\,e^{-2it}\int_t^\infty\frac{\dd s}{s}\,e^{2is}\right]\\[.15truecm]\notag
&\quad\quad\left[y_0^*(u)y_0(x)-y_0^*(x)y_0(u)\right]\\[.15truecm]\notag
&\quad-\frac{g_1g_2}{6}\int_x^\infty\frac{\dd u}{u^2}\left[\frac{8e^{iu}}{3u}+\frac{2e^{iu}\ln u}{u}+\frac{7i}{3}\,y_0^*(u)\int_u^\infty\frac{\dd t}{t}\,e^{2it}+i\,y_0^*(u)\int_u^\infty\frac{\dd t}{t}\,e^{2it}\ln t\right]\\[.15truecm]\notag
&\quad\quad\left[y_0^*(u)y_0(x)-y_0^*(x)y_0(u)\right]\\[.2truecm]
&\equiv y_{311} + y_{312} \,.
\end{align}
\end{subequations}
These two terms correspond to
\begin{subequations}
\begin{align}
y_{311} =& \frac{g_1^3}{27}\left[\frac{4 i e^{i x}}{9 x}-\left(\frac{2}{9}+\frac{10 i}{9 x}\right) e^{-i x}\int_x^\infty\,\frac{\dd u}{u}\,e^{2iu} +\frac{2}{3} \left(-1+\frac{2 i}{x}\right) e^{i x}\int_x^\infty\,\frac{\dd u}{u}\,e^{-2iu}\int_u^\infty\,\frac{\dd t}{t}\,e^{2it} \right. \notag\\
&\left.-\left(1-\frac{i}{x}\right) e^{-i x}\int_x^\infty\,\frac{\dd u}{u}\,e^{2iu}\int_u^\infty\,\frac{\dd t}{t}\,e^{-2it}\int_t^\infty\,\frac{\dd s}{s}\,e^{2is}\right]
\end{align}
and
\begin{align}
y_{312}=&\frac{i\,g_1g_2}{9}\left[-\left(\frac{28}{9x}+\frac{2\ln x}{3x}\right)e^{ix}
-\left(\frac{26i}{9}-\frac{16}{9x}\right)e^{-ix}\int_x^\infty\,\frac{\dd u}{u}\,e^{2iu} \right. \notag\\
&+\left(\frac{5}{3x}-\frac{i}{3}\right)e^{-ix}\int_x^\infty\,\frac{\dd u}{u}\,e^{2iu}\ln u
+\frac{7}{3}\left(\frac{1}{x}-i\right)e^{ix}\int_x^\infty\,\frac{\dd u}{u}\,e^{-2iu}\int_u^\infty\frac{\dd t}{t}\,e^{2it} \notag\\
&+\left.\left(\frac{1}{x}-i\right)e^{ix}\int_x^\infty\,\frac{\dd u}{u}\,e^{-2iu}\int_u^\infty\frac{\dd t}{t}\,e^{2it}\ln t\right] \,.
\end{align}
\end{subequations}
Here, it is convenient to rewrite the last integral as 
\begin{equation}\label{eqn:F01}
    \int_x^\infty\,\frac{\dd u}{u}\,e^{-2iu}\int_u^\infty\frac{\dd t}{t}\,e^{2it}\ln t = 
    \int_x^\infty\,\frac{\dd u}{u}\,e^{-2iu} \int_x^\infty\frac{\dd t}{t}\,e^{2it}\ln t 
    - \int_x^\infty\,\frac{\dd u}{u}\,e^{2iu}\ln u\int_u^\infty\frac{\dd t}{t}\,e^{-2it} \,.
\end{equation}

For the second contribution to the third-order correction, taking into account~\cref{eqn:y_1}, we have
\begin{equation} \label{eqn:y_32}
y_{32}(x)=-\frac{g_1g_2}{6}\int_x^\infty\frac{\dd u}{u^2}\left[\frac{2e^{iu}}{u}+i\,y_0^*(u)\int_u^\infty\frac{\dd t}{t}\,e^{2it}\right]\ln u\left[y_0^*(u)y_0(x)-y_0^*(x)y_0(u)\right] \,.
\end{equation}
The result is
\begin{align}\notag
y_{32}(x)=&\frac{ig_1g_2}{9}\left[
-\left(\frac{28}{9x}+\frac{2\ln x}{3x}\right)e^{i x}-\left(\frac{26i}{9}+\frac{2}{9 x}-\frac{2 \ln x}{x}\right)e^{-i x} \int_x^\infty\,\frac{\dd u}{u}\,e^{2iu}\right.\\ \notag
&\left.-\frac{1}{3}\left(\frac{1}{x}+i\right)e^{-ix}\int_x^\infty\,\frac{\dd u}{u}\,e^{2iu}\ln u+\frac{7}{3}\left(\frac{1}{x}-i\right)e^{ix}\int_x^\infty\,\frac{\dd u}{u}\,e^{-2iu}\int_u^\infty\frac{\dd t}{t}\,e^{2it}\right.\\
&\left.+\left(\frac{1}{x}-i\right)e^{ix}\int_x^\infty\,\frac{\dd u}{u}\,e^{-2iu}\ln u\int_u^\infty\frac{\dd t}{t}\,e^{2it}\right] \,. 
\end{align}

The third contribution to the third-order correction, taking into account~\cref{eqn:y_0}, reads 
\begin{equation} \label{eqn:y_33}
    y_{33}(x) = \frac{i\,g_3}{4}\int_x^\infty\frac{\dd u}{u^2}\left(1+\frac{i}{u}\right)e^{iu}\ln^2u\left[y_0^*(u)y_0(x)-y_0^*(x)y_0(u)\right]\,.
\end{equation}
The result is 
\begin{align}
y_{33}(x) =& \frac{ig_3}{6}\left[\left(\frac{52}{9x}+\frac{16\ln x}{3x}+\frac{2\ln^2x}{x}\right)e^{i x}
+\frac{50}{9}\left(\frac{1}{x}+i\right)e^{-ix}\int_x^\infty\,\frac{\dd u}{u}e^{2iu} \right. \notag\\
&\left.+\frac{14}{3}\left(\frac{1}{x}+i\right)e^{-ix}\int_x^\infty\,\frac{\dd u}{u}\,e^{2iu}\ln u 
+\left(\frac{1}{x}+i\right)e^{-ix}\int_x^\infty\,\frac{\dd u}{u}\,e^{2iu}\ln^2u\right] \,.
\end{align}

We have collected in~\ref{sec:app_int} the integral manipulations, iterating integration by parts, used to obtain the 
results reported here. 

We are interested in the asymptotic form of $y(x)$ in the super-Hubble limit, that is $x \to 0$; in this limit the 
asymptotic forms for $y_0(x)$, $y_0(x)$, $y_{21}(x)$, $y_{22}(x)$, $y_{31}(x)$, $y_{32}(x)$, and $y_{33}(x)$ are 
\begin{subequations}\label{eqn:y_asymptotic}
\begin{align} 
    y_0(x) \to &i x^{-1} \\
    y_1(x) \to &\frac{i g_1}{3} \left[\left(\alpha + \frac{i\pi}{2}\right)x^{-1} - \frac{\ln x}{x}\right] \\
    y_{21}(x) \to &\frac{i g^2_1}{18}\left[\left(i\pi\alpha - \frac{i\pi}{3} + \frac{\pi^2}{4}+\alpha^2-\frac{2}{3}\alpha-4\right)x^{-1}-2\left(\alpha-\frac{1}{3}+\frac{i\pi}{2}\right)\frac{\ln x}{x}+\frac{\ln^2 x}{x}\right] \\
    y_{22}(x) \to &\frac{i g_2}{6}\left[\left(\frac{i\pi}{3}-\frac{\pi^2}{12}+\frac{2\alpha}{3}+i\pi\alpha+\alpha^2\right)x^{-1}
    -\frac{2\ln x}{3x}-\frac{\ln^2 x}{x}\right] \\
    y_{31}(x) \to &\frac{i g_1^3}{27}\left[
    \left(\frac{\alpha ^3}{6} - \frac{\alpha ^2}{3} + \frac{i\pi\alpha ^2}{4} + \frac{\pi ^2 \alpha}{8} - \frac{16 \alpha }{9} - \frac{i\pi  \alpha }{3} - \frac{7\zeta (3)}{3} + 4 - \frac{\pi ^2}{12} + \frac{5i \pi ^3}{48} - \frac{8 i\pi }{9}\right)x^{-1} \right. \notag\\
    &\left. +\left(- \frac{\alpha^2}{2} - \frac{i\pi\alpha}{2} + \frac{2\alpha}{3} + \frac{i\pi}{3} - \frac{\pi^2}{8} + \frac{16}{9} \right) \frac{\ln x}{x} \right.\notag\\
    &\left. +\left(\frac{\alpha }{2} - \frac{1}{3} + \frac{i \pi}{4}\right) \frac{\ln^2 x}{x}  - \frac{\ln^3 x}{6 x}\right] \notag\\
    &+\frac{i g_1 g_2}{6}\left[
    \left(\frac{\alpha^3}{3} - \frac{2\alpha^2}{3} + \frac{i\pi\alpha^2}{6} + \frac{5\pi^2\alpha}{36} + \frac{4i\alpha}{3} - \frac{4\alpha}{27} 
    + \frac{2i\pi\alpha}{3} + \frac{4i}{3} - \frac{16}{9} \right.\right. \notag\\
    &\left.\left. - \frac{7\pi^2}{54} + \frac{i\pi^3}{72} - \frac{38i\pi}{27} + \frac{2\pi}{3}\right)x^{-1}
    - \left(\frac{\alpha ^2}{3} + \frac{2\alpha }{9} + \frac{i\pi\alpha}{3} - \frac{\pi^2}{36} - \frac{4}{27} + \frac{i\pi}{9}\right) \frac{\ln x}{x}  \right. \notag\\
    &\left. + \frac{2\ln ^2 x}{9 x} + \frac{\ln ^3 x}{9 x}\right] \\
    y_{32}(x) \to &\frac{i g_1 g_2}{6}\left[\left(\frac{2\alpha ^2}{3} + \frac{4i\alpha}{3} - \frac{76\alpha}{27}
    + \frac{2i\pi\alpha}{3} + \frac{4i}{3} + \frac{8}{9} - \frac{38i\pi}{27} - \frac{2\pi}{3} + \frac{11\pi^2}{54}\right)x^{-1} \right. \notag\\
    &\left.- \left(\frac{2\alpha}{9} + \frac{i\pi}{9} - \frac{4}{27}\right)\frac{\ln x}{x}
    - \left(\frac{\alpha}{3} + \frac{i\pi}{6} - \frac{2}{9}\right)\frac{\ln^2 x}{x} 
    + \frac{2}{9}\frac{\ln^3 x}{x}\right] \\
  y_{33}(x) \to &\frac{i g_3}{6}\left[\left(\frac{\alpha ^3}{3}+\frac{\alpha ^2}{3}+\frac{i\pi  \alpha ^2}{2}+\frac{2 \alpha}{9}+\frac{i\pi   \alpha}{3}-\frac{\pi ^2 \alpha}{12}-\frac{2 \zeta (3)}{3}\right.\right.\notag\\
  &\left.\left.+\frac{i\pi}{9}+\frac{4}{3}+\frac{i\pi ^3}{24}-\frac{\pi ^2}{36}\right)x^{-1} - \frac{2\ln x}{9 x} - \frac{\ln ^2x}{3 x} - \frac{\ln ^3x}{3 x}\right]  
\end{align}
\end{subequations}
where $\alpha \equiv 2 -\ln 2 -\gamma$ and $\gamma \simeq 0.5772$ is the Euler-Mascheroni constant. We have described 
in~\ref{sec:app_int2} the procedure adopted to derive the super-Hubble solutions of the integrals needed to derive the 
asymptotic solution above.

\subsubsection{The Primordial Power Spectrum of Scalar Perturbations}
The dimensionless PPS of the comoving curvature perturbation $\zeta_k$ is defined as
\begin{equation} \label{eqn:Pzeta}
     {\cal P}_\zeta(k) = \frac{k^3}{2\pi^2} \lim_{x \to 0} \left|\zeta_k\right|^2 
     = \frac{k^3}{2\pi^2} \lim_{x \to 0} \left|\frac{v_k^{\rm (s)}}{z}\right|^2
     = \frac{k^2}{4\pi^2} \lim_{x \to 0} \left|\frac{y^{\rm (s)}}{z}\right|^2 \,.
\end{equation}
We are interested in the evaluation of the PPS at horizon crossing. While we have already expanded the $g_n$ around $x = 1$, 
we have to perform the same expansion for the squared modulus of $1 / (xz)$ entering~\cref{eqn:Pzeta} when writing explicitly 
$y^{\rm (s)}$, that corresponds to  
\begin{align} \label{eqn:xz_exp}
    \left|\frac{1}{xz}\right|^2 \simeq\, & \left(\frac{H^2}{k \dot\phi}\right)^2  
    \left[1 - 2\epsilon_1 - 2\epsilon_1 \epsilon_2 + \epsilon_1^2 - 2\epsilon_1 \epsilon_2^2 - 2\epsilon_1 \epsilon_2 \epsilon_3 \right. \notag\\
    &\left. + \left(2\epsilon_1  + \epsilon_2  + \epsilon_1 \epsilon_2  - 2\epsilon_1^2 - 2\epsilon_1^2\epsilon_2  + \epsilon_1 \epsilon_2^2  + 2\epsilon_1 \epsilon_2 \epsilon_3\right) \ln x \right. \notag \\
    &\left. + \left(2\epsilon_1^2  + \epsilon_1 \epsilon_2 + \frac{\epsilon_2^2}{2} - \frac{\epsilon_2\epsilon_3}{2} + 3\epsilon_1^2 \epsilon_2 + \frac{\epsilon_1\epsilon_2^2}{2}    - \epsilon_1 \epsilon_2 \epsilon_3\right) \ln^2 x \right. \notag \\
    &\left. + \left( \frac{4\epsilon_1^3}{3} + \frac{\epsilon_1\epsilon_2^2}{3}  + \frac{\epsilon_2^3}{6}  -  \frac{2\epsilon_1 \epsilon_2 \epsilon_3}{3}  - \frac{\epsilon_2^2 \epsilon_3}{2}  + \frac{\epsilon_2 \epsilon_3^2}{6}  + \frac{\epsilon_2 \epsilon_3 \epsilon_4}{6} \right)\ln^3 x \right].
\end{align}
Finally, we expand the HFF and the Hubble parameter around a conformal time $\tau_*$ as 
\begin{align} \label{eqn:eps_exp}
\epsilon_i(\tau) \simeq\, & \epsilon_{i*}\left[1 - \epsilon_{i+1*} (1+\epsilon_{1*} + \epsilon^2_{1*} + \epsilon_{1*}\epsilon_{2*}) \ln\left(\frac{\tau}{\tau_*}\right) \right.\notag \\ 
&\left. + \frac{\epsilon_{i+1*}}{2} (\epsilon_{i+1*} + \epsilon_{i+2*} +\epsilon_{1*}\epsilon_{2*} + 2\epsilon_{1*}\epsilon_{i+1*} + 2\epsilon_{1*}\epsilon_{i+2*})\ln^2\left(\frac{\tau}{\tau_*}\right) \right.\notag \\
&\left. -\frac{\epsilon_{i+1*}}{6} \left(\epsilon_{i+1*}^2 + \epsilon_{i+2*}^2 + 3\epsilon_{i+1*}\epsilon_{i+2*} + \epsilon_{i+2*}\epsilon_{i+3*}\right) \ln^3\left(\frac{\tau}{\tau_*}\right)\right] \,, \\
H(\tau) \simeq\, & H_* \Bigg[1 + \left(\epsilon_{1*} + \epsilon_{1*}^2 + \epsilon_{1*}^3 + \epsilon_{1*}^2 \epsilon_{2*}\right) \ln\left(\frac{\tau}{\tau_*}\right) \notag \\
&+ \left(\frac{\epsilon_{1*}^2}{2} + \epsilon_{1*}^3 - \frac{1}{2} \epsilon_{1*} \epsilon_{2*} - \frac{3}{2} \epsilon_{1*}^2 \epsilon_{2*}\right) \ln^2\left(\frac{\tau}{\tau_*}\right) \notag \\
&+ \left(\frac{\epsilon_{1*}^3}{6} - \frac{1}{2} \epsilon_{1*}^2 \epsilon_{2*} + \frac{1}{6} \epsilon_{1*} \epsilon_{2*}^2 + \frac{1}{6} \epsilon_{1*} \epsilon_{2*} \epsilon_{3*}\right) \ln^3\left(\frac{\tau}{\tau_*}\right) \Bigg] \,.
\end{align}
Actually, in order to compute the scalar PPS, we need the squared expansion of the Hubble parameter, which reads
\begin{align} \label{eqn:H2_exp}
H^2(\tau) \simeq H^2_*\Bigg[1 &+ \left(2 \epsilon_{1*} + 2 \epsilon_{1*}^2 + 2 \epsilon_{1*}^3 + 2 \epsilon_{1*}^2 \epsilon_{2*}\right) \ln\left(\frac{\tau}{\tau_*}\right) \notag \\
&+ \left(2 \epsilon_{1*}^2 + 4 \epsilon_{1*}^3 - \epsilon_{1*} \epsilon_{2*} - 3 \epsilon_{1*}^2 \epsilon_{2*}\right) \ln^2\left(\frac{\tau}{\tau_*}\right) \notag \\
&+ \left(\frac{4 \epsilon_{1*}^3}{3} - 2 \epsilon_{1*}^2 \epsilon_{2*} + \frac{1}{3} \epsilon_{1*} \epsilon_{2*}^2 + \frac{1}{3} \epsilon_{1*} \epsilon_{2*} \epsilon_{3*}\right) \ln^3\left(\frac{\tau}{\tau_*}\right) \Bigg] \,.
\end{align}

By combining~\cref{eqn:Pzeta} with~\cref{eqn:y,eqn:g_s,eqn:xz_exp,eqn:eps_exp,eqn:H2_exp}, the dimensionless PPS for scalar 
fluctuation expanded around the pivot scale $k_*$ reads
\begin{align} \label{eqn:Pzeta_sol}
{\cal P}_\zeta(k) = & \frac{H^2_*}{8\pi^2 \epsilon_{1*}}\biggl\{ \biggl[ 1 - 2 (1 - \alpha) \epsilon_{1*} + \left( -3 - 2 \alpha + 2 \alpha^2 + \frac{\pi^2}{2} \right) \epsilon_{1*}^2 + \alpha \epsilon_{2*} \notag \\
& + \left( -6 + \alpha + \alpha^2 + \frac{7 \pi^2}{12} \right) \epsilon_{1*} \epsilon_{2*} + \frac{1}{8} \left( -8 + 4 \alpha^2 + \pi^2 \right) \epsilon_{2*}^2 \notag\\
& + \frac{1}{24} \left( -12 \alpha^2 + \pi^2 \right) \epsilon_{2*} \epsilon_{3*} - \frac{1}{24} \left( -16 + 24 \alpha - 4 \alpha^3 - 3 \alpha \pi^2 + 6{\cal Z} \right) (8 \epsilon_{1*}^3 + \epsilon_{2*}^3)\notag \\
& + \frac{1}{12} \left( -72 \alpha + 36 \alpha^2 + 13 \pi^2 + 8 \alpha \pi^2 - 36{\cal Z} \right) \epsilon_{1*}^2 \epsilon_{2*} \notag\\
& - \frac{1}{24} \left( 16 + 24 \alpha - 12 \alpha^2 - 8 \alpha^3 - 15 \pi^2 - 6 \alpha \pi^2 + 84{\cal Z} \right) \epsilon_{1*} \epsilon_{2*}^2 \notag\\
& + \frac{1}{24} \left( 16 + 4 \alpha^3 - \alpha \pi^2 - 24{\cal Z} \right) ( \epsilon_{2*} \epsilon_{3*}^2 + \epsilon_{2*} \epsilon_{3*} \epsilon_{4*}) \notag\\
& + \frac{1}{24} \left( 48 \alpha - 12 \alpha^3 - 5\alpha \pi^2 \right) \epsilon_{2*}^2 \epsilon_{3*} \notag\\
& + \frac{1}{12} \left( -8 + 72 \alpha - 12 \alpha^2 - 8\alpha^3 + \pi^2 - 6\alpha \pi^2 - 24{\cal Z} \right) \epsilon_{1*} \epsilon_{2*} \epsilon_{3*} \biggr] \notag \\
&+ \biggl[-2 \epsilon_{1*} + 2 (-2\alpha + 1) \epsilon_{1*}^2 - \epsilon_{2*} + (-2\alpha - 1) \epsilon_{1*} \epsilon_{2*} - \alpha \epsilon_{2*}^2 + \alpha \epsilon_{2*} \epsilon_{3*} \notag\\
& - \frac{1}{8} \left( -8 + 4 \alpha^2 + \pi^2 \right) (8 \epsilon_{1*}^3 + \epsilon_{2*}^3) - \frac{2}{3} \left( -9 + 9\alpha + \pi^2 \right) \epsilon_{1*}^2 \epsilon_{2*} \notag\\
& - \frac{1}{4} \left( -4 + 4 \alpha + 4 \alpha^2 + \pi^2 \right) \epsilon_{1*} \epsilon_{2*}^2 + \frac{1}{2} \left( -12 + 4\alpha + 4\alpha^2 + \pi^2 \right) \epsilon_{1*} \epsilon_{2*} \epsilon_{3*} \notag\\
& + \frac{1}{24} \left( -12\alpha^2 + \pi^2 \right) ( \epsilon_{2*} \epsilon_{3*}^2 + \epsilon_{2*} \epsilon_{3*} \epsilon_{4*}) + \frac{1}{24} \left( -48 + 36\alpha^2 + 5 \pi^2 \right) \epsilon_{2*}^2 \epsilon_{3*}  \biggr]  \ln\left(\frac{k}{k_*}\right) \notag \\
& +\frac{1}{2}\biggl[4 \epsilon_{1*}^2 + 2 \epsilon_{1*} \epsilon_{2*} + 6 \epsilon_{1*}^2 \epsilon_{2*} + \epsilon_{2*}^2 - \epsilon_{2*} \epsilon_{3*}  + (1 + 2\alpha) (\epsilon_{1*} \epsilon_{2*}^2 - 2 \epsilon_{1*} \epsilon_{2*} \epsilon_{3*})\notag \\
& + \alpha (8 \epsilon_{1*}^3 + \epsilon_{2*}^3 - 3 \epsilon_{2*}^2 \epsilon_{3*} + \epsilon_{2*} \epsilon_{3*}^2 + \epsilon_{2*} \epsilon_{3*} \epsilon_{4*}) \biggr]\ln^2\left(\frac{k}{k_*}\right) \notag \\
& +\frac{1}{6} \left(-8 \epsilon_{1*}^3 - 2 \epsilon_{1*} \epsilon_{2*}^2 - \epsilon_{2*}^3 + 4 \epsilon_{1*} \epsilon_{2*} \epsilon_{3*}  + 3 \epsilon_{2*}^2 \epsilon_{3*} - \epsilon_{2*} \epsilon_{3*}^2 - \epsilon_{2*} \epsilon_{3*} \epsilon_{4*}\right) \ln^3\left(\frac{k}{k_*}\right)
\biggr\}
\end{align}
where all the terms divergent in~\cref{eqn:y_asymptotic} proportional to $\ln x$ cancel exactly. ${\cal Z}$ is a constant encoding 
the difference due to the approximation scheme used to calculate the triple integral appearing in~\cref{eqn:y_31}, see~\ref{sec:app_int2}. Here, we have written the scalar PPS with respect to a fixed pivot scale $k_*$, using 
$k_* \tau_* = -1$ leading to 
\begin{equation}
    \frac{\tau}{\tau_*} = - \frac{k}{k_*} \,.
\end{equation}

From~\cref{eqn:Pzeta_sol}, we can derive the third-order slow-roll expansion for the scalar spectral index $n_{\rm s}$ and 
its runnings $\alpha_{\rm s}$ and $\beta_{\rm s}$. We can define the scalar spectral index as
\begin{align}
    n_{\rm s}(k) \equiv\,& 1 + \frac{\dd \ln {\cal P}_\zeta}{\dd \ln k} \\
    =\,& \biggl[1-2 \epsilon_{1*} - \epsilon_{2*} - 2 \epsilon_{1*}^2 - (3 - 2\alpha) \epsilon_{1*} \epsilon_{2*} + \alpha \epsilon_{2*} \epsilon_{3*} - 2 \epsilon_{1*}^3 \notag \\
    & + (-15 + 6\alpha + \pi^2) \epsilon_{1*}^2 \epsilon_{2*} + \frac{1}{12} (-84 + 36\alpha - 12\alpha^2 + 7\pi^2) \epsilon_{1*} \epsilon_{2*}^2 \notag \\
    & + \frac{1}{12} (-72 + 48\alpha - 12\alpha^2 + 7\pi^2) \epsilon_{1*} \epsilon_{2*} \epsilon_{3*} + \frac{1}{4} (-8 + \pi^2) \epsilon_{2*}^2 \epsilon_{3*} \notag \\
& + \left( -\frac{\alpha^2}{2} + \frac{\pi^2}{24} \right) \epsilon_{2*} \epsilon_{3*}^2 + \left( -\frac{\alpha^2}{2} + \frac{\pi^2}{24} \right) \epsilon_{2*} \epsilon_{3*} \epsilon_{4*} \biggr] \notag \\
& + \left( -2 \epsilon_{1*} \epsilon_{2*} - 6 \epsilon_{1*}^2 \epsilon_{2*} - 3 \epsilon_{1*} \epsilon_{2*}^2 + 2\alpha \epsilon_{1*} \epsilon_{2*}^2 - \epsilon_{2*} \epsilon_{3*} - 4 \epsilon_{1*} \epsilon_{2*} \epsilon_{3*} \right. \notag \\
&\left. +  2\alpha \epsilon_{1*} \epsilon_{2*} \epsilon_{3*} + \alpha \epsilon_{2*} \epsilon_{3*}^2 + \alpha \epsilon_{2*} \epsilon_{3*} \epsilon_{4*} \right) \ln\left(\frac{k}{k_*}\right) \notag \\
& + \left( -\epsilon_{1*} \epsilon_{2*}^2 - \epsilon_{1*} \epsilon_{2*} \epsilon_{3*} - \frac{1}{2} \epsilon_{2*} \epsilon_{3*}^2 - \frac{1}{2} \epsilon_{2*} \epsilon_{3*} \epsilon_{4*}  \right) \ln^2\left(\frac{k}{k_*}\right) \,.
\end{align}
The running of the scalar spectral index \cite{Kosowsky:1995aa} reads
\begin{align}
    \alpha_{\rm s}(k) \equiv\, & \frac{\dd n_{\rm s}}{\dd \ln k} \\
    =\, & \left[-2 \epsilon_{1*} \epsilon_{2*} - \epsilon_{2*} \epsilon_{3*} - 6 \epsilon_{1*}^2 \epsilon_{2*}  + (-3 + 2\alpha) \epsilon_{1*} \epsilon_{2*}^2 - 2(2 - \alpha) \epsilon_{1*} \epsilon_{2*} \epsilon_{3*} + \alpha \epsilon_{2*} \epsilon_{3*}^2 \right. \notag \\
&\left. + \alpha \epsilon_{2*} \epsilon_{3*} \epsilon_{4*} \right]+ \left(-2\epsilon_{1*} \epsilon_{2*}^2 - 2\epsilon_{1*} \epsilon_{2*} \epsilon_{3*} - \epsilon_{2*} \epsilon_{3*}^2 - \epsilon_{2*} \epsilon_{3*} \epsilon_{4*}  \right) \ln\left(\frac{k}{k_*}\right) \,.
\end{align}
The running of the running of the scalar spectral index is given by
\begin{align}
    \beta_{\rm s}(k) \equiv\, & \frac{\dd \alpha_{\rm s}}{\dd \ln k} \\
    =\, & -2 \epsilon_{1*} \epsilon_{2*}^2 - 2 \epsilon_{1*} \epsilon_{2*} \epsilon_{3*}  - \epsilon_{2*} \epsilon_{3*}^2 - \epsilon_{2*} \epsilon_{3*} \epsilon_{4*} \,.
\end{align}

\subsubsection{The Primordial Power Spectrum of Tensor Perturbations}
The dimensionless PPS for the two polarisation of gravitational waves generated during inflation, defined as
\begin{equation}
    {\cal P}_{\rm t}(k) \equiv 2{\cal P}_{\rm h}(k) = \frac{2k^3}{\pi^2} \lim_{x \to 0} \left| h_k \right|^2 = \frac{2k^3}{\pi^2} \lim_{x \to 0} \left| \frac{v_k^{(\rm t)}}{a} \right|^2 
    = \frac{k^2}{\pi^2} \lim_{x \to 0} \left| \frac{y^{(\rm t)}}{a} \right|^2 \,,
\end{equation}
can be easily calculated following the same procedure described before and substituting~\cref{eqn:g_t} to the asymptotic 
solutions in~\cref{eqn:y_asymptotic}, where for primordial tensor fluctuations we need 
\begin{align}
    \left|\frac{1}{xa}\right|^2 \simeq\, & \left(\frac{H}{k}\right)^2 \left[1 - 2\epsilon_1 - 2\epsilon_1 \epsilon_2 + \epsilon_1^2 - 2\epsilon_1 \epsilon_2^2 - 2\epsilon_1 \epsilon_2 \epsilon_3 \right. \notag \\ 
    &\left. +\left( 2\epsilon_1  +2 \epsilon_1 \epsilon_2  - 2\epsilon_1^2 - 2\epsilon_1^2\epsilon_2  + 2\epsilon_1 \epsilon_2^2  + 2\epsilon_1 \epsilon_2 \epsilon_3 \right)\ln x \right.\notag \\
    &\left. + \left(  2\epsilon_1^2  -  \epsilon_1 \epsilon_2 +3\epsilon_1^2\epsilon_2 -  \epsilon_1 \epsilon_2^2   -  \epsilon_1 \epsilon_2 \epsilon_3 \right)\ln^2 x \right.\notag \\ 
    &\left. + \frac{1}{3}\left( 4\epsilon_1^3  - 6\epsilon_1^2 \epsilon_2  +  \epsilon_1 \epsilon_2^2  +  \epsilon_1 \epsilon_2 \epsilon_3 \right)\ln^3 x \right] \,.
\end{align}
Results for the dimensionless PPS, tensor spectral index, running of the tensor spectral index, and running of the running 
of the tensor spectral index, expanded around a pivot scale $k_*$, are reported below. We find respectively 
\begin{align} \label{eqn:Pt_sol}
     {\cal P}_{\rm t}(k) =\, & \frac{2H^2_*}{\pi^2} \biggl\{
\biggl[ 1 + (-2 + 2\alpha) \epsilon_{1*} 
+ \frac{1}{2} (-6 - 4\alpha + 4\alpha^2 + \pi^2) \epsilon_{1*}^2  \notag \\ 
& + \left(-2 + 2\alpha - \alpha^2 + \frac{\pi^2}{12}\right) \epsilon_{1*} \epsilon_{2*} 
- \frac{1}{3} (-16 + 24\alpha - 4\alpha^3 - 3\alpha \pi^2 + 6{\cal Z}) \epsilon_{1*}^3 \notag \\ 
& + \frac{1}{12} (-96 + 72\alpha + 36\alpha^2 - 24\alpha^3 + 13 \pi^2 - 10\alpha \pi^2) \epsilon_{1*}^2 \epsilon_{2*}  \notag \\ 
&  - \frac{1}{12} (8 - 24\alpha + 12\alpha^2 - 4\alpha^3 - \pi^2 +\alpha \pi^2 + 24{\cal Z}) (\epsilon_{1*} \epsilon_{2*}^2 + \epsilon_{1*} \epsilon_{2*} \epsilon_{3*}) \biggr]  \notag \\  
&+ \biggl[-2 \epsilon_{1*} + 2 \epsilon_{1*}^2 - 
   4\alpha \epsilon_{1*}^2 + (-2 + 2\alpha) \epsilon_{1*} \epsilon_{2*} - (-8 + 4\alpha^2 + \pi^2) \epsilon_{1*}^3 \notag \\ & 
   + \frac{1}{6} (-36 - 36\alpha + 36\alpha^2 + 5 \pi^2) \epsilon_{1*}^2 \epsilon_{2*} \notag\\
   &+ 
   \frac{1}{12} (-24 + 24\alpha - 12\alpha^2 + \pi^2) (\epsilon_{1*} \epsilon_{2*}^2 + \epsilon_{1*} \epsilon_{2*} \epsilon_{3*})\biggr] \ln\left(\frac{k}{k_*}\right)  \notag \\
  & +\left[2 \epsilon_{1*}^2 + 4\alpha \epsilon_{1*}^3 - 
   \epsilon_{1*} \epsilon_{2*} + (3 - 6\alpha) \epsilon_{1*}^2 \epsilon_{2*} - (-\alpha + 1) (\epsilon_{1*} \epsilon_{2*}^2 + \epsilon_{1*} \epsilon_{2*} \epsilon_{3*})\right] \ln^2\left(\frac{k}{k_*}\right)  \notag \\
   &+ \frac{1}{3} \left(-4 \epsilon_{1*}^3 + 6 \epsilon_{1*}^2 \epsilon_{2*} - \epsilon_{1*} \epsilon_{2*}^2 - \epsilon_{1*} \epsilon_{2*} \epsilon_{3*}\right) \ln^3\left(\frac{k}{k_*}\right)\biggr\} \,,
\end{align}

\begin{align}
    n_{\rm t}(k) =\, & -2 \epsilon_{1*} - 2 \epsilon_{1*}^2 - 2 (1 - \alpha) \epsilon_{1*} \epsilon_{2*} - 2 \epsilon_{1*}^3 + (-14 + 6\alpha + \pi^2) \epsilon_{1*}^2 \epsilon_{2*} \notag \\
& + \left(-2 + 2\alpha - \alpha^2 + \frac{\pi^2}{12}\right) (\epsilon_{1*} \epsilon_{2*}^2 + \epsilon_{1*} \epsilon_{2*} \epsilon_{3*}) \notag \\
& + \left(-2 \epsilon_{1*} \epsilon_{2*} - 6 \epsilon_{1*}^2 \epsilon_{2*} - 2 \epsilon_{1*} \epsilon_{2*}^2 + 2\alpha \epsilon_{1*} \epsilon_{2*}^2 - 2 \epsilon_{1*} \epsilon_{2*} \epsilon_{3*} + 2\alpha \epsilon_{1*} \epsilon_{2*} \epsilon_{3*}\right) \ln\left(\frac{k}{k_*}\right) \notag \\
& + \left(-\epsilon_{1*} \epsilon_{2*}^2 - \epsilon_{1*} \epsilon_{2*} \epsilon_{3*}\right) \ln^2\left(\frac{k}{k_*}\right) \,,
\end{align}

\begin{align}
\alpha_{\rm t}(k) =\, & -2 \epsilon_{1*} \epsilon_{2*} - 6 \epsilon_{1*}^2 \epsilon_{2*} - 2 (1 - \alpha) \epsilon_{1*} \epsilon_{2*}^2 - 2 (1 - \alpha) \epsilon_{1*} \epsilon_{2*} \epsilon_{3*} \notag \\
& + 2 \left(-\epsilon_{1*} \epsilon_{2*}^2 - \epsilon_{1*} \epsilon_{2*} \epsilon_{3*}\right) \ln\left(\frac{k}{k_*}\right) \,,
\end{align}

\begin{equation}
    \beta_{\rm t}(k) = -2 \epsilon_{1*} \epsilon_{2*}^2 - 2 \epsilon_{1*} \epsilon_{2*} \epsilon_{3*} \,.
\end{equation}

Our results for the dimensionless PPS of scalar and tensor perturbations calculated at third order in the slow-roll 
expansion, obtained using the integrals and their super-Hubble limits computed and reported in~\ref{sec:app_int} and 
in~\ref{sec:app_int2} respectively, agree with the ones previously obtained in Ref.~\cite{Auclair:2022yxs} with some 
negligible differences on the numerical coefficients of the constant part of~\cref{eqn:Pzeta_sol,eqn:Pt_sol}. 
In~\ref{app:an_bn}, we report alternative expressions for the PPS.

\section{Accuracy of Slow-Roll Analytic Power Spectra Against the Exact Numerical Solution} \label{sec:numeric}
In this section, we present a comparison between the analytical results obtained above in~\cref{sec:analytic} at different 
orders in the slow-roll expansion to the numerical solution for the PPS obtained for two different single-field slow-roll 
inflationary models.

In order to do so, we numerically solve the coupled system of background and perturbation equations corresponding the the 
Friedmann equation for a spatially flat universe dominated by a scalar field $\phi$, the Klein-Gordon equation governing the 
background dynamics of a scalar field with standard kinetic term and minimally coupled to gravity, the Mukhanov-Sasaki and 
the tensor perturbation equations describing the evolution of scalar and tensor primordial perturbations. 
We describe in the following the adopted strategy to ensure numerical stability.

\subsection{Background Dynamics}
We time-evolve the background equations in~\cref{eqn:dynamicalsystem} in number of $e$-folds 
\begin{equation}
    N = \int_{t_i}^t \dd t \  H(t) = \ln \left[\frac{a(t)}{a(t_i)}\right] \,.
\end{equation}
In addition, we rewrite the system of background equations adopting the field redefinition $\psi = \dot\phi$. This allows to 
reduce the background to a system of first-order differential equations. Under these assumptions, the background system becomes
\begin{equation}
    \begin{cases}
\phi_N \equiv \frac{\psi }{H} = \psi M_{\text{Pl}} \sqrt{\frac{3}{\frac{\psi^2}{2}+ V(\phi)}} \\ \psi_N = -3\psi - V_\phi M_{\text{Pl}}\sqrt{\frac{3}{\frac{\psi^2}{2}+ V(\phi)}}
\end{cases} \ ,
\end{equation}
where the subscript $N$ indicates the derivative with respect to $N$.

\subsection{Perturbation Equations}
We also solve the Mukhanov-Sasaki equation and the tensor perturbation equation in number of $e$-folds. 
Eqs.~\cref{eqn:mukhanov_1} become 
\begin{equation}
    v^{\rm (s,t)}_{k\,NN} = [\epsilon_1(N) - 1] v^{\rm (s,t)}_{k\,N} + \frac{1}{a^2H^2}\left[U^{\rm (s,t)}(N) - k^2\right] v^{\rm (s,t)}_k \,,
\end{equation}
where
\begin{align}
    &U^{\rm (s)}(N) = \frac{z''}{z} = a^2H^2 \left( 2-\epsilon_1 +\frac{3\epsilon_2}{2} -\frac{\epsilon_1 \epsilon_2}{2}  + \frac{\epsilon^2_2}{4}  + \frac{\epsilon_2\epsilon_3}{2}\right) \,,\\
    &U^{\rm (t)}(N) = \frac{a''}{a} = a^2H^2 \left( 2-\epsilon_1\right) \,.
\end{align}
We solve separately the perturbation equations for the real and imaginary part of the mode functions as 
$v^{\rm (s,t)}_k = {\rm Re}\left[v^{\rm (s,t)}_k\right] + i\, {\rm Im}\left[v^{\rm (s,t)}_k\right]$ and we combine them 
back only at the end.

\subsection{Initial Conditions}
Concerning the initial conditions, for the background we set the initial value of the scalar field $\phi_i$ to ensure that 
there are more than 55 $e$-folds between the mode $k_* = 0.05\, {\rm Mpc}^{-1}$ crosses the Hubble radius and the end of 
inflation. The initial value of $\dot\phi$ is determined using slow-roll initial condition corresponding to 
$\phi_N|_i \simeq -M_{\rm Pl}^2 V_\phi/V|_i$.

For the perturbation equations, we impose Bunch-Davies initial conditions a mode at a time when $k = 10^2\,a H$. The value 
$10^2$ is just a choice to ensure that all the modes evolved are well within the Hubble horizon at the beginning of their 
evolution, meaning that we can safety impose the Bunch-Davies vacuum. Modes are then evolved until they became super Hubble 
for $k = 10^{-3}\, a H$. 
The value $10^{-3}$ ensures that all the modes evolved leave the horizon before their evolution ends. These values define 
the interval of integration for the number of $e$-folds. Bunch-Davies initial conditions for the real and the imaginary part 
of the mode function reads
\begin{equation}
   {\rm Re}\left[v^{\rm (s,t)}_k\right] = \frac{1}{\sqrt{2k}} \,, \qquad {\rm Re}\left[v^{\rm(s,t)}_{k\,N}\right] = 0 \,,
\end{equation}
and
\begin{equation}
   {\rm Im}\left[v^{\rm(s,t)}_k\right] = 0 \,, \qquad {\rm Im}\left[v^{\rm(s,t)}_{k\,N}\right] = -\frac{1}{aH}\sqrt{\frac{k}{2}} \,.
\end{equation}

Since curvature and tensor perturbations freeze at super-Hubble scales, it is sufficient to calculate the power spectrum for 
the modes when they are at super-Hubble scales (after horizon crossing) instead of calculating it at the end of inflation.

\subsection{Inflationary Models}
We study the numerical solution for the scalar and tensor perturbations for the two following single-field slow-roll 
inflationary potentials
\begin{align}
    &V_{\rm T-model}(\phi) = V_0 \tanh^2\left({\frac{\phi}{\sqrt{6\alpha}}}\right) \,,
    &V_{\rm KKLT}(\phi) = V_0 \left(1 + \frac{m}{|\phi|} \right)^{-1} \,.
\end{align}
The first one corresponds to the T-model of $\alpha$-attractor inflation 
\cite{Kallosh:2013tua,Ferrara:2013rsa,Kallosh:2013yoa,Kallosh:2014rga,Kallosh:2014laa,Galante:2014ifa} and the second 
corresponds to the the inverse linear case of KKLT inflation \cite{Kachru:2003aw} associated to D6-$\overline{\rm D6}$ 
potential in type IIA string theory \cite{Kallosh:2018nrk,Blaback:2018hdo}.

For the numerical solution, we solve the background equations for a given potential and we insert the solution for the scalar 
field background into the HFFs entering the perturbation equations 
\begin{align}\label{eqn:epsilon_num1}
    &\epsilon_1(N) = \frac{\phi^2_N}{2 M^2_{\text{Pl}}} \,,\\
    &\epsilon_2(N) = \frac{\dd\ln |\epsilon_1|}{\dd N}=  \frac{2\phi_{NN}}{\phi_N} \,,\\
    &\epsilon_3(N) = \frac{\dd\ln |\epsilon_2|}{\dd N}= \frac{\phi_{NNN}}{\phi_{NN}}-\frac{\phi_{NN}}{\phi_{N}} \,.  
    \label{eqn:epsilon_num4}
\end{align}

For the analytical solution, we calculate the HFFs as functions of the potential slow-roll parameters \cite{Liddle:1994dx} 
as explained in details in Refs.~\cite{Vennin:2014xta,Martin:2024qnn}. We start from the LO expressions 
\begin{align}
    \epsilon_1^{\rm LO} \simeq&\, \frac{M_{\text{Pl}}^2}{2} \left( \frac{V_\phi}{V} \right)^2 \,, \label{eqn:eps1LO} \\
    \epsilon_2^{\rm LO} \simeq&\, 2 M_{\text{Pl}}^2 \left( \frac{V_\phi^2}{V^2} - \frac{V_{\phi\phi}}{V} \right) \,, \label{eqn:eps2LO} \\
    \epsilon_3^{\rm LO} \simeq&\, \frac{2 M_{\text{Pl}}^4}{\epsilon_2} \left(
\frac{2 V_\phi^4}{V^4}
- \frac{3 V_\phi^2 V_{\phi\phi}}{V^3}
+ \frac{V_\phi V_{\phi\phi\phi}}{V^2}
\right) \,,\label{eqn:eps3LO} \\
    \epsilon_4^{\rm LO} \simeq&\, \frac{2 M_{\text{Pl}}^6}{\epsilon_2 \epsilon_3} \bigg[ 
     \frac{8 V_\phi^6}{V^6} - \frac{17 V_\phi^4 V_{\phi\phi}}{V^5} + \frac{6 V_\phi^2 V_{\phi\phi}^2}{V^4} 
     + \frac{5 V_\phi^3 V_{\phi\phi\phi}}{V^4} - \frac{V_\phi V_{\phi\phi} V_{\phi\phi\phi}}{V^3} - \frac{V_\phi^2 V_{\phi\phi\phi\phi}}{V^3} + \nonumber \\
    & + \frac{2 M_{\text{Pl}}^2}{\epsilon_2} \bigg( - \frac{4 V_\phi^8}{V^8} + \frac{12 V_\phi^6 V_{\phi\phi}}{V^7} 
     - \frac{9 V_\phi^4 V_{\phi\phi}^2}{V^6} - \frac{4 V_\phi^5 V_{\phi\phi\phi}}{V^6} + \frac{6 V_\phi^3 V_{\phi\phi} V_{\phi\phi\phi}}{V^5} 
     - \frac{V_\phi^2 V_{\phi\phi\phi}^2}{V^4} \bigg) \bigg] \,. \label{eqn:eps4LO}
\end{align}
These equations have been derived by approximating at LO the Hubble parameter as $H^2 \simeq V/(3M_{\rm Pl})$, which leads 
to the LO formula of the derivative with respect to $N$ \cite{Vennin:2014xta}
\begin{equation}
    \frac{\dd}{\dd N}\Bigg|_{\rm LO} = - M_{\rm Pl}^2 \frac{V_\phi}{V} \frac{\dd}{\dd \phi} \,.
\end{equation}
To compare the analytical and numerical solutions, we need to include corrections up to the third order in terms of potential 
slow-roll parameters. 
The equations at third order have been already calculated in Ref.~\cite{Martin:2024qnn}, we report here the results 
in terms of the LO HFFs
\begin{align} \label{eqn:eps1N3LO}
\epsilon_1^{\rm N3LO} &=\left.\left( \epsilon_1 - \frac{1}{3} \epsilon_1 \epsilon_2 - \frac{1}{9} \epsilon_1^{2} \epsilon_2 + \frac{5}{36} \epsilon_1 \epsilon_2^{2} + \frac{1}{9} \epsilon_1 \epsilon_2 \epsilon_3 \right)\right|_{\rm LO} \,,
\end{align}
\begin{align}
\epsilon_2^{\rm N3LO} &=\left.\left( \epsilon_2 - \frac{1}{6} \epsilon_2^{2} - \frac{1}{3} \epsilon_2 \epsilon_3 - \frac{1}{6} \epsilon_1 \epsilon_2^{2}  + \frac{1}{18} \epsilon_2^{3} - \frac{1}{9} \epsilon_1 \epsilon_2 \epsilon_3 + \frac{5}{18} \epsilon_2^{2} \epsilon_3  + \frac{1}{9} \epsilon_2 \epsilon_3^{2} + \frac{1}{9} \epsilon_2 \epsilon_3 \epsilon_4 \right)\right|_{\rm LO} \,, \label{eqn:eps2N3LO}
\end{align}
\begin{align}
\epsilon_3^{\rm N3LO} &=\left( \epsilon_3 - \frac{1}{3} \epsilon_2 \epsilon_3 - \frac{1}{3} \epsilon_3 \epsilon_4 - \frac{1}{6} \epsilon_1 \epsilon_2^{2} - \frac{1}{3} \epsilon_1 \epsilon_2 \epsilon_3 + \frac{1}{6} \epsilon_2^{2} \epsilon_3 + \frac{5}{18} \epsilon_2 \epsilon_3^{2} - \frac{1}{9} \epsilon_1 \epsilon_3 \epsilon_4 \right.\nonumber \\
& \left.\left.\quad + \frac{5}{18} \epsilon_2 \epsilon_3 \epsilon_4 + \frac{1}{9} \epsilon_3^{2} \epsilon_4  + \frac{1}{9} \epsilon_3 \epsilon_4^{2} + \frac{1}{9} \epsilon_3 \epsilon_4 \epsilon_5 \right)\right|_{\rm LO} \,, \label{eqn:eps3N3LO}
\end{align}
\begin{align} \label{eqn:eps4N3LO}
\epsilon_4^{\rm N3LO} &=\left.\left( \epsilon_4 - \frac{1}{3} \epsilon_2 \epsilon_3 - \frac{1}{6} \epsilon_2 \epsilon_4 - \frac{1}{3} \epsilon_4 \epsilon_5 \right)\right|_{\rm LO} \,,
\end{align}
where all the HFFs on the right hand side have been calculated at LO using~\cref{eqn:eps1LO,eqn:eps2LO,eqn:eps3LO,eqn:eps4LO} 
and we set in our analysis $\epsilon_5 = 0$.

In order to relate the HFFs in terms of number of $e$-folds rather than values of the scalar field, we need to solve 
analytically the expression of the classical inflationary trajectory $\phi(N)$. This is done at LO by solving 
\begin{equation}
    N(\phi) \simeq -\frac{1}{M^2_{\text{Pl}} }\int_{\phi}^{\phi_{\rm end}} \frac{V}{V_\phi} \dd \phi \,,
\end{equation}
and by inverting this relation. $\phi_{\rm end}$ is the value of the field at the end of inflation, computed by imposing 
$\epsilon_{1,\,{\rm end}} \equiv \epsilon_1(\phi_{\rm end}) = 1$, namely when the kinetic term start to dominate over the 
potential energy. For T-model inflation we found
\begin{align}
    \phi_*^{\rm T}(N,\,\alpha) =&\, \sqrt{\frac{3\alpha}{2}} {\rm sech}^{-1}\left(\frac{3 \alpha}{\alpha \sqrt{\frac{12}{\alpha}+9} + 4N}\right) \,,\\
    \phi_{\rm end}^{\rm T}(\alpha) =&\, \sqrt{\frac{3 \alpha}{2}} \sinh ^{-1}\left(\frac{2}{\sqrt{3\alpha}}\right) \,,
\end{align}
and for KKLT inflation we find
\begin{align}
    &\phi_*^{\rm KKLT}(N,\,m) = \frac{1}{2} \Bigg\{\Bigg[ 12 N m + \left(\sqrt{2} - m\right) m \sqrt{m \left(2 \sqrt{2} + m\right)} \notag\\
    &\qquad + \sqrt{2m^2 \left[\left(2 \sqrt{2} - 3 m\right) m + 12 N \left(6 N + \left(\sqrt{2} - m\right) \sqrt{m \left(2 \sqrt{2} + m\right)}\right) \right]} \Bigg]^{1/3} \notag\\
    &\qquad - m + m^2 \Bigg[12 N m + \left(\sqrt{2} - m\right) m \sqrt{m \left(2 \sqrt{2} + m\right)} \notag\\
    &\qquad + \sqrt{2m^2 \left[\left(2 \sqrt{2} - 3 m\right) m + 12 N \left(6 N + \left(\sqrt{2} - m\right) \sqrt{m \left(2 \sqrt{2} + m\right)}\right) \right]}\Bigg]^{-1/3}\Bigg\}  \,,\\
    &\phi_{\rm end}^{\rm KKLT}(m) = \frac{1}{2}\left[\sqrt{m\left(2\sqrt{2} + m\right)} - m\right] \,.
\end{align}
Once obtained the trajectory $\phi(N)$, it is possible to get the analytical HFFs at LO written with respect to the number 
of $e$-folds, namely $\epsilon^{\rm LO}_i(N)$. Therefore, from these equations we can obtain the ones at N3LO using 
the~\cref{eqn:eps1N3LO,eqn:eps2N3LO,eqn:eps3N3LO,eqn:eps4N3LO}. Finally, we are able to calculate the scalar and tensor 
PPS by using~\cref{eqn:Pzeta_sol,eqn:Pt_sol}.

We fix the potential amplitude $V_0$ in order to normalise the dimensionless scalar PPS to 
${\cal P}_\zeta(k_*) = 2\times 10^{-9}$ at $k_* = 0.05 \, \text{Mpc}^{-1}$.

The code and the pipeline developed for this section are publicly 
available.\footnote{\href{https://github.com/SirlettiSS/PyPPSinflation.git}{https://github.com/SirlettiSS/PyPPSinflation.git}}

\subsection{Accuracy of Slow-Roll Analytic Spectra}
\begin{figure}[!ht]
    \centering
    \includegraphics[width=\textwidth]{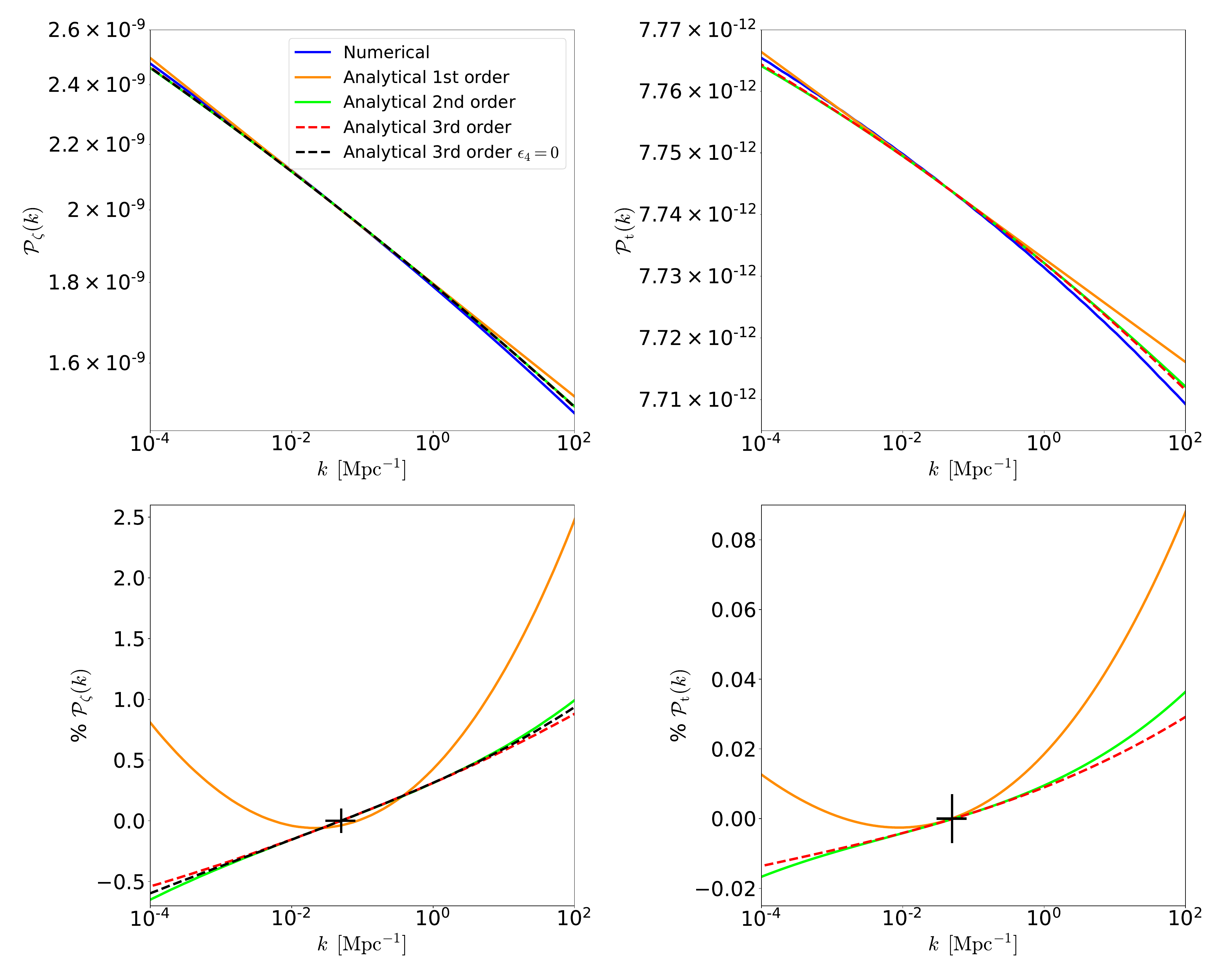}
    \caption{Scalar (left panels) and tensor (right panels) error curves for the T-model $\alpha$-attractor inflation with 
    $\alpha = 1$. The pivot scale crosses the Hubble horizon 55 $e$-folds before the end of inflation. We show on the upper 
    panels the dimensionless PPS and on the lower panels the relative differences of the analytic spectra compared to the 
    numerical solution.}\label{fig:Tmodel_spectra}
\end{figure}

\begin{figure}[!ht]
    \centering
    \includegraphics[width=\textwidth]{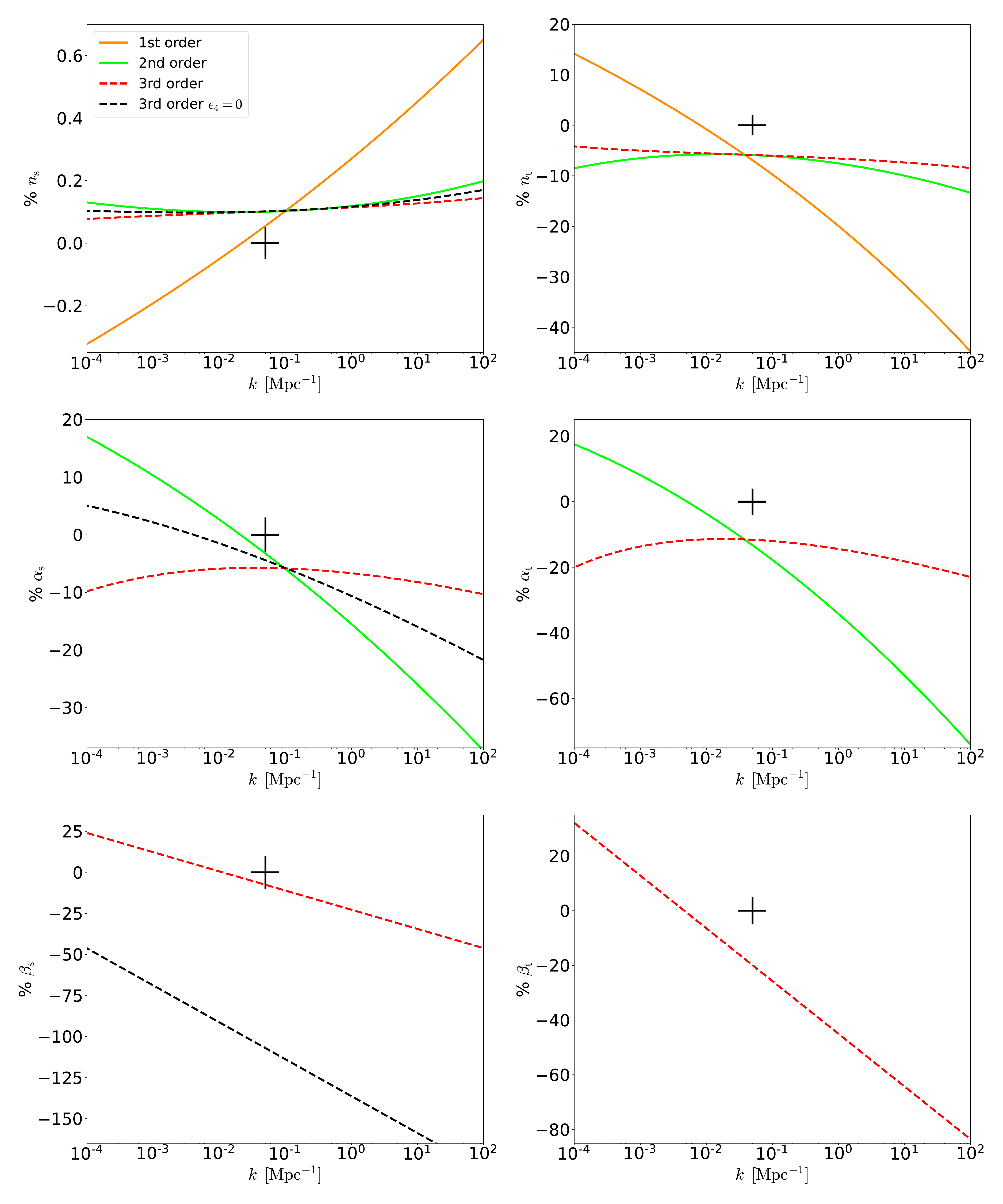}
    \caption{Relative differences with respect to the numerical solution for the spectral index (upper panels), running of 
    the spectral index (central panels), and running of the running of the spectral index (lower panels) for the T-model 
    $\alpha$-attractor inflation with $\alpha = 1$.}\label{fig:Tmodel_parameters}
\end{figure}

\begin{figure}[!ht]
    \centering
    \includegraphics[width=\textwidth]{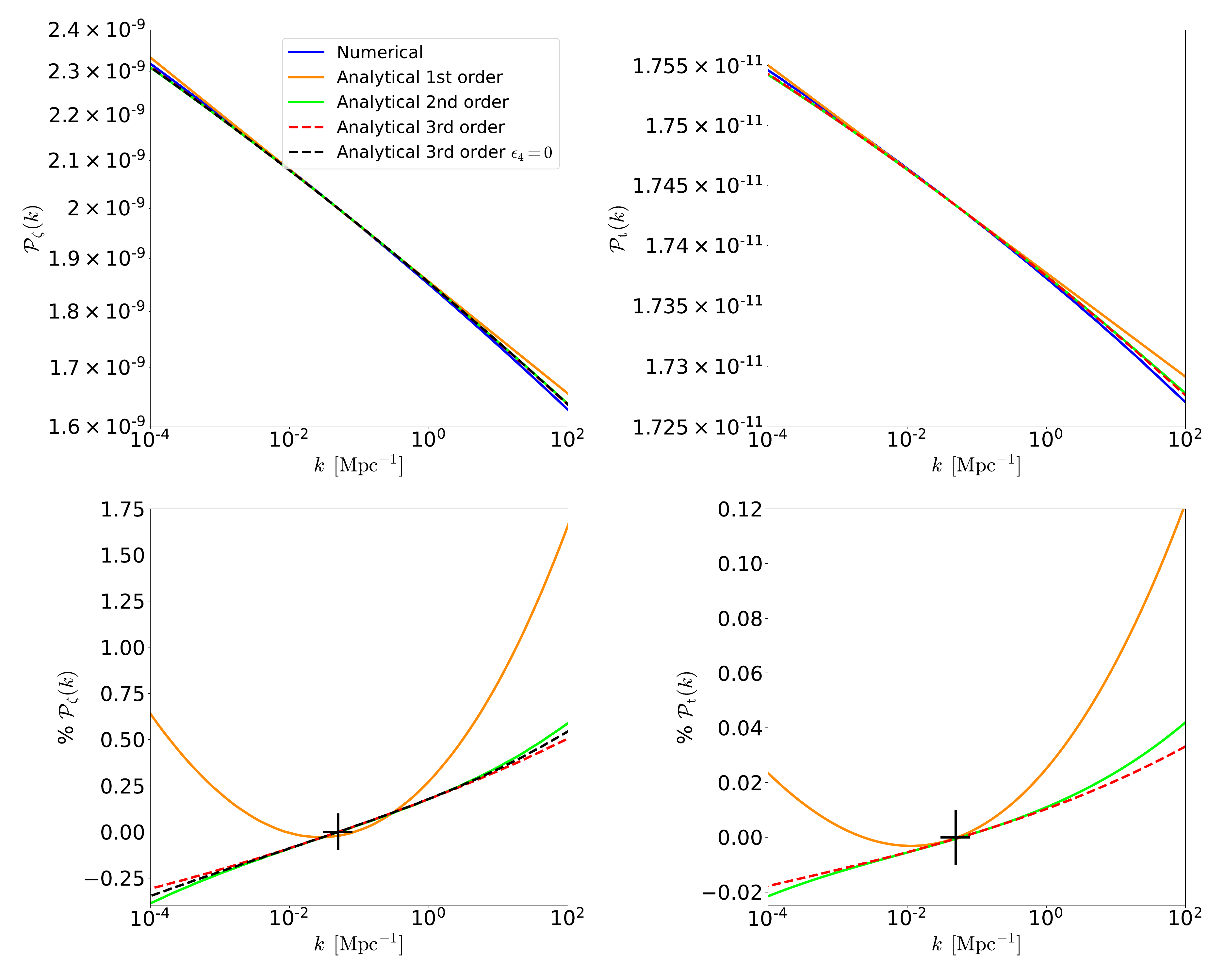}
    \caption{Same as~\cref{fig:Tmodel_spectra} but for KKLT inflation for $m = 1$.}\label{fig:KKLT_spectra}
\end{figure}

\begin{figure}[!ht]
    \centering
    \includegraphics[width=\textwidth]{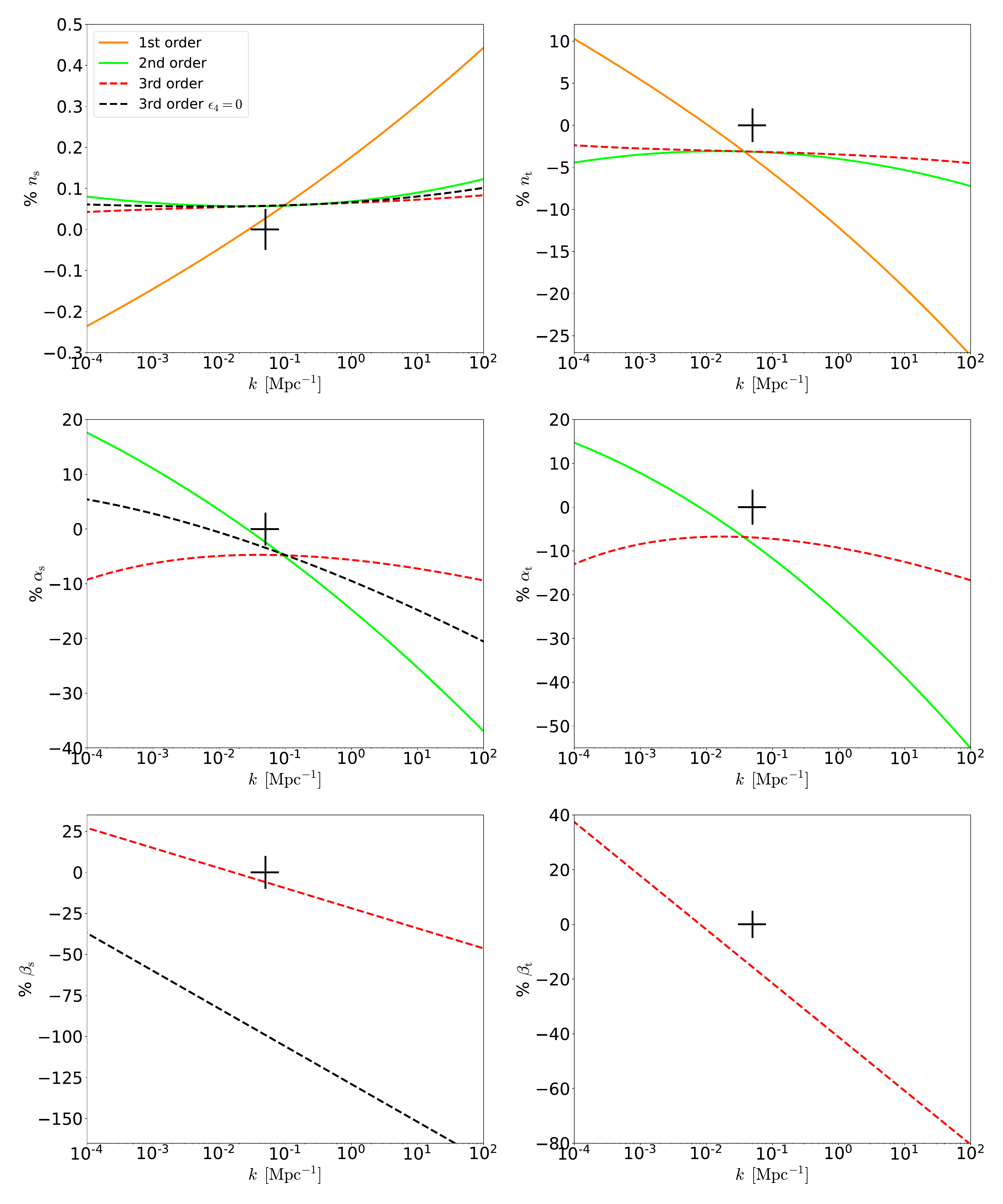}
    \caption{Same as~\cref{fig:Tmodel_parameters} but for KKLT inflation for $m = 1$.}\label{fig:KKLT_parameters}
\end{figure}

In~\cref{fig:Tmodel_spectra,fig:Tmodel_parameters}, we present the results for the T-model of $\alpha$-attractor inflation 
with $\alpha=1$, while in~\cref{fig:KKLT_spectra,fig:KKLT_parameters} we present the results for KKLT inflation with $m = 1$. 
We compare the numerical results with respect to the analytical results calculated at first-, second-, and third-order slow-roll 
expansion in the range $k \in \left[10^{-4}, 10^2\right]\, \text{Mpc}^{-1}$. Moreover, we also show the comparison for 
third-order with $\epsilon_4=0$. Such comparison can be found in the literature between the first- and the second-order 
corrections, see for instance Refs.~\cite{Leach:2002ar,Habib:2005mh}.

In~\cref{fig:Tmodel_spectra,fig:KKLT_spectra}, we present the power spectra comparisons together with the corresponding 
relative differences of the analytical results with respect to the numerical solutions. In~\cref{fig:Tmodel_parameters,fig:KKLT_parameters}, we present the relative differences of the spectral indices $\ns$ and 
$\nt$, their runnings $\as$ and $\at$, and the runnings of the running $\bs$ and $\bt$. 
For a certain quantity $X$, where 
$X = \{ \mathcal{P}_\zeta,\, \mathcal{P}_{\rm t},\, \ns,\, \nt,\, \as,\, \at,\, \bs,\, \bt\}$, the corresponding percentage 
difference $\% X$ is defined as follows 
\begin{equation}
    \% X = \left(1 - \frac{X_{\text{analytical}}}{X_{\text{numerical}}}\right) \times 100 \,.
\end{equation}

As we can observe in~\cref{fig:Tmodel_spectra,fig:KKLT_spectra}, the relative differences on the tensor PPS are one order of 
magnitude lower than those of the scalar PPS. We also observe that, on the entire $k$ range, the relative errors follow the 
expected perturbative hierarchy, i.e. the higher the order, the lower the errors. It is also useful to point out that the 
third-order case with $\epsilon_4 =0$ acts as a mid-case between the second- and pure third-order. For both models, we observe 
that around the pivot scale $k_*$ the second- and third-order PPS relative differences are nearly identical and match the 
numerical results thanks to the normalisation performed. Small differences between the second and third order start to appear 
around $k\sim 10 \, \text{Mpc}^{-1}$ of the order of 0.1\% and the curves visibly separate from each other at 
$k\sim 10^2\, \text{Mpc}^{-1}$ reaching a maximum difference of 0.5\% and 0.3\%, for T-model and KKLT inflation respectively. 
For the tensor PPS, small differences between second and third order start to arise at $k \sim 1\, \text{Mpc}^{-1}$, where the 
relative errors are of the order of 0.01\%, and they increase at higher $k$ reaching 0.02\% at $k \sim 10^2\, \text{Mpc}^{-1}$. 

In~\cref{fig:Tmodel_parameters,fig:KKLT_parameters}, we present the relative differences for $n_{\rm s/t}$, $\alpha_{\rm s/t}$, 
and $\beta_{\rm s/t}$, for T-model and KKLT inflation, respectively. In general, we find larger differences for the tensor 
quantities, however, associated usually to larger observational uncertainties. 
For the spectral indexes $\ns$ and $\nt$, we find that around the pivot scale the relative differences are nearly identical, 
they start to differ from $k\sim 10\, \text{Mpc}^{-1}$. On the other hand, when $k$ is lower than the pivot scale, the 
differences start to appear from $k\sim 10^{-2}\, \text{Mpc}^{-1}$. Including corrections at the second and third order, the 
relative differences for $\ns$ in both the models are always at most equal or lower than 0.1\%, while for $\nt$ they are always 
at most equal or lower than 10\% (in terms of absolute value). For the runnings of the spectral indexes $\as$ and $\at$, we 
observe increasing differences between the second and the third order moving away from the pivot scale; in both cases these 
differences are always of the order of $\sim 1\%$ around the pivot scale and they increase reaching the order of $\sim 10 \%$. 

The third-order relative differences for $\bs$ and $\bt$ are consistently around $\sim 10\%$ for both the models on the range 
of scales probed by CMB measurements; far from the pivot scale the errors are larger. When we consider the case with 
third-order corrections and $\epsilon_4=0$, $\bs$ exhibits significantly larger differences compared to the numerical result: 
these errors start to be approximately $50\%$ at $k \sim 10^{-4}\, \text{Mpc}^{-1}$ and increase up to $150\%$ at 
$k \sim 10^2\, \text{Mpc}^{-1}$.

Moreover, we also repeated the comparison for larger values of the inflationary parameters, $\alpha = 100$ for T-model 
inflation and $m = 100$ for KKLT inflation. We observe no significant difference with respect the case presented here for 
$\alpha = 1$ and $m = 1$. 

\begin{figure}[!ht]
    \centering
    \includegraphics[width=\textwidth]{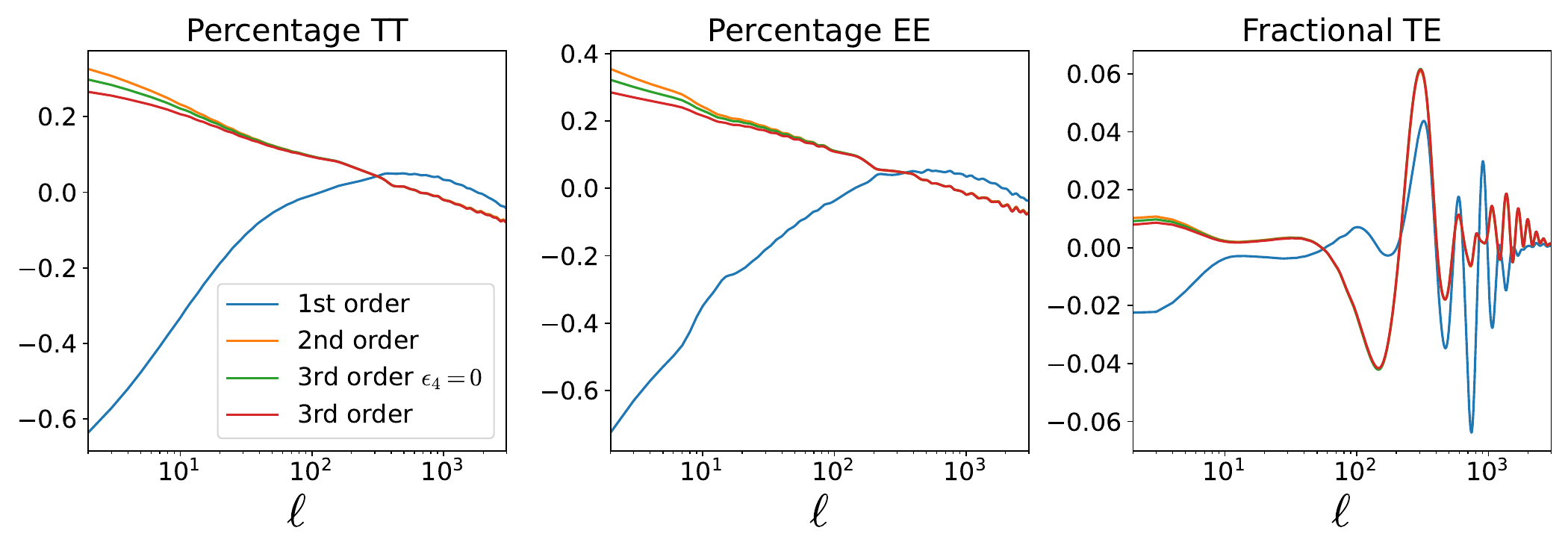}
    \caption{Differences with respect to the numerical solution for the CMB temperature and E-mode polarisation 
    lensed angular power spectra for the T-model $\alpha$-attractor inflation with $\alpha = 1$.}\label{fig:CMB_diff}
\end{figure}

We conclude this section showing in~\cref{fig:CMB_diff} the percentage differences on the TT and EE CMB lensed angular power spectra 
and the absolute differences on the TE spectrum for T-model $\alpha$-attractor inflation with $\alpha = 1$. The size of the 
differences is compatible to the one of the scalar PPS shown in~\cref{fig:Tmodel_spectra}. 

\section{Data Analysis and Cosmological Constraints} \label{sec:results}
In this section, we calculate the uncertainties on the HFFs considering different truncation of the PPS expansion at first, 
second, and third order. 
We use {\tt CosmoMC} \cite{Lewis:2013hha}\footnote{\url{https://github.com/cmbant/CosmoMC}} connected to our modified 
version of the code {\tt CAMB} \cite{Lewis:1999bs,Howlett:2012mh}.\footnote{\url{https://github.com/cmbant/CAMB}}
Mean values and uncertainties on the parameters reported, as well as the contours plotted, have been obtained using 
{\tt GetDist} \cite{Lewis:2019xzd}.\footnote{\url{https://github.com/cmbant/getdist}}
For the Markov chain Monte Carlo (MCMC) analysis, we vary the standard cosmological parameters for a flat $\Lambda$CDM 
concordance model $\omega_{\rm b}$, $\omega_{\rm c}$, $H_0$, $\tau$, and $\ln \left(10^{10}A_{\rm s}\right)$,\footnote{Note 
that sampling on the amplitude of the scalar PPS $\ln \left(10^{10}A_{\rm s}\right)$ rather that sampling directly on the 
amplitude of the scalar field potential $V_0$ allows to overcome some uncertainties connected to the determination of the end of 
inflation for which the slow-roll approximation breaks down; see Ref.~\cite{Auclair:2024udj} for analytic advances.} plus 
the HFFs $\epsilon_i$. We vary also nuisance and foreground parameters for the likelihood considered. We assume two massless 
neutrino with $N_{\rm eff} = 2.046$, and a massive one with fixed minimum mass $m_\nu = 0.06\, {\rm eV}$. 

We focus our analysis on comparing different cosmological datasets of primary CMB measurements, i.e. temperature and 
polarisation anisotropies. We also consider the case adding non-primary CMB datasets. We therefore consider the following 
datasets:
\begin{itemize}
    \item P18 refers to \Planck\ PR3 primary CMB data \cite{Planck:2019nip}. Low-multipole data ($\ell < 30$) consists to 
    the {\tt commander} likelihood for temperature and {\tt SimAll} for the E-mode polarisation. On high multipoles 
    ($\ell \geq 30$), we use the {\tt Plik} likelihood including CMB temperature up to $\ell_{\rm max} = 2508$, E-mode 
    polarisation and temperature-polarisation cross correlation up to 
    $\ell_{\rm max} = 1996$.\footnote{\url{https://pla.esac.esa.int/pla/\#cosmology}} 
    \item ACT refers to ACT DR4 TT, EE, TE power spectra \cite{ACT:2020frw,ACT:2020gnv} covering the multipole range 
    $[350,\,4000]$. When combining ACT with \Planck\ data we remove temperature data at a certain threshold to avoid for 
    the unaccounted correlation between the two datasets. We consider two multipole cuts: in the first case we remove 
    \Planck\ data in temperature above $\ell > 650$ and in the second case we remove ACT data in temperature below 
    $\ell < 1800$. 
    \item SPT refers to SPT-3G 2018 TT, EE, TE power spectra covering the angular multipole range $750 < \ell < 3000$ 
    \cite{SPT-3G:2022hvq}. Combining SPT and \Planck\ data we do not consider any cut on the multipole range since data 
    \Planck\ cover a large amount of sky not observed by SPT.
    \item BK18 refers B-mode polarisation spectrum for $20 < \ell < 330$ from BICEP2, Keck Array, and BICEP3 observations 
    up to 2018 \cite{BICEP:2021xfz}. 
    \item {\em ext} (external) refers to the combination of late-time probes and non-primary CMB data. We include measurements 
    of baryon acoustic oscillations (BAO) and redshift space distortions (RSD) at low redshift $0.07 < z < 0.2$ from SDSS-I 
    and -II sample as {\em Main Galaxy Sample} (MGS), BOSS DR12 galaxies over the redshift interval $0.2 < z < 0.6$, eBOSS 
    luminous red galaxies (LRG) and quasars $0.6 < z < 2.2$, and Lyman-$\alpha$ forest samples $1.8 < z < 3.5$ 
    \cite[see][for details and references]{eBOSS:2020yzd}.\footnote{\url{https://svn.sdss.org/public/data/eboss/DR16cosmo/tags/v1\_0\_1/likelihoods/BAO-plus/}} 
    We include the Pantheon catalogue of uncalibrated Type Ia Supernovae (SNe) over the redshift range $0.01 < z < 2.3$ 
    \cite{Pan-STARRS1:2017jku}.\footnote{\url{https://github.com/dscolnic/Pantheon/}} Finally we also include CMB lensing 
    data from \Planck\ PR3 considering the conservative multipole range $8<\ell<400$ \cite{Planck:2018lbu}.
\end{itemize}
When considering ACT and SPT data not combined with \Planck\ data, we include the \Planck\ low-$\ell$ E-mode polarisation 
likelihood {\tt SimAll} in order to provide information on the optical depth $\tau$ to avoid the use of a Gaussian 
\Planck-inspired prior on it. We consider P18+BK18, ACT+BK18, SPT+BK18, P18+ACT+BK18, P18+SPT+BK18, and all these combinations 
also adding the external datasets. See Refs.~\cite{Galli:2021mxk,Smith:2022hwi,FrancoAbellan:2023gec,Antony:2024vrx} for the combination of \Planck\, ACT, and SPT data for models beyond the $\Lambda$CDM one.

Prior ranges on the cosmological parameters are collected on~\cref{tab:prior}. We assume uniform priors an all the HFF 
parameters $\epsilon_i$. In Refs.~\cite{Martin:2010kz,Martin:2013tda,Ringeval:2013lea,Martin:2024nlo} $\epsilon_1$ has been 
sampled with a log-uniform prior corresponding effectively to sampling the tensor-to-scalar ratio $r$ with a logarithmic prior; 
see Refs.~\cite{Hergt:2021qlh,Galloni:2024lre} for an extended discussion on the use of uniform or logarithmic priors for the 
tensor-to-scalar ratio. Note also that in Refs.~\cite{Martin:2010kz,Martin:2013tda,Ringeval:2013lea,Martin:2024nlo} a tighter 
flat prior was adopted for the second, third, and forth HFFs, corresponding to $\epsilon_{2,3,4} \in [-0.2,0.2]$. To ensure the 
validity of the slow-roll equations derived in \cref{sec:analytic} and adopted here, the HFFs should be less 
than one. We show in~\ref{app:an_prior} a comparison of the posterior distribution on the sampled and derived inflationary 
parameters by adopting a larger prior with $\epsilon_{2,3,4} \in [-1,1]$.\footnote{A detailed comparison of the effect of the 
size of the priors on the slow-roll parameters was done in Ref.~\cite{Martin:2024nlo} for both second- and third-order results 
but for a different combination of datasets.}
\begin{table*}
\centering
\begin{tabular}{l|c}
\hline
\hline
Parameter & Uniform prior \\
\hline
$\Omega_{\rm b} h^2$                & [0.019, 0.025] \\
$\Omega_{\rm c} h^2$                & [0.095, 0.145] \\
$100\theta_{\rm MC}$                & [1.03, 1.05]   \\
$\tau$                              & [0.01, 0.4]    \\
$\ln\left(10^{10}A_{\rm s}\right)$  & [2.5, 3.7]     \\
$\epsilon_1$               & [0, 0.1]        \\
$\epsilon_2$               & [-0.5, 0.5]        \\
$\epsilon_3$               & [-0.5, 0.5]        \\
$\epsilon_4$               & [-0.5, 0.5]        \\
\hline
\end{tabular}
\caption{Prior ranges for cosmological parameters used in the Bayesian analysis.}\label{tab:prior}
\end{table*}

\subsection{Results} 
We first present the constraints on slow-roll parameters obtained through the analytic perturbative expansion in terms of the 
HFFs $\epsilon_i$ for the primordial spectra of cosmological fluctuations during slow-roll inflation. When restricting to 
first-order expansion, we obtain
\begin{alignat}{2}
    &\epsilon_1 < 0.0022 &&({\rm 95\%\ CL,\ P18 + BK18}) \,,\\
    &\epsilon_2 = 0.0332 \pm 0.0046 \qquad &&({\rm 68\%\ CL,\ P18 + BK18}) \,,
\end{alignat}
\begin{alignat}{2}
    &\epsilon_1 < 0.0023 &&({\rm 95\%\ CL,\ ACT + BK18}) \,,\\
    &\epsilon_2 = -0.005 \pm 0.014 \qquad &&({\rm 68\%\ CL,\ ACT + BK18}) \,,
\end{alignat}
\begin{alignat}{2}
    &\epsilon_1 < 0.0024 &&({\rm 95\%\ CL,\ SPT + BK18}) \,,\\
    &\epsilon_2 = 0.032 \pm 0.015 \qquad &&({\rm 68\%\ CL,\ SPT + BK18}) \,.
\end{alignat}
When second-order contributions in the HFFs are included, we obtain
\begin{alignat}{2}
    &\epsilon_1 < 0.0023 &&({\rm 95\%\ CL,\ P18 + BK18}) \,,\\
    &\epsilon_2 = 0.0379 \pm 0.0078 \qquad &&({\rm 68\%\ CL,\ P18 + BK18}) \,,\\
    &\epsilon_3 = 0.13^{+0.19}_{-0.12} \qquad &&({\rm 68\%\ CL,\ P18 + BK18}) \,,
\end{alignat}
\begin{alignat}{2}
    &\epsilon_1 < 0.0023 &&({\rm 95\%\ CL,\ ACT + BK18}) \,,\\
    &\epsilon_2 = -0.007^{+0.016}_{-0.014} \qquad &&({\rm 68\%\ CL,\ ACT + BK18}) \,,\\
    &\epsilon_3\ {\rm unconstrained} \qquad &&({\rm 95\%\ CL,\ ACT + BK18}) \,,
\end{alignat}
\begin{alignat}{2}
    &\epsilon_1 < 0.0025 &&({\rm 95\%\ CL,\ SPT + BK18}) \,,\\
    &\epsilon_2 = 0.035 \pm 0.017 \qquad &&({\rm 68\%\ CL,\ SPT + BK18}) \,,\\
    &\epsilon_3 {\rm \ unconstrained} \qquad &&({\rm 95\%\ CL,\ SPT + BK18}) \,.
\end{alignat}
Including third-order contributions in the HFFs, we find close results to the second-order ones with $\epsilon_4$ 
unconstrained, see~\cref{tab:eps}. The addition of external datasets, which are BAO and RSD measurements, SNe, and CMB 
lensing, leads to slightly tighter uncertainties and consistent mean values for $\epsilon_2$ and $\epsilon_3$, 
see~\cref{tab:eps_ext}.

\begin{table*}[!ht]
\centering
{\small
\centering
\begin{tabular}{|l|ccc|}
\hline
                                        & P18 + BK18            & ACT + BK18         & SPT + BK18 \\
\hline
                                        & \multicolumn{3}{c|}{FIRST ORDER} \\
\hline
$\ln \left( 10^{10} A_{\rm s} \right)$ & $3.046\pm 0.015$      & $3.030\pm 0.019$      & $3.038\pm 0.016$ \\
$\epsilon_1$  (at 95\% CL)              & $< 0.0022$             & $< 0.0023$            & $< 0.0024$  \\
$\epsilon_2$                            & $0.0332\pm 0.0046$    & $-0.005\pm 0.014$     & $0.032\pm 0.015$ \\
\hline
$n_{\rm s,\,0.05}$                      & $0.9647\pm 0.0043$    & $1.003\pm 0.014$      & $0.966\pm 0.015$ \\
$r_{0.05}$  (at 95\% CL)                & $< 0.035$          & $< 0.037$             & $< 0.039$  \\
$n_{\rm t,\,0.05}$                      & $-0.0020^{+0.0015}_{-0.0009}$    & $-0.0022^{+0.0015}_{-0.0009}$      & $-0.0024^{+0.0017}_{-0.0010}$ \\
\hline
                                        & \multicolumn{3}{c|}{SECOND ORDER} \\
\hline
$\ln \left( 10^{10} A_{\rm s} \right)$ & $3.050\pm 0.017$        & $3.029\pm 0.019$     & $3.040\pm 0.017$ \\
$\epsilon_1$ (at 95\% CL)               & $< 0.0023$             & $< 0.0023$                     & $< 0.0025$  \\
$\epsilon_2$                            & $0.0379\pm 0.0078$     & $-0.007^{+0.016}_{-0.014}$    & $0.035\pm 0.017$ \\
$\epsilon_3$ (at 95\% CL)               & $0.13^{+0.30}_{-0.35}$ & $-$         & $-$ \\
\hline
$n_{\rm s,\,0.05}$                      & $0.9641\pm 0.0046$     & $1.003^{+0.015}_{-0.013}$              & $0.967\pm 0.014$ \\
$\alpha_{\rm s,\,0.05}$                 & $-0.0059\pm 0.0067$    & $0.0020^{+0.0029}_{-0.0051}$     & $-0.006^{+0.012}_{-0.010}$ \\
$r_{0.05}$ (at 95\% CL)                 & $< 0.036$              & $< 0.038$                      & $< 0.038$  \\
$n_{\rm t,\,0.05}$                      & $-0.0022^{+0.0016}_{-0.0010}$     & $-0.0022^{+0.0016}_{-0.0010}$              & $-0.0024^{+0.0017}_{-0.0010}$ \\
$10^5\alpha_{\rm t,\,0.05}$             & $-8.1^{+5.9}_{-3.3}$   & $1.9^{+2.3}_{-4.4}$     & $-7.8^{+6.8}_{-3.4}$ \\
\hline
\end{tabular}}
\caption{Constraints on the main (inflationary related) parameters and derived ones (at 68\% CL if not otherwise stated) 
considering P18+BK18, ACT+BK18, and SPT+BK18.\label{tab:eps}}
\end{table*}

\begin{table*}[!ht]
\centering
{\small
\centering
\begin{tabular}{|l|ccc|}
\hline
                                        & P18 + BK18 + ext           & ACT + BK18 + ext  & SPT + BK18 + ext \\
\hline
                                        & \multicolumn{3}{c|}{FIRST ORDER} \\
\hline
$\ln \left( 10^{10} A_{\rm s} \right)$ & $3.052^{+0.013}_{-0.015}$  & $3.042\pm 0.014$  & $3.043\pm 0.011$ \\
$\epsilon_1$ (at 95\% CL)               & $< 0.0022$                 & $< 0.0023$         & $< 0.0025$  \\
$\epsilon_2$                            & $0.0306\pm 0.0039$         & $0.003\pm 0.010$  & $0.034\pm 0.012$ \\
\hline
$n_{\rm s,\,0.05}$                      & $0.9673\pm 0.0036$         & $0.995\pm 0.010$  & $0.963\pm 0.012$ \\
$r_{0.05}$ (at 95\% CL)                 & $< 0.036$                   & $< 0.036$          & $< 0.039$  \\
$n_{\rm t,\,0.05}$                      & $-0.0021^{+0.0015}_{-0.0009}$    & $-0.0022^{+0.0015}_{-0.0009}$    & $-0.0024^{+0.0016}_{-0.0010}$ \\
\hline
                                        & \multicolumn{3}{c|}{SECOND ORDER} \\
\hline
$\ln \left( 10^{10} A_{\rm s} \right)$  & $3.055\pm 0.015$        & $3.041\pm 0.014$       & $3.045\pm 0.012$ \\
$\epsilon_1$  (at 95\% CL)              & $< 0.0023$              & $< 0.0023$                       & $< 0.0024$  \\
$\epsilon_2$                            & $0.0342\pm 0.0071$      & $0.002^{+0.012}_{-0.010}$    & $0.036^{+0.012}_{-0.014}$ \\
$\epsilon_3$  (at 95\% CL)              & $0.11^{+0.33}_{-0.38}$  & $-$                           & $-$ \\
\hline
$n_{\rm s,\,0.05}$                      & $0.9669\pm 0.0038$      & $0.995^{+0.011}_{-0.010}$                & $0.965\pm 0.012$ \\
$\alpha_{\rm s,\,0.05}$                 & $-0.0048\pm 0.0065$     & $0.0011^{+0.0022}_{-0.0031}$       & $-0.005^{+0.012}_{-0.010}$ \\
$r_{0.05}$ (at 95\% CL)                 & $< 0.037$               & $< 0.037$                        & $< 0.038$ \\
$n_{\rm t,\,0.05}$                      & $-0.0023^{+0.0016}_{-0.0010}$     & $-0.0022^{+0.0016}_{-0.0010}$              & $-0.0024^{+0.0016}_{-0.0010}$ \\
$10^5\alpha_{\rm t,\,0.05}$      & $-7.6^{+5.4}_{-3.1}$    & $-0.3^{+2.0}_{-2.4}$     & $-8.4^{+6.6}_{-3.3}$ \\
\hline
\end{tabular}}
\caption{Same as~\cref{tab:eps} in combination with the external datasets, which are BAO and RSD measurements, SNe, and CMB 
lensing.\label{tab:eps_ext}}
\end{table*}

The upper limit on $\epsilon_1$, as well as the derived constraint on $r$ ($\lesssim 0.04$ at 95\% CL), is almost unchanged 
among the different combination of datasets and different truncation being dominated by BK18 data which are always included. 
On the other hand, we find different results for $\epsilon_2$ and $\epsilon_3$ moving from \Planck\ to ACT data. This 
difference comes from the almost $3\sigma$ tension between \Planck\ and ACT data on the inferred value of the scalar spectral 
index $\ns$. While the result from \Planck\ (combining temperature, E-mode polarisation, and lensing) $\ns = 0.9649 \pm 0.0044$ 
\cite{Planck:2018vyg} agrees with the prediction of simplest single-field slow-roll inflationary models, the result from ACT 
\cite{ACT:2020gnv} points to $\ns = 1.008 \pm 0.015$ ($2.8\sigma$\footnote{Quantified as 
$|\ns^{(\rm P18)}-\ns^{(\rm ACT)}| / \sqrt{\sigma\left(\ns^{(\rm P18)}\right)^2+\sigma\left(\ns^{(\rm ACT)}\right)^2}$; 
we perform the analogous estimation for the running of the scalar spectral index $\as$.}) compatible with a scale invariant 
Harrison-Zel'dovich (HZ) primordial power spectrum \cite{Jiang:2022uyg,DiValentino:2022rdg,Giare:2022rvg}. We find that 
\Planck\ ($\ns = 0.9647 \pm 0.0043$ at 68\% CL) and ACT ($\ns = 1.003 \pm 0.014$ at 68\% CL) results have a $2.6\sigma$ on 
the inferred value of $\ns$ when considering primary CMB only at first order in slow-roll expansion, see~\cref{tab:eps}. 
This number does not change when the external data are included, see~\cref{tab:eps_ext}.

The discrepancy between \Planck\ and ACT data at the level of the scalar PPS parameters persists even when allowing a running 
of the scalar spectral index $\as$ \cite{DiValentino:2022rdg,Giare:2022rvg}; in our case when we move to the second-order 
expansion case. Despite the discrepant results, while \Planck\ is consistent with a zero running $\as = -0.0045 \pm 0.0067$ 
at 68\% CL \cite{Planck:2018vyg} (and with slow-roll single-field inflation predictions, see Ref.~\cite{Martin:2024nlo}), on 
the contrary ACT data point to a $2.5\sigma$ preference for a positive running $\as = 0.069 \pm 0.029$ \cite{ACT:2020gnv}. We 
find a discrepancy on the inferred value of $\ns$ of $3.8\sigma$ with primary CMB alone and $3.2\sigma$ in combination with 
external datasets; larger numbers compared to the ones obtained considering only second-order terms. We find that \Planck\ 
($\as = -0.0059 \pm 0.0067$ at 68\% CL) and ACT ($\as = 0.0020^{+0.0029}_{-0.0051}$  at 68\% CL) results have a $0.9\sigma$ 
discrepancy on the inferred value of $\as$ both considering primary CMB only and in combination with external datasets at 
second order in slow-roll expansion, see~\cref{tab:eps,tab:eps_ext}. 
SPT data agrees with \Planck\ findings but with lager error bars.

\subsection{Combined Results}
We present here the results for the two combined cases P18+ACT+BK18, assuming two different multiple cuts, and the combination 
P18+SPT+BK18. When restricting to first-order expansion, we obtain
\begin{alignat}{2}
    &\epsilon_1 < 0.0024 &&{\rm 95\%\ CL,\ P18\ (\ell_{\rm TT} < 650) + ACT + BK18} \,,\\
    &\epsilon_2 = 0.0183 \pm 0.0059 \qquad &&{\rm 68\%\ CL,\ P18\ (\ell_{\rm TT} < 650) + ACT + BK18} \,,
\end{alignat}
\begin{alignat}{2}
    &\epsilon_1 < 0.0022 &&{\rm 95\%\ CL,\ P18 + ACT\ (\ell_{\rm TT} > 1800) + BK18} \,,\\
    &\epsilon_2 = 0.0312 \pm 0.0044 \qquad &&{\rm 68\%\ CL,\ P18 + ACT\ (\ell_{\rm TT} > 1800) + BK18} \,,
\end{alignat}
\begin{alignat}{2}
    &\epsilon_1 < 0.0022 &&{\rm 95\%\ CL,\ P18 + SPT + BK18} \,,\\
    &\epsilon_2 = 0.0327 \pm 0.0044 \qquad &&{\rm 68\%\ CL,\ P18 + SPT + BK18} \,.
\end{alignat}
When second-order contributions in the HFFs are included, we obtain
\begin{alignat}{2}
    &\epsilon_1 < 0.0023 &&{\rm 95\%\ CL,\ P18\ (\ell_{\rm TT} < 650) + ACT + BK18} \,,\\
    &\epsilon_2 = 0.0166^{+0.0045}_{-0.0057} \qquad &&{\rm 68\%\ CL,\ P18\ (\ell_{\rm TT} < 650) + ACT + BK18} \,,\\
    &\epsilon_3 < - 0.053 \qquad &&{\rm 68\%\ CL,\ P18\ (\ell_{\rm TT} < 650) + ACT + BK18} \,,
\end{alignat}
\begin{alignat}{2}
    &\epsilon_1 < 0.0023 &&{\rm 95\%\ CL,\ P18 + ACT\ (\ell_{\rm TT} > 1800) + BK18} \,,\\
    &\epsilon_2 = 0.0274^{+0.0047}_{-0.0063} \qquad &&{\rm 68\%\ CL,\ P18 + ACT\ (\ell_{\rm TT} > 1800) + BK18} \,,\\
    &\epsilon_3 = -0.08^{+0.23}_{-0.19} \qquad &&{\rm 68\%\ CL,\ P18 + ACT\ (\ell_{\rm TT} > 1800) + BK18} \,,
\end{alignat}
\begin{alignat}{2}
    &\epsilon_1 < 0.0024 &&{\rm 95\%\ CL,\ P18 + SPT + BK18} \,,\\
    &\epsilon_2 = 0.0354\pm 0.0066 \qquad &&{\rm 68\%\ CL,\ P18 + SPT + BK18} \,,\\
    &\epsilon_3 = 0.13^{+0.19}_{-0.12} \qquad &&{\rm 68\%\ CL,\ P18 + SPT + BK18} \,.
\end{alignat}
Also in this case, we find small differences between the second-order and the third-order case and again with $\epsilon_4$ 
unconstrained, see~\cref{tab:eps_comb} and~\cref{tab:eps_com_ext} for the results including the external datasets. We report 
the full results for the third-order case only for the combined datasets with the addition of external data; 
see~\cref{tab:eps_com_ext}.

\begin{table*}[!ht]
\centering
{\small
\centering
\begin{tabular}{|l|ccc|}
\hline
                                        & P18 ($\ell_{\rm TT} < 650$) + ACT & P18 + ACT ($\ell_{\rm TT} > 1800$) & P18 + SPT \\
                                        & + BK18 &  + BK18 &  + BK18 \\
\hline
                                        & \multicolumn{3}{c|}{FIRST ORDER} \\
\hline
$\ln \left( 10^{10} A_{\rm s} \right)$  & $3.041\pm 0.016$      & $3.048\pm 0.015$      & $3.045^{+0.010}_{-0.015}$ \\
$\epsilon_1$ (at 95\% CL)               & $< 0.0024$            & $< 0.0022$            & $< 0.0022$  \\
$\epsilon_2$                            & $0.0183\pm 0.0059$    & $0.0312\pm 0.0044$    & $0.0327\pm 0.0044$  \\
\hline
$n_{\rm s,\,0.05}$                      & $0.9794\pm 0.0056$    & $0.9667\pm 0.0040$    & $0.9651\pm 0.0041$ \\
$r_{0.05}$ (at 95\% CL)                             & $< 0.038$            & $< 0.035$             & $< 0.035$  \\
$n_{\rm t,\,0.05}$                      & $-0.0023^{+0.0017}_{-0.0010}$    & $-0.0021^{+0.0015}_{-0.0009}$    & $-0.0021^{+0.0015}_{-0.0009}$ \\
\hline
                                        & \multicolumn{3}{c|}{SECOND ORDER} \\
\hline
$\ln \left( 10^{10} A_{\rm s} \right)$  & $3.040\pm 0.016$      & $3.047\pm 0.016$              & $3.046\pm 0.012$ \\
$\epsilon_1$ (at 95\% CL)               & $< 0.0024$            & $< 0.0022$                    & $< 0.0023$ \\
$\epsilon_2$                            & $0.0166^{+0.0050}_{-0.0063}$    & $0.0297^{+0.0055}_{-0.006}$            & $0.0383\pm 0.0071$  \\
$\epsilon_3$ (at 95\% CL)               & $< 0.27$         & $-0.06^{+0.33}_{-0.38}$       & $0.15^{+0.27}_{-0.31}$ \\
\hline
$n_{\rm s,\,0.05}$                      & $0.9796\pm 0.0055$   & $0.9674\pm 0.0040$   & $0.9640\pm 0.0041$ \\
$\alpha_{\rm s,\,0.05}$                 & $0.0020^{+0.0045}_{-0.0027}$             & $0.0010^{+0.0063}_{-0.0045}$   & $-0.0064\pm 0.0062$ \\
$r_{0.05}$ (at 95\% CL)                 & $< 0.038$                      & $< 0.035$            & $< 0.036$ \\
$n_{\rm t,\,0.05}$                      & $-0.0023^{+0.0016}_{-0.0010}$     & $-0.0021^{+0.0016}_{-0.0010}$              & $-0.0023^{+0.0016}_{-0.0010}$ \\
$10^5\alpha_{\rm t,\,0.05}$             & $-3.6^{+2.7}_{-1.4}$    & $-6.1^{+4.4}_{-2.5}$     & $-8.4^{+5.9}_{-3.5}$ \\
\hline
\end{tabular}}
\caption{Constraints on the main (inflationary related) parameters and derived ones (at 68\% CL if not otherwise stated) 
considering the combination P18+ACT+BK18 with two different multipole cuts, and P18+SPT-3G+BK18.\label{tab:eps_comb}} 
\end{table*}

\begin{table*}[!ht]
\centering
{\small
\centering
\begin{tabular}{|l|ccc|}
\hline
                                        & P18 ($\ell_{\rm TT} < 650$) + ACT & P18 + ACT ($\ell_{\rm TT} > 1800$) & P18 + SPT \\
                                        & + BK18 + ext &  + BK18 + ext &  + BK18 + ext \\
\hline
                                        & \multicolumn{3}{c|}{FIRST ORDER} \\
\hline
$\ln \left( 10^{10} A_{\rm s} \right)$  & $3.048\pm 0.014$      & $3.052\pm 0.013$  & $3.052^{+0.026}_{-0.020}$ \\
$\epsilon_1$ (at 95\% CL)                & $< 0.0024$            & $< 0.0023$        & $< 0.0022$  \\
$\epsilon_2$                            & $0.0183\pm 0.0051$    & $0.0292\pm 0.0037$  & $0.0311\pm 0.0036$  \\
\hline
$n_{\rm s,\,0.05}$                      & $0.9794\pm 0.0049$    & $0.9687\pm 0.0035$    & $0.9667\pm 0.0033$ \\
$r_{0.05}$ (at 95\% CL)                  & $< 0.038$             & $< 0.036$  & $< 0.036$  \\
$n_{\rm t,\,0.05}$                      & $-0.0023^{+0.0016}_{-0.0010}$    & $-0.0021^{+0.0016}_{-0.0009}$    & $-0.0022^{+0.0016}_{-0.0009}$ \\
\hline
                                        & \multicolumn{3}{c|}{SECOND ORDER} \\
\hline
$\ln \left( 10^{10} A_{\rm s} \right)$  & $3.047\pm 0.014$                & $3.051\pm 0.014$              & $3.048\pm 0.011$ \\
$\epsilon_1$  (at 95\% CL)              & $< 0.0023$                      & $< 0.0023$                    & $< 0.0024$  \\
$\epsilon_2$                            & $0.0166^{+0.0045}_{-0.0057}$    & $0.0274^{+0.0047}_{-0.0063}$  & $0.0354\pm 0.0066$  \\
$\epsilon_3$ (at 95\% CL)               & $<0.26$         & $-0.08^{+0.32}_{-0.40}$       & $0.13^{+0.30}_{-0.34}$ \\
\hline
$n_{\rm s,\,0.05}$                      & $0.9796\pm 0.0049$    & $0.9694\pm 0.0034$  & $0.9661\pm 0.0035$ \\
$\alpha_{\rm s,\,0.05}$                 & $0.0021^{+0.0045}_{-0.0026}$    & $0.0014^{+0.0062}_{-0.0040}$  & $-0.0055\pm 0.0062$ \\
$r_{0.05}$  (at 95\% CL)                & $< 0.037$                      & $< 0.035$                      & $< 0.037$  \\
$n_{\rm t,\,0.05}$                      & $-0.0023^{+0.0016}_{-0.0010}$     & $-0.0022^{+0.0016}_{-0.0010}$              & $-0.0023^{+0.0016}_{-0.0010}$ \\
$10^5\alpha_{\rm t,\,0.05}$             & $-3.6^{+2.6}_{-1.4}$    & $-5.7^{+4.0}_{-2.4}$     & $-8.0^{+5.5}_{-3.3}$ \\
\hline
                                        & \multicolumn{3}{c|}{THIRD ORDER} \\
\hline
$\ln \left( 10^{10} A_{\rm s} \right)$  & $3.046\pm 0.014$        & $3.052\pm 0.014$       & $3.048\pm 0.011$ \\
$\epsilon_1$ (at 95\% CL)               & $< 0.0024$ & $< 0.0022$                    & $< 0.0023$  \\
$\epsilon_2$                            & $0.0166^{+0.0047}_{-0.0063}$      & $0.0291^{+0.0049}_{-0.0066}$    & $0.0363^{+0.0068}_{-0.0085}$ \\
$\epsilon_3$ (at 95\% CL)               & $< 0.34$  & $0.00^{+0.37}_{-0.35}$          & $> -0.13$ \\
$\epsilon_4$                            & $-$  & $-$          & $-$ \\
\hline
$n_{\rm s,\,0.05}$                      & $0.9804\pm 0.0049$      & $0.9695\pm 0.0035$                & $0.9673\pm 0.0038$ \\
$\alpha_{\rm s,\,0.05}$                 & $0.0019^{+0.0046}_{-0.0039}$     & $-0.0002^{+0.0054}_{-0.0047}$       & $-0.0056^{+0.0064}_{-0.0051}$ \\
$\beta_{\rm s,\,0.05}$                 & $-0.0011^{+0.0018}_{-0.0007}$     & $-0.0010^{+0.0020}_{-0.0006}$       & $-0.0025^{+0.0043}_{-0.0012}$ \\
$r_{0.05}$ (at 95\% CL)                   & $< 0.037$             & $< 0.035$                      & $< 0.036$  \\
$n_{\rm t,\,0.05}$                      & $-0.0023^{+0.0016}_{-0.0010}$     & $-0.0022^{+0.0015}_{-0.0010}$              & $-0.0023^{+0.0016}_{-0.0010}$ \\
$10^5\alpha_{\rm t,\,0.05}$      & $-3.6^{+2.7}_{-1.3}$    & $-6.2^{+4.5}_{-2.5}$     & $-8.9^{+6.4}_{-3.7}$ \\
$10^5\beta_{\rm t,\,0.05}$      & $0.2^{+1.0}_{-0.6}$    & $-0.4^{+1.4}_{-0.7}$     & $-2.2^{+2.6}_{-1.1}$ \\
\hline
\end{tabular}}
\caption{Same as~\cref{tab:eps_comb} in combination with the external datasets, which are BAO and RSD measurements, SNe, and 
CMB lensing.\label{tab:eps_com_ext}}
\end{table*}

The combination P18+ACT is closer to the \Planck\ alone results for both the two multipole cuts applied in temperature. While 
the case with the cut on ACT data, with $\ell_{\rm TT}^{\rm ACT} > 1800$, $\ns = 0.9694\pm 0.0034$ at 68\% CL ($0.3\sigma$) 
agrees well with \Planck\ results, the case cutting \Planck\ temperature data, with $\ell_{\rm TT}^{\rm P18} < 650$, gives 
$\ns =  0.9796\pm 0.0049$ at 68\% CL ($2.1\sigma$) when considering primary CMB only; we find $0.6\sigma$ and $2.3\sigma$ 
including second-order corrections, respectively. 
Results for the scalar running for which the combination P18+ACT with $\ell_{\rm TT}^{\rm ACT} > 1800$ are loser to the 
constraints obtained with \Planck\ data alone; we find $\as = 0.0014^{+0.0062}_{-0.0040}$ with the cut on ACT data and 
$\as = 0.0021^{+0.0045}_{-0.0026}$ by cutting \Planck\ data, both at 68\% CL. 
The combination P18+SPT agrees with \Planck\ findings but with tighter error bars.

\begin{figure}[!ht]
\centering
\includegraphics[width=0.49\textwidth]{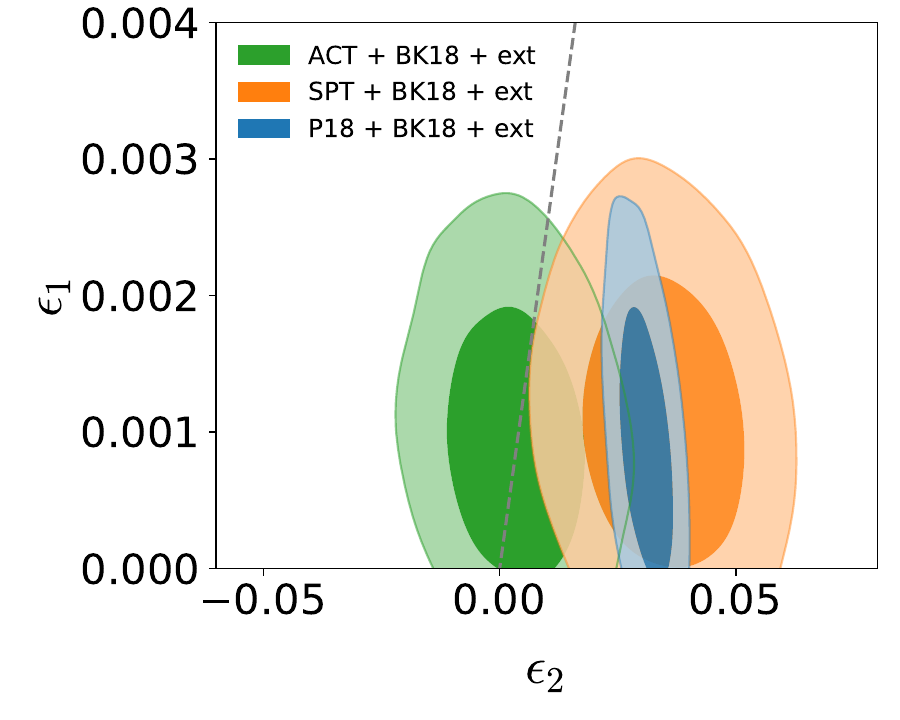}
\includegraphics[width=0.49\textwidth]{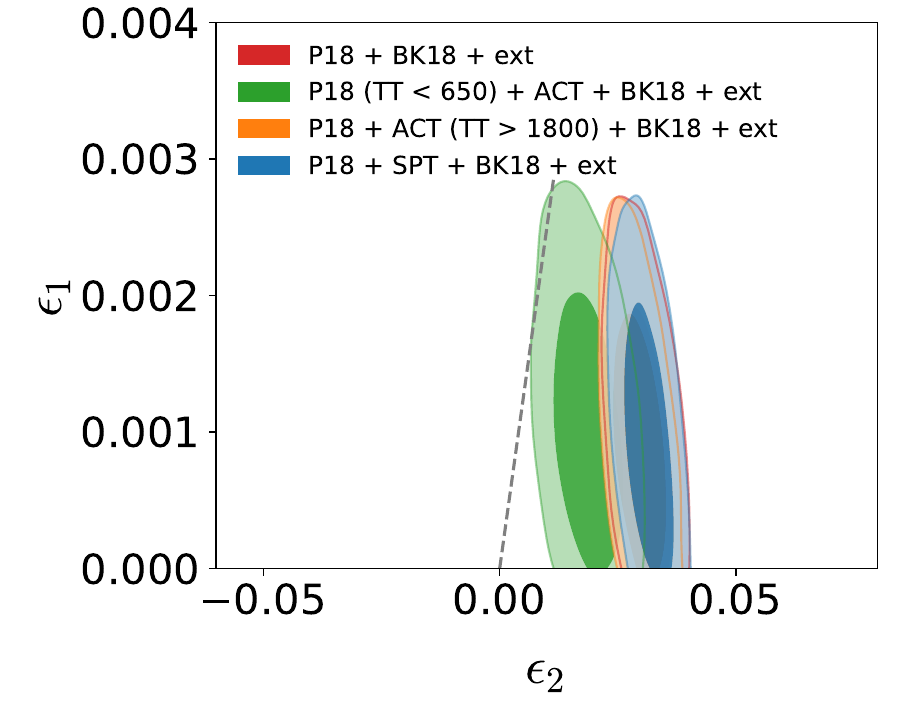}
\caption{Marginalised joint confidence contours for the first two HFF parameters $\epsilon_1$ and $\epsilon_2$ assuming 
first-order slow-roll predictions. The grey dashed line divide the parameter space from convex (left side) to concave (right 
side) single-field slow-roll potentials.\label{fig:Eps12}}
\end{figure}

\begin{figure}[!hb]
\centering
\includegraphics[width=\textwidth]{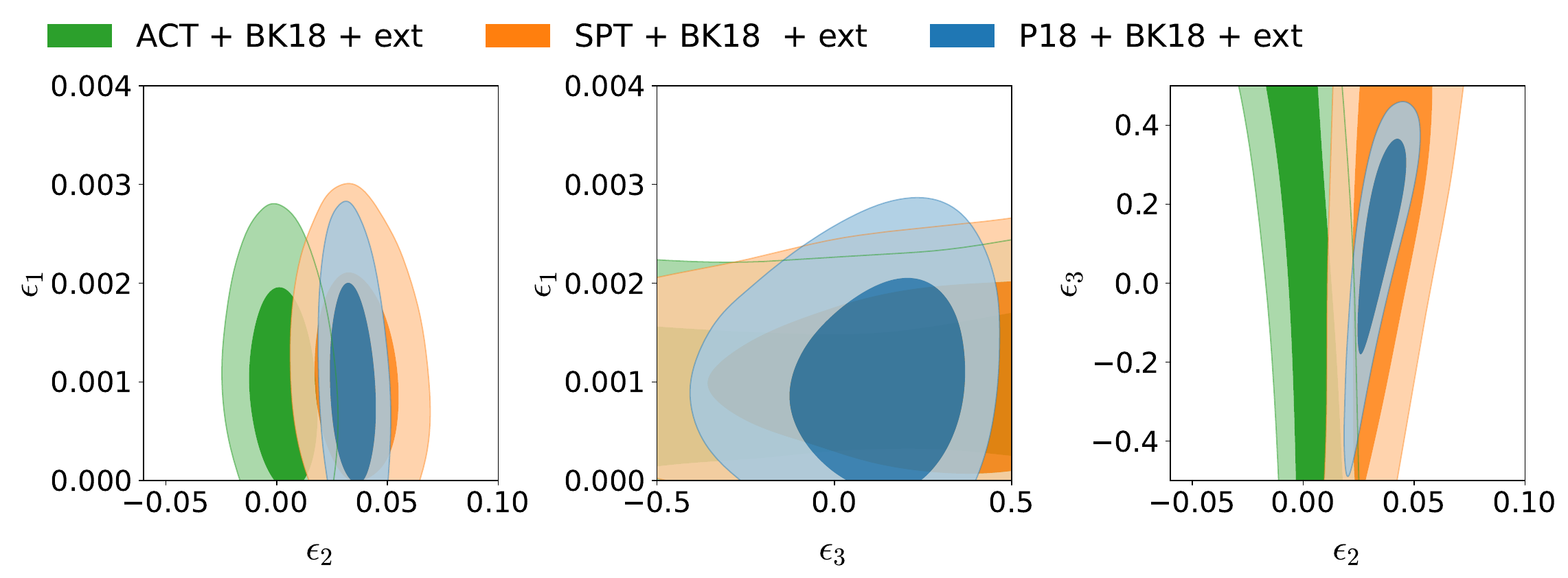}
\includegraphics[width=\textwidth]{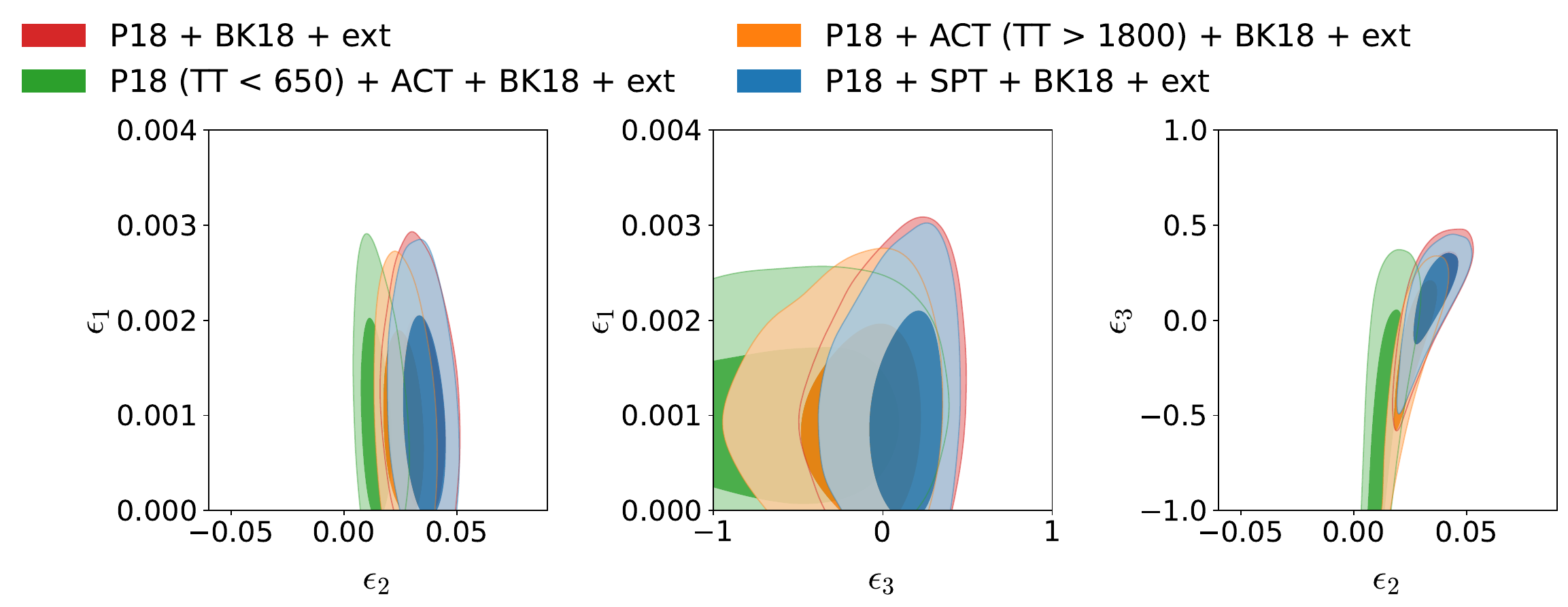}
\caption{Marginalised joint confidence contours for the first three HFF parameters $\epsilon_1$, 
$\epsilon_2$, and $\epsilon_3$ assuming second-order slow-roll predictions.\label{fig:Eps123}}
\end{figure}

In~\cref{fig:Eps12}, we show the 68\% CL and 95\% CL of the HFFs obtained for the first-order analysis both for \Planck, ACT, 
and SPT alone and their combinations. Our findings are consistent with the global picture that CMB data prefer potentials which 
are concave, i.e. $V_{\phi\phi} < 0$, in the observable window, with exception of the ACT+BK18 case.
In~\cref{fig:Eps123}, we show the 68\% CL and 95\% CL of the HFFs obtained for the second-order analysis both for \Planck, 
ACT, and SPT alone on the left and their combinations on the right.

\begin{figure}[!ht]
\centering
\includegraphics[width=0.49\textwidth]{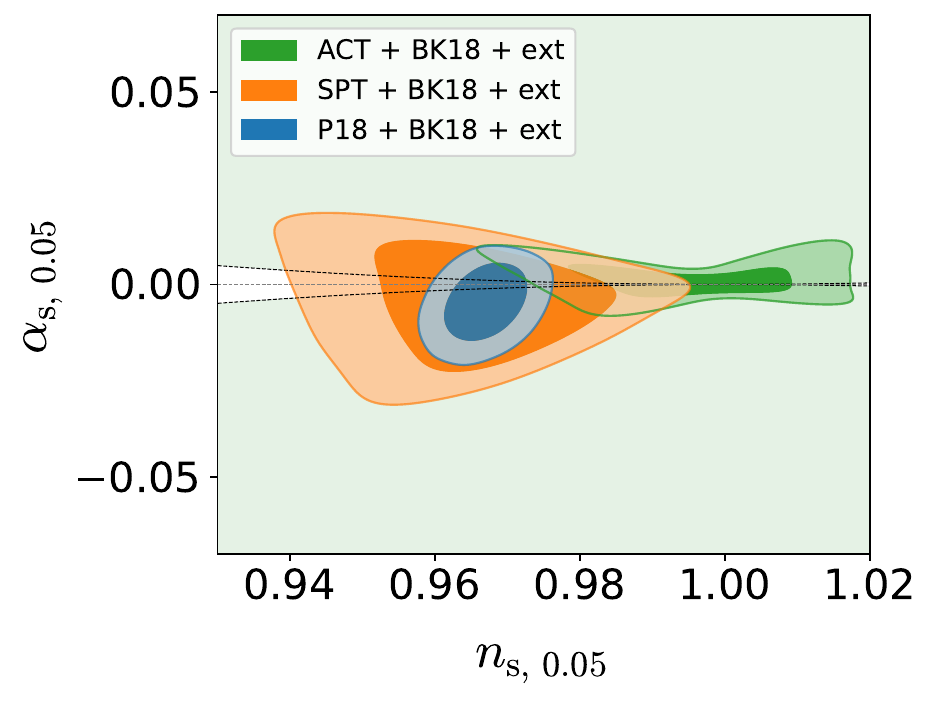}
\includegraphics[width=0.49\textwidth]{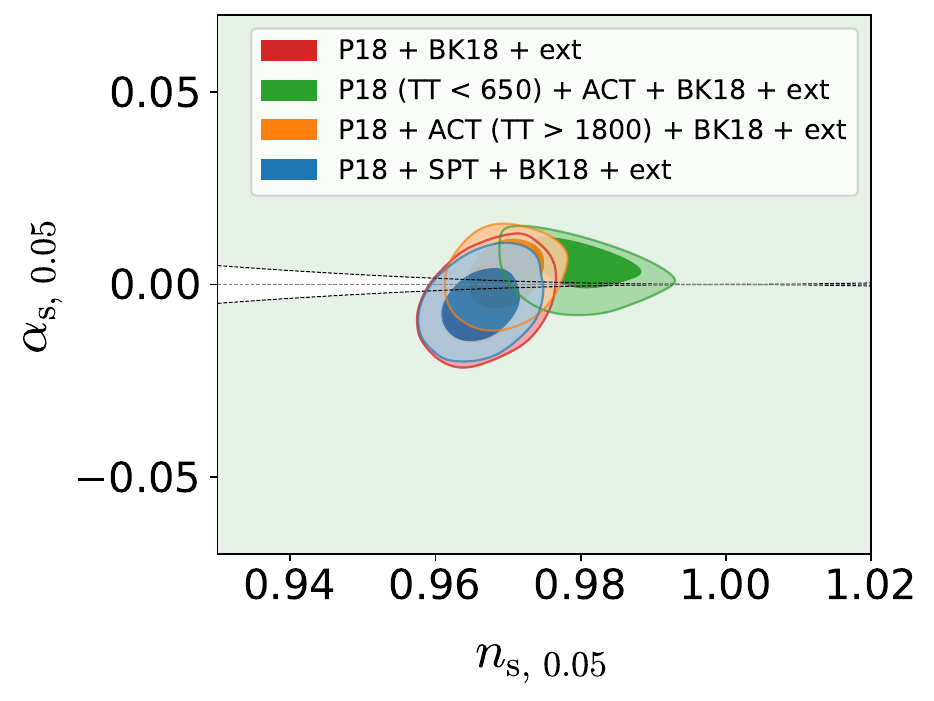}
\includegraphics[width=0.49\textwidth]{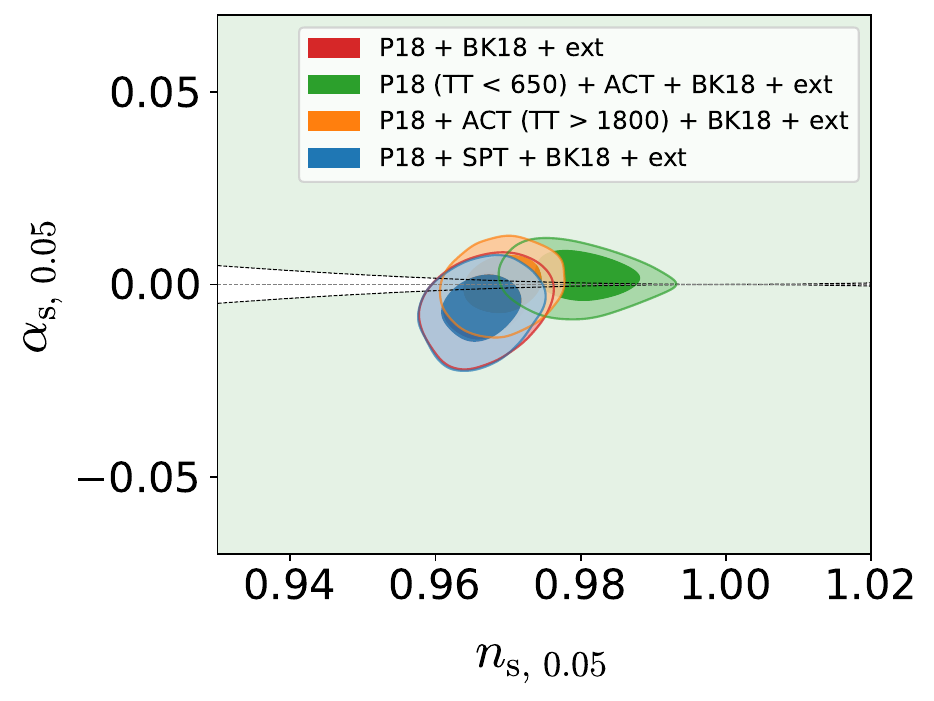}
\includegraphics[width=0.49\textwidth]{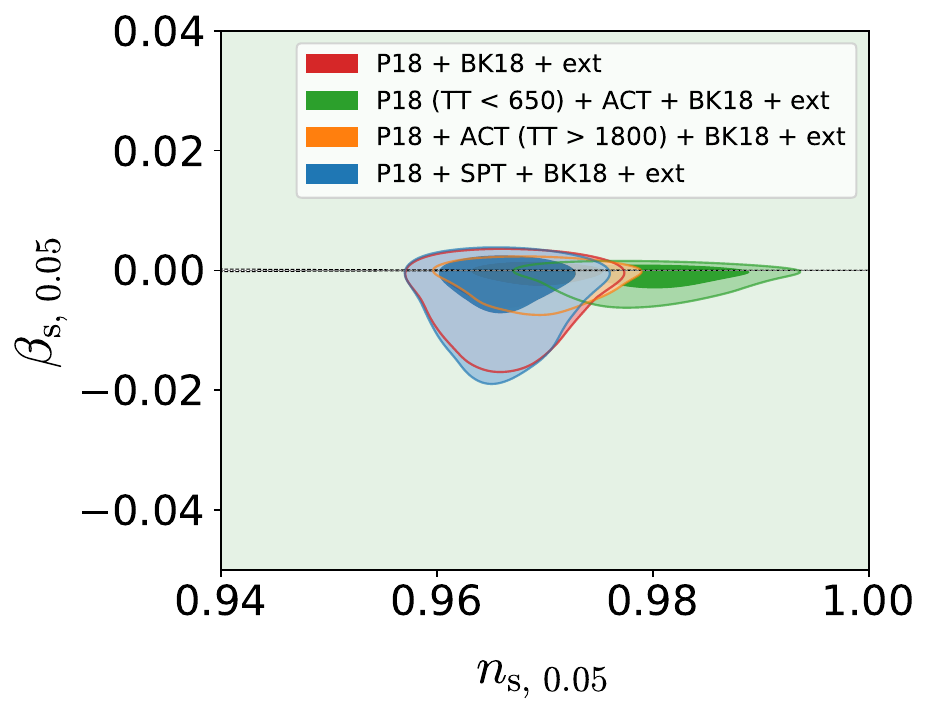}
\caption{Marginalised joint confidence contours for the scalar spectral index $\ns$ and its running $\as$ assuming 
second-order slow-roll predictions (upper panels) and for the scalar spectral index $\ns$, its running $\as$, and the running 
of the running $\bs$ assuming third-order slow-roll predictions (lower panels).}\label{fig:Eps1234_running}
\end{figure}

Finally, we show the  68\% CL and 95\% CL marginalised constraints on the scalar inflationary parameters, derived 
using~\cref{eqn:ns,eqn:as,eqn:bs}, that are for the second-order HFF expansion the scalar spectral index $\ns$ and its running 
$\as$ in~\cref{fig:Eps1234_running} while for the third-order expansion include also the running of the running of the scalar 
spectral index $\bs$, see~\cref{fig:Eps1234_running}. 
Although current constraints are consistent with small values for $1-\ns$, $\as$, and $\bs$ as predicted by single-field 
slow-roll inflationary models, much larger values are still allowed by current CMB measurements. Here the light shaded region 
corresponds to $|\as| > |\ns-1|^2$ for which we have qualitatively violation of the single-field slow-roll predictions and 
analogously $|\bs| > |\ns-1|^3$, as presented in Ref.~\cite{2018JCAP...01..019V}. 

\subsection{Future CMB Constraints}
In this subsection, we explore the forecasted capabilities of a concept for a future CMB space mission. Assuming that CMB 
anisotropies follow a Gaussian distribution and are statistically isotropic, we use the following Gaussian likelihood 
${\cal L}$ \cite{Knox:1995dq,Hamimeche:2008ai}
\begin{equation}
    - 2 \ln {\cal L} = \sum_{\ell} (2\ell+1) f_\mathrm{sky} \left( \mathrm{Tr} [ \hat{{\bf C}}_\ell \bar{{\bf C}}_\ell^{-1}] +
    \ln |\hat{{\bf C}}_\ell \bar{{\bf C}}_\ell^{-1}| - 4 \right),
\label{like}
\end{equation}
where $\bar{{\bf C}}_\ell$ denote the theoretical data covariance matrices 
\begin{equation}
\label{eq:covariance_definition}
\bar{\boldsymbol{C}}_\ell \equiv
 \begin{bmatrix}
\bar{C}_\ell^{TT} + N_\ell^{TT} & \bar{C}_\ell^{TE} & 0 & 0 \\
\bar{C}_\ell^{TE} & \bar{C}_\ell^{EE} + N_\ell^{EE} & 0 & 0 \\
0 & 0 & \bar{C}_\ell^{BB} + N_\ell^{BB} & 0  \\
0  & 0 & 0 & \bar{C}_\ell^{\phi\phi} + N_\ell^{\phi\phi}
\end{bmatrix} \,,
\end{equation}
and $\hat{{\bf C}}_\ell$ are the fiducial data covariance matrices
\begin{equation}
\hat{\boldsymbol{C}}_\ell \equiv
 \begin{bmatrix}
\hat{C}_\ell^{TT} & \hat{C}_\ell^{TE} & 0 & 0 \\
\hat{C}_\ell^{TE} & \hat{C}_\ell^{EE} & 0 & 0 \\
0 & 0 & \hat{C}_\ell^{BB} & 0  \\
0  & 0 & 0 & \hat{C}_\ell^{\phi\phi} 
\end{bmatrix} \,.
\end{equation}

Noise power spectra for the temperature and polarisation angular power spectra account for isotropic noise deconvolved with 
the instrumental Gaussian beam as
\begin{equation}
    N_\ell^{\mathrm XX} = w_{XX}^{-1} e^{ \ell (\ell+1) \frac{\theta_{\rm FWHM}}{8 \ln 2}} \,.
\end{equation}
We assume an effective noise variance $w_{\rm TT}^{-1/2} = 1.2\, \mu{\rm K} \, {\rm arcmin}$ and 
$w_{\rm EE}^{-1/2} = w_{\rm BB}^{-1/2} = \sqrt{2}\, w_{\rm TT}^{-1/2}$ and an effective beam resolution of 
$\theta_{\rm FWHM} = 5.5 \, {\rm arcmin}$ over 70\% of the sky considering the multipole range $2 \leq \ell \leq 3000$. 
These instrumental specifications correspond to a CMB experiment cosmic-variance limited up to $\ell = 2800$ in temperature, 
$\ell = 2000$ in E-mode polarisation, and $\ell = 800$ in the gravitational lensing over almost the full sky.

Together with the primary temperature and polarisation anisotropy signal, we also take information from CMB weak 
lensing into account, considering the power spectrum of the CMB lensing potential $C_\ell^{\phi\phi}$. For the CMB lensing noise power 
spectrum, we adopt the minimum-variance quadratic estimator for the lensing reconstruction \cite{Hu:2001kj} in the range 
$30 \leq \ell \leq 3000$, combining the TT, EE, BB, TE, TB, and EB estimators and applying iterative lensing reconstruction 
according to Refs.~\cite{Hirata:2003ka,Smith:2010gu}. 

Finally, we consider internal delensing \cite{2002PhRvL..89a1303K,2002PhRvL..89a1304K,2004PhRvD..69d3005S}, with the above 
specifications, of the B-mode angular power spectrum in order to reduce the non-primordial contribution induced by lensing. 
We implement the delensing removing the lensing contribution to the B-mode angular power spectra and adding an error 
contribution to the instrumental noise $N_\ell^{\rm BB}$.

\begin{figure}[!hb]
\centering
\includegraphics[width=\textwidth]{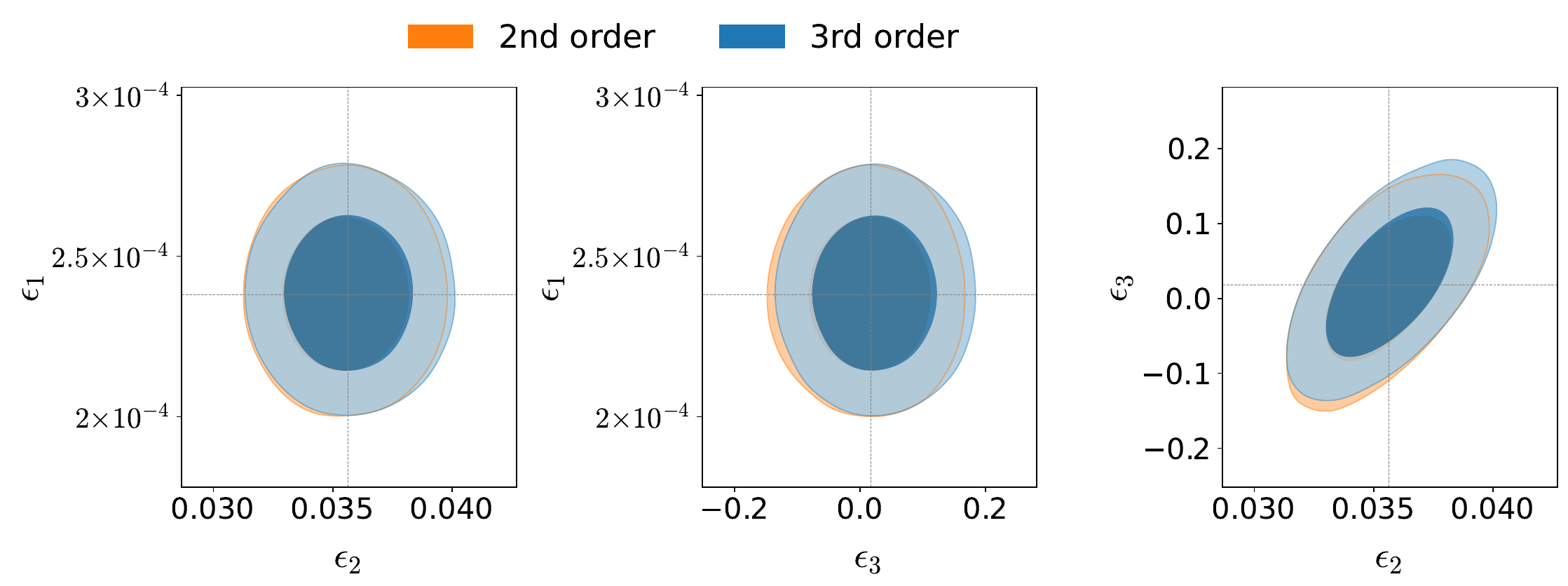}
\includegraphics[width=0.49\textwidth]{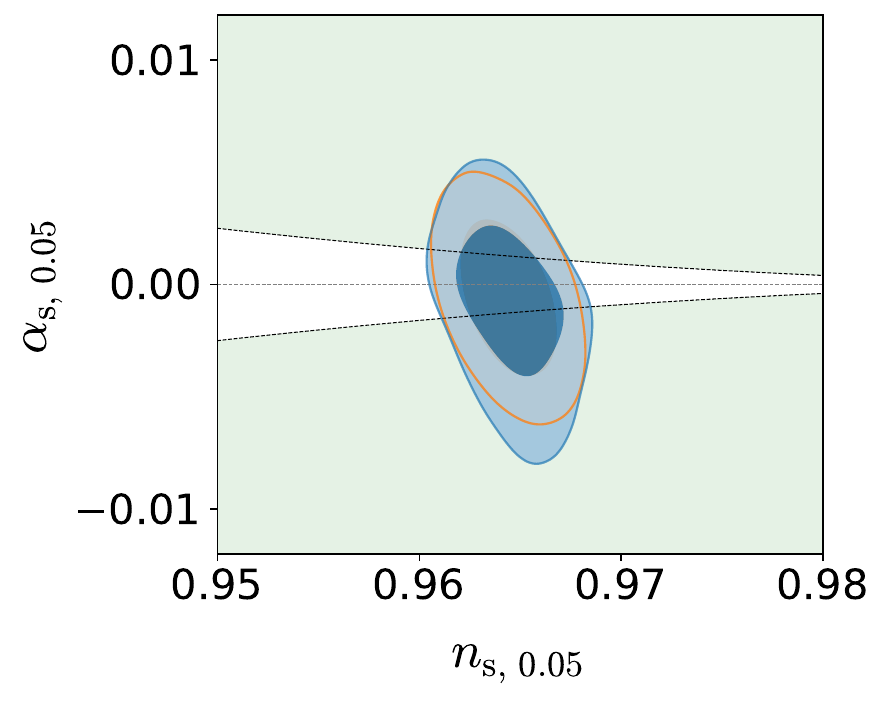}
\includegraphics[width=0.49\textwidth]{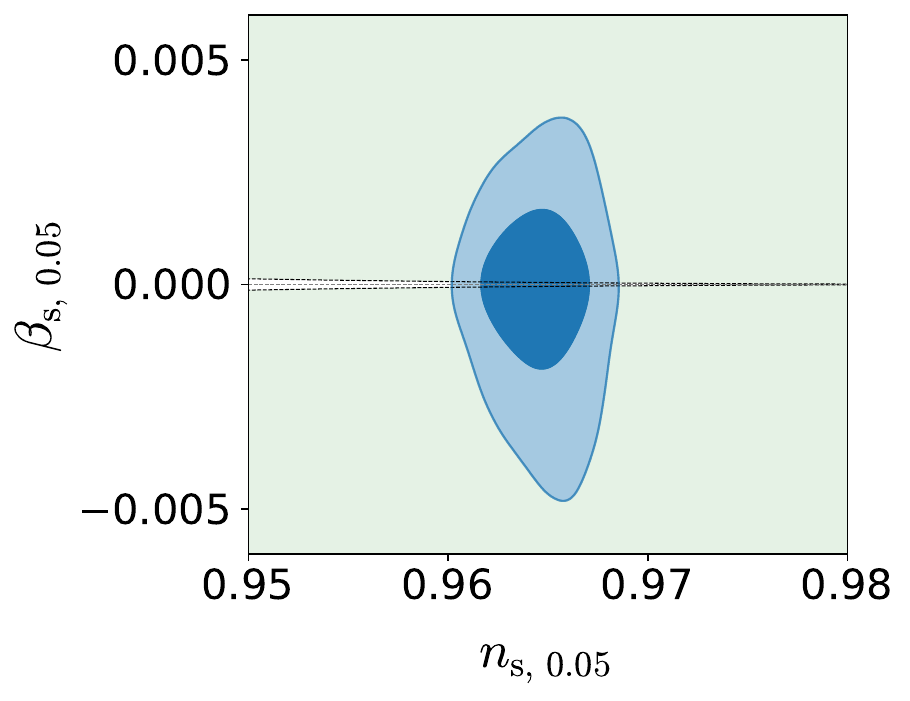}
\caption{Marginalised joint confidence contours for the first two HFF parameters $\epsilon_1$, $\epsilon_2$, and $\epsilon_3$ 
assuming second- and third-order slow-roll predictions (upper panels). Marginalised joint confidence contours for the scalar 
spectral index $\ns$ and its running $\as$ assuming second- and third-order slow-roll predictions (lower left panel) and for 
the scalar spectral index $\ns$ and its running of the running $\bs$ assuming third-order slow-roll predictions (lower right 
panel).}\label{fig:Eps1234_forecast}
\end{figure}
We assume a flat Universe with a cosmological constant, two massless neutrino with $N_{\rm eff} = 2.046$, and a massive one 
with fixed minimum mass $m_\nu = 0.06\, {\rm eV}$. To generate the fiducial angular power spectra, we have fixed the HFFs to 
the expected numbers for T-model of $\alpha$-attractor inflation with $\alpha = 1$ assuming $N_* = 55$ at 
$k_* = 0.05\,{\rm Mpc}^{-1}$. 
This corresponds to $\epsilon_1 = 0.000235$, $\epsilon_2 = 0.0352$, $\epsilon_3 = 0.0175$, and $\epsilon_4 = 0.0175$. 
On these simulated measurements, we analyse the PPS equations at second order, with $\epsilon_4 = 0$ and neglecting 
third-order corrections, and the full third-order expressions. The results are shown in~\cref{fig:Eps1234_forecast}. 
We conclude that for a futuristic CMB experiment alone:
\begin{itemize}
    \item HFFs are well recovered without any bias or significant change in the uncertainties stopping at second order; 
    \item the fourth HFF $\epsilon_4$ remains unconstrained considering CMB experiments only;
    \item uncertainties on the HFFs are significantly reduced with a high significance statistical detection of $\epsilon_2$ 
    (for this specific fiducial also $\epsilon_1$ is well measured). For these case, we obtain 
    $\epsilon_1 = 0.000239\pm 0.000016$, $\epsilon_2 = 0.0357\pm 0.0018$, and $\epsilon_3 = 0.023\pm 0.066$ at 68\% CL 
    consistently with previous forecast studies, see Ref.~\cite{CORE:2016ymi}.
\end{itemize}

\section{Conclusions} \label{sec:conclusions}
In this paper, we conducted an extensive analysis of the primordial power spectra (PPS) for both scalar and tensor 
perturbations, focusing on the higher-order corrections within the slow-roll inflationary framework. By utilising Green's 
function techniques to solve the Mukhanov-Sasaki equation and its tensor counterpart, we were able to extend our analytical 
calculations up to third-order corrections, thereby refining the perturbative expansion in terms of slow-roll parameters. 
Third-order corrections for both scalar and tensor PPS have already been calculated in Ref.~\cite{Auclair:2022yxs}. 
Here we have used a different strategy to solve the integrals, and found some, but small, differences in some of the coefficients 
which multiply the third-order terms. These different coefficients only appear in the constant part of the PPS without affecting the results 
obtained for the spectral indices and their slopes. We have verified that the differences in PPS are numerically negligible.

Our results demonstrate that higher-order corrections significantly enhance the accuracy of the predicted power spectra, 
spectral indices, and their derivatives. However, in going from the second to the third-order expansion, the improvement becomes 
appreciable at very small scales, $k \gtrsim 10\, {\rm Mpc}^{-1}$ leaving almost unaffected the CMB angular power spectra, 
see~\cref{fig:Tmodel_spectra,fig:Tmodel_parameters,fig:KKLT_spectra,fig:KKLT_parameters,fig:CMB_diff}. 
Since the accuracy requirement for tensor quantities is less than on the scalars, it is the scalars upon which attention 
should be focused.

We investigated the constraints on the first four HFFs $\epsilon_i$ obtained from CMB anisotropy measurements in combination 
with late-time cosmological observations such as uncalibrated Type Ia Supernovae from the Pantheon catalogue, baryon acoustic 
oscillations and redshift space distortions from SDSS/BOSS/eBOSS. 
Regarding the CMB datasets considered, we study the impact of different combination of temperature and E-mode polarisation data 
from \Planck, ACT, and SPT. We always included in our analysis B-mode polarisation measurements from BICEP/Keck. Our analysis 
yields a stringent upper limit for the first HFF, $\epsilon_1 \lesssim 0.002$ at 95\% CL, primarily constrained by BICEP/Keck 
data. This result underscores the robustness of $\epsilon_1$ across various combinations of observational datasets. By 
combining data from \Planck\ with late-time cosmological observations, we derived $\epsilon_2 \simeq 0.031 \pm 0.004$ at 68\% 
CL when considering only first-order corrections. Including second- and third-order corrections broadens this constraint to 
$\epsilon_2 \simeq 0.034 \pm 0.007$ at 68\% CL. For the third HFF, our findings indicate a value of 
$\epsilon_3 \simeq 0.1 \pm 0.4$ at 95\% CL, which remains consistent across second- and third-order corrections. The fourth 
HFF $\epsilon_4$, however, remains unconstrained by current data as shown also in Ref.~\cite{Martin:2024qnn}.

We then add small scale CMB measurements from ACT and SPT to the \Planck\ data. Our combination of \Planck\ data with 
measurements from ACT (were we removed the temperature data below $\ell = 1800$ as recommended from the ACT collaboration to 
avoid correlations between the two datasets) and from SPT led to consistent results with slightly more stringent constraints on 
$\epsilon_2$ and $\epsilon_3$. ACT data alone results and the combination of ACT with \Planck\ temperature data removed above 
$\ell = 650$, lead to shifts in the mean values of $\epsilon_2$ and $\epsilon_3$, which suggest a preference for higher values 
of the scalar spectral index and positive values for the running of the scalar spectral index.

We studied the case of a futuristic CMB experiments with realistic and almost cosmic-variance limited specifications. 
These results show that second-order equations are accurate enough to describe current and future CMB observations, 
see~\cref{fig:Eps1234_forecast}. $\epsilon_4$ remains unconstrained also for future CMB measurements.

In conclusion, future small-scale CMB measurements from ACT, SPT, and Simons Observatory will be crucial to further test 
high-order terms in the slow-roll expansion and the validity of single-field slow-roll predictions. In 
addition, future data over a wider range of scales such as large-scale structure (LSS) measurements from {\em Euclid} 
\cite{Ballardini:2016hpi}, CMB spectral distorsions \cite{Cabass:2016ldu}, 21 cm experiments \cite{Munoz:2016owz}, and on 
smaller scales abundance of primordial black holes can offer useful constraints on the primordial curvature power spectrum 
\cite{Sato-Polito:2019hws}. These additional cosmological probes will enable us to confirm single-field slow-roll inflation, 
for instance through a detection of the running of the scalar spectral index 
\cite{Ballardini:2016hpi,Bahr-Kalus:2022prj,Easther:2021rdg}, or to falsify it.

\section*{Acknowledgements}
MB and SSS acknowledge financial support from the INFN InDark initiative. 
MB also acknowledges financial support from the COSMOS network ({\tt www.cosmosnet.it}) through the ASI (Italian Space Agency) 
Grants 2016-24-H.0, 2016-24-H.1-2018, 2020-9-HH.0 (participation in LiteBIRD phase A) and by ``Bando Giovani anno 2023 per 
progetti di ricerca finanziati con il contributo 5x1000 anno 2021''. 
SSS acknowledges that this publication was produced while attending the PhD program in Space Science and Technology at the University of Trento, Cycle XXXVIII, with the support of a scholarship co-financed by the Ministerial Decree no. 351 of 9th April 2022, based on the NRRP - funded by the European Union - NextGenerationEU - Mission 4 "Education and Research", Component 2 "From Research to Business", Investment 3.3. SSS acknowledges that this publication is based upon work from COST Action CA21136 – “Addressing observational tensions in cosmology with systematics and fundamental physics (CosmoVerse)”, supported by COST (European Cooperation in Science and Technology).

\newpage
\appendix

\section{Useful Integrals} \label{sec:app_int}
Here some useful integrals used to manipulate the solutions entering in the third-order Green's function solution~\cref{eqn:y}. 
These integrals have been computed by repeatedly applying integration by parts.

\begin{subequations}
\begin{align}
\int_x^\infty\,\frac{\dd u}{u^2}\,e^{2iu}&=\frac{e^{2ix}}{x}+2i\int_x^\infty\,\frac{\dd u}{u}\,e^{2iu}\\[.2truecm]
\int_x^\infty\,\frac{\dd u}{u^3}\,e^{2iu}&=\left(\frac{1}{2x^2}+\frac{i}{x}\right)e^{2ix}-2\int_x^\infty\,\frac{\dd u}{u}\,e^{2iu}\\[.2truecm]
\int_x^\infty\,\frac{\dd u}{u^4}\,e^{2iu}&=\left(\frac{1}{3x^3}+\frac{i}{3x^2}-\frac{2}{3x}\right)e^{2ix}-\frac{4i}{3}\int_x^\infty\,\frac{\dd u}{u}\,e^{2iu}\\[.2truecm]\notag
\end{align}
\end{subequations}

\begin{subequations}
\begin{align}\notag
\int_x^\infty\,\frac{\dd u}{u^2}\,e^{-2iu}\int_u^\infty\,\frac{\dd t}{t}\,e^{2it}&=-\frac{1}{x}+\frac{e^{-2ix}}{x}\int_x^\infty\,\frac{\dd u}{u}\,e^{2iu}\\
&\quad-2i\int_x^\infty\,\frac{\dd u}{u}\,e^{-2iu}\int_u^\infty\,\frac{\dd t}{t}\,e^{2it}\\[.2truecm]\notag
\int_x^\infty\,\frac{\dd u}{u^3}\,e^{-2iu}\int_u^\infty\,\frac{\dd t}{t}\,e^{2it}&=-\left(\frac{1}{4x^2}-\frac{i}{x}\right)+\left(\frac{1}{2x^2}-\frac{i}{x}\right)e^{-2ix}\int_x^\infty\,\frac{\dd u}{u}\,e^{2iu}\\
&\quad-2\int_x^\infty\,\frac{\dd u}{u}\,e^{-2iu}\int_u^\infty\,\frac{\dd t}{t}\,e^{2it}\\[.2truecm]\notag
\int_x^\infty\,\frac{\dd u}{u^4}\,e^{-2iu}\int_u^\infty\,\frac{\dd t}{t}\,e^{2it}&=-\left(\frac{1}{9x^3}-\frac{i}{6x^2}-\frac{2}{3x}\right)+\left(\frac{1}{3x^3}-\frac{i}{3x^2}-\frac{2}{3x}\right)e^{-2ix}\int_x^\infty\,\frac{\dd u}{u}\,e^{2iu}\\
&\quad+\frac{4i}{3}\int_x^\infty\,\frac{\dd u}{u}\,e^{-2iu}\int_u^\infty\,\frac{\dd t}{t}\,e^{2it}\\[.2truecm]\notag
\end{align}
\end{subequations}

\begin{subequations}
\begin{align}
\int_x^\infty\,\frac{\dd u}{u^2}\,e^{2iu}\ln u&=\frac{e^{2ix}}{x}+\frac{e^{2ix}\ln x}{x}+2i\int_x^\infty\,\frac{\dd u}{u}\,e^{2iu}+2i\int_x^\infty\,\frac{\dd u}{u}\,e^{2iu}\ln u\\[.2truecm]\notag
\int_x^\infty\,\frac{\dd u}{u^3}\,e^{2iu}\ln u&=\left(\frac{1}{4x^2}+\frac{3i}{2x}\right)e^{2ix}+\left(\frac{1}{2x^2}+\frac{i}{x}\right)e^{2ix}\ln x\\
&\quad-3\int_x^\infty\,\frac{\dd u}{u}\,e^{2iu}-2\int_x^\infty\,\frac{\dd u}{u}\,e^{2iu}\ln u\\[.2truecm]\notag
\int_x^\infty\,\frac{\dd u}{u^4}\,e^{2iu}\ln u&=\left(\frac{1}{9x^3}+\frac{5i}{18x^2}-\frac{11}{9x}\right)e^{2ix}+\left(\frac{1}{3x^3}+\frac{i}{3x^2}-\frac{2}{3x}\right)e^{2ix}\ln x\\
&\quad-\frac{22i}{9}\int_x^\infty\,\frac{\dd u}{u}\,e^{2iu}-\frac{4i}{3}\int_x^\infty\,\frac{\dd u}{u}\,e^{2iu}\ln u\\[.2truecm]\notag
\end{align}
\end{subequations}

\begin{subequations}
\begin{align}\notag
\int_x^\infty\,\frac{\dd u}{u^2}\int_u^\infty\frac{\dd t}{t}\,e^{-2it}\int_t^\infty\frac{\dd s}{s}\,e^{2is}&=\frac{1}{x}-\frac{e^{-2ix}}{x}\int_x^\infty\,\frac{\dd u}{u}\,e^{2iu}\\
&\quad+\left(\frac{1}{x}+2i\right)\int_x^\infty\frac{\dd u}{u}\,e^{-2iu}\int_u^\infty\frac{\dd t}{t}\,e^{2it}\\[.2truecm]\notag
\int_x^\infty\,\frac{\dd u}{u^4}\int_u^\infty\frac{\dd t}{t}\,e^{-2it}\int_t^\infty\frac{\dd s}{s}\,e^{2is}&=\left(\frac{1}{27x^3}-\frac{i}{18x^2}-\frac{2}{9x}\right)\\\notag
&\quad-\left(\frac{1}{9x^3}-\frac{i}{9x^2}-\frac{2}{9x}\right)e^{-2ix}\int_x^\infty\,\frac{\dd u}{u}\,e^{2iu}\\
&\quad+\left(\frac{1}{3x^3}-\frac{4i}{9}\right)\int_x^\infty\frac{\dd u}{u}\,e^{-2iu}\int_u^\infty\frac{\dd t}{t}\,e^{2it}\\[.2truecm]\notag
\end{align}
\end{subequations}

\begin{subequations}
\begin{align}\notag
\int_x^\infty\,\frac{\dd u}{u^2}\,e^{2iu}\int_u^\infty\frac{\dd t}{t}\,e^{-2it}\int_t^\infty\frac{\dd s}{s}\,e^{2is}&=\frac{e^{2ix}}{x}-\left(\frac{1}{x}-2i\right)\int_x^\infty\frac{\dd u}{u}\,e^{2iu}\\\notag
&\quad+\frac{e^{2ix}}{x}\int_x^\infty\frac{\dd u}{u}\,e^{-2iu}\int_u^\infty\frac{\dd t}{t}\,e^{2it}\\
&\quad+2i\int_x^\infty\,\frac{\dd u}{u}\,e^{2iu}\int_u^\infty\frac{\dd t}{t}\,e^{-2it}\int_t^\infty\frac{\dd s}{s}\,e^{2is}\\[.2truecm]\notag
\int_x^\infty\,\frac{\dd u}{u^3}\,e^{2iu}\int_u^\infty\frac{\dd t}{t}\,e^{-2it}\int_t^\infty\frac{\dd s}{s}\,e^{2is}&=\left(\frac{1}{8x^2}+\frac{5i}{4x}\right)e^{2ix}-\left(\frac{1}{4x^2}+\frac{i}{x}+\frac{5}{2}\right)\int_x^\infty\frac{\dd u}{u}\,e^{2iu}\\\notag
&\quad+\left(\frac{1}{2x^2}+\frac{i}{x}\right)e^{2ix}\int_x^\infty\frac{\dd u}{u}\,e^{-2iu}\int_u^\infty\frac{\dd t}{t}\,e^{2it}\\
&\quad-2\int_x^\infty\,\frac{\dd u}{u}\,e^{2iu}\int_u^\infty\frac{\dd t}{t}\,e^{-2it}\int_t^\infty\frac{\dd s}{s}\,e^{2is}\\[.2truecm]\notag
\int_x^\infty\,\frac{\dd u}{u^4}\,e^{2iu}\int_u^\infty\frac{\dd t}{t}\,e^{-2it}\int_t^\infty\frac{\dd s}{s}\,e^{2is}&=\left(\frac{1}{27x^3}+\frac{13i}{108x^2}-\frac{49}{54x}\right)e^{2ix}\\\notag
&\quad-\left(\frac{1}{9x^3}+\frac{i}{6x^2}-\frac{2}{3x}+\frac{49i}{27}\right)\int_x^\infty\frac{\dd u}{u}\,e^{2iu}\\\notag
&\quad+\left(\frac{1}{3x^3}+\frac{i}{3x^2}-\frac{2}{3x}\right)e^{2ix}\int_x^\infty\frac{\dd u}{u}\,e^{-2iu}\int_u^\infty\frac{\dd t}{t}\,e^{2it}\\
&\quad-\frac{4i}{3}\int_x^\infty\,\frac{\dd u}{u}\,e^{2iu}\int_u^\infty\frac{\dd t}{t}\,e^{-2it}\int_t^\infty\frac{\dd s}{s}\,e^{2is}\\[.2truecm]\notag
\end{align}
\end{subequations}

\begin{subequations}
\begin{align}\notag
\int_x^\infty\,\frac{\dd u}{u^2}\,e^{-2iu}\int_u^\infty\frac{\dd t}{t}\,e^{2it}\ln t&=-\frac{1}{x}-\frac{\ln x}{x}+\frac{e^{-2ix}}{x}\int_x^\infty\frac{\dd u}{u}\,e^{2iu}\ln u\\
&\quad-2i\int_x^\infty\,\frac{\dd u}{u}\,e^{-2iu}\int_u^\infty\frac{\dd t}{t}\,e^{2it}\ln t\\[.2truecm]\notag
\int_x^\infty\,\frac{\dd u}{u^3}\,e^{-2iu}\int_u^\infty\frac{\dd t}{t}\,e^{2it}\ln t&=-\left(\frac{1}{8x^2}-\frac{i}{x}\right)-\left(\frac{1}{4x^2}-\frac{i}{x}\right)\ln x\\\notag
&\quad+\left(\frac{1}{2x^2}-\frac{i}{x}\right)e^{-2ix}\int_x^\infty\frac{\dd u}{u}\,e^{2iu}\ln u\\
&\quad-2\int_x^\infty\,\frac{\dd u}{u}\,e^{-2iu}\int_u^\infty\frac{\dd t}{t}\,e^{2it}\ln t\\[.2truecm]\notag
\int_x^\infty\,\frac{\dd u}{u^4}\,e^{-2iu}\int_u^\infty\frac{\dd t}{t}\,e^{2it}\ln t&=-\left(\frac{1}{27x^3}-\frac{i}{12x^2}-\frac{2}{3x}\right)-\left(\frac{1}{9x^3}-\frac{i}{6x^2}-\frac{2}{3x}\right)\ln x\\\notag
&\quad+\left(\frac{1}{3x^3}-\frac{i}{3x^2}-\frac{2}{3x}\right)e^{-2ix}\int_x^\infty\frac{\dd u}{u}\,e^{2iu}\ln u\\
&\quad+\frac{4i}{3}\int_x^\infty\,\frac{\dd u}{u}\,e^{-2iu}\int_u^\infty\frac{\dd t}{t}\,e^{2it}\ln t\\[.2truecm]\notag
\end{align}
\end{subequations}

\begin{subequations}
\begin{align}
\int_x^\infty\,\frac{\dd u}{u^2}\int_u^\infty\frac{\dd t}{t}\,e^{2it}&=-\frac{e^{2ix}}{x}+\left(\frac{1}{x}-2i\right)\int_x^\infty\frac{\dd u}{u}\,e^{2iu}\\[.2truecm]
\int_x^\infty\,\frac{\dd u}{u^3}\int_u^\infty\frac{\dd t}{t}\,e^{2it}&=-\left(\frac{1}{4x^2}+\frac{i}{2x}\right)e^{2ix}+\left(\frac{1}{2x^2}+1\right)\int_x^\infty\frac{\dd u}{u}\,e^{2iu}\\[.2truecm]
\int_x^\infty\,\frac{\dd u}{u^4}\int_u^\infty\frac{\dd t}{t}\,e^{2it}&=-\left(\frac{1}{9x^3}+\frac{i}{9x^2}-\frac{2}{9x}\right)e^{2ix}+\left(\frac{1}{3x^3}+\frac{4i}{9}\right)\int_x^\infty\frac{\dd u}{u}\,e^{2iu}\\[.2truecm]\notag
\end{align}
\end{subequations}

\begin{subequations}
\begin{align}\notag
\int_x^\infty\,\frac{\dd u}{u^2}\int_u^\infty\frac{\dd t}{t}\,e^{2it}\ln t&=-\frac{e^{2ix}}{x}-\frac{e^{2ix}\ln x}{x}\\
&\quad-2i\int_x^\infty\frac{\dd u}{u}\,e^{2iu}+\left(\frac{1}{x}-2i\right)\int_x^\infty\frac{\dd u}{u}\,e^{2iu}\ln u\\[.2truecm]\notag
\int_x^\infty\,\frac{\dd u}{u^4}\int_u^\infty\frac{\dd t}{t}\,e^{2it}\ln t&=-\left(\frac{1}{27x^3}+\frac{5i}{54x^2}-\frac{11}{27x}\right)e^{2ix}-\left(\frac{1}{9x^3}+\frac{i}{9x^2}-\frac{2}{9x}\right)e^{2ix}\ln x\\
&\quad+\frac{22i}{27}\int_x^\infty\frac{\dd u}{u}\,e^{2iu}+\left(\frac{1}{3x^3}+\frac{4i}{9}\right)\int_x^\infty\frac{\dd u}{u}\,e^{2iu}\ln u\\[.2truecm]\notag
\end{align}
\end{subequations}

\begin{subequations}
\begin{align}\notag
\int_x^\infty\,\frac{\dd u}{u^2}\,e^{-2iu}\ln u\int_u^\infty\frac{\dd t}{t}\,e^{2it}&=-\frac{2}{x}-\frac{\ln x}{x}+\frac{e^{-2ix}}{x}\int_x^\infty\frac{\dd u}{u}\,e^{2iu}\\\notag
&\quad+\frac{e^{-2ix}\ln x}{x}\int_x^\infty\frac{\dd u}{u}\,e^{2iu}-2i\int_x^\infty\frac{\dd u}{u}\,e^{-2iu}\int_u^\infty\frac{\dd t}{t}\,e^{2it}\\
&\quad-2i\int_x^\infty\frac{\dd u}{u}\,e^{-2iu}\ln u\int_u^\infty\frac{\dd t}{t}\,e^{2it}\\[.2truecm]\notag
\int_x^\infty\,\frac{\dd u}{u^3}\,e^{-2iu}\ln u\int_u^\infty\frac{\dd t}{t}\,e^{2it}&=-\left(\frac{1}{4x^2}-\frac{5i}{2x}\right)-\left(\frac{1}{4x^2}-\frac{i}{x}\right)\ln x\\\notag
&\quad+\left(\frac{1}{4x^2}-\frac{3i}{2x}\right)e^{-2ix}\int_x^\infty\frac{\dd u}{u}\,e^{2iu}\\\notag
&\quad+\left(\frac{1}{2x^2}-\frac{i}{x}\right)e^{-2ix}\ln x\int_x^\infty\frac{\dd u}{u}\,e^{2iu}\\\notag
&\quad-3\int_x^\infty\frac{\dd u}{u}\,e^{-2iu}\int_u^\infty\frac{\dd t}{t}\,e^{2it}\\
&\quad-2\int_x^\infty\frac{\dd u}{u}\,e^{-2iu}\ln u\int_u^\infty\frac{\dd t}{t}\,e^{2it}\\[.2truecm]\notag
\int_x^\infty\,\frac{\dd u}{u^4}\,e^{-2iu}\ln u\int_u^\infty\frac{\dd t}{t}\,e^{2it}&=-\left(\frac{2}{27x^3}-\frac{2i}{9x^2}-\frac{17}{9x}\right)-\left(\frac{1}{9x^3}-\frac{i}{6x^2}-\frac{2}{3x}\right)\ln x\\\notag
&\quad+\left(\frac{1}{9x^3}-\frac{5i}{18x^2}-\frac{11}{9x}\right)e^{-2ix}\int_x^\infty\frac{\dd u}{u}\,e^{2iu}\\\notag
&\quad+\left(\frac{1}{3x^3}-\frac{i}{3x^2}-\frac{2}{3x}\right)e^{-2ix}\ln x\int_x^\infty\frac{\dd u}{u}\,e^{2iu}\\\notag
&\quad+\frac{22i}{9}\int_x^\infty\frac{\dd u}{u}\,e^{-2iu}\int_u^\infty\frac{\dd t}{t}\,e^{2it}\\
&\quad+\frac{4i}{3}\int_x^\infty\frac{\dd u}{u}\,e^{-2iu}\ln u\int_u^\infty\frac{\dd t}{t}\,e^{2it}
\end{align}
\end{subequations}

\begin{subequations}
\begin{align}\notag
\int_x^\infty\,\frac{\dd u}{u^2}\,\ln u\int_u^\infty\frac{\dd t}{t}\,e^{2it}&=-\frac{2e^{2ix}}{x}-\frac{e^{2ix}\ln x}{x}+\left(\frac{1}{x}-4i\right)\int_x^\infty\frac{\dd u}{u}\,e^{2iu}\\
&\quad+\frac{\ln x}{x}\int_x^\infty\frac{\dd u}{u}\,e^{2iu}-2i\int_x^\infty\frac{\dd u}{u}\,e^{2iu}\ln u\\[.2truecm]\notag
\int_x^\infty\,\frac{\dd u}{u^4}\,\ln u\int_u^\infty\frac{\dd t}{t}\,e^{2it}&=-\left(\frac{2}{27x^3}+\frac{7i}{54x^2}-\frac{13}{27x}\right)e^{2ix}-\left(\frac{1}{9x^3}-\frac{i}{9x^2}-\frac{2}{9x}\right)e^{2ix}\ln x\\\notag
&\quad+\left(\frac{1}{9x^3}+\frac{26i}{27}\right)\int_x^\infty\frac{\dd u}{u}\,e^{2iu}\\\notag
&\quad+\frac{\ln x}{3x^3}\int_x^\infty\frac{\dd u}{u}\,e^{2iu}\\
&\quad+\frac{4i}{9}\int_x^\infty\frac{\dd u}{u}\,e^{2iu}\ln u
\end{align}
\end{subequations}

\begin{subequations}
\begin{align}\notag
\int_x^\infty\,\frac{\dd u}{u^2}\,e^{2iu}\ln^2 u&=\frac{2e^{2ix}}{x}+\frac{2e^{2ix}\ln x}{x}+\frac{e^{2ix}\ln^2x}{x}\\\notag
&\quad+4i\int_x^\infty\,\frac{\dd u}{u}\,e^{2iu}+4i\int_x^\infty\,\frac{\dd u}{u}\,e^{2iu}\ln u\\
&\quad+2i\int_x^\infty\,\frac{\dd u}{u}\,e^{2iu}\ln^2u\\[.2truecm]\notag
\int_x^\infty\,\frac{\dd u}{u^3}\,e^{2iu}\ln^2 u&=\left(\frac{1}{4x^2}+\frac{7i}{2x}\right)e^{2ix}+\left(\frac{1}{2x^2}+\frac{3i}{x}\right)e^{2ix}\ln x\\\notag
&\quad+\left(\frac{1}{2x^2}+\frac{i}{x}\right)e^{2ix}\ln^2x-7\int_x^\infty\,\frac{\dd u}{u}\,e^{2iu}\\
&\quad-6\int_x^\infty\,\frac{\dd u}{u}\,e^{2iu}\ln u-2\int_x^\infty\,\frac{\dd u}{u}\,e^{2iu}\ln^2u\\[.2truecm]\notag
\int_x^\infty\,\frac{\dd u}{u^4}\,e^{2iu}\ln^2 u&=\left(\frac{2}{27x^3}+\frac{19i}{54x^2}-\frac{85}{27x}\right)e^{2ix}+\left(\frac{2}{9x^3}+\frac{5i}{9x^2}-\frac{22}{9x}\right)e^{2ix}\ln x\\\notag
&\quad+\left(\frac{1}{3x^3}+\frac{i}{3x^2}-\frac{2}{3x}\right)e^{2ix}\ln^2x\\\notag
&\quad-\frac{170i}{27}\int_x^\infty\,\frac{\dd u}{u}\,e^{2iu}-\frac{44i}{9}\int_x^\infty\,\frac{\dd u}{u}\,e^{2iu}\ln u\\
&\quad-\frac{4i}{3}\int_x^\infty\,\frac{\dd u}{u}\,e^{2iu}\ln^2u
\end{align}
\end{subequations}

\section{Super-Hubble Limits for the Integrals} \label{sec:app_int2}
We report here the calculation of the integrals appearing in~\cref{eqn:y_1,eqn:y_21,eqn:y_22,eqn:y_31,eqn:y_32,eqn:y_33} and their 
asymptotic solutions, calculated in the super-Hubble limit, required to calculate the PPS of scalar and tensor perturbations.

We start recognising that~\cref{eqn:y_1} corresponds to the exponential integral
\begin{equation}
    {\rm Ei}(-2 i x) = \int_x^\infty\frac{\dd u}{u}\,e^{2iu} \,,
\end{equation}
defined for $x > 0$ whose series is
\begin{equation}
   {\rm Ei}(z) \equiv \gamma + \ln (-z) + \sum_{n=1}^{+\infty} \frac{(z)^n}{n! n } \,.
\end{equation}
Afterwards, in order to compute the super-Hubble limit for $x\to 0$, we consider terms up to the linear one, obtaining
\begin{equation}
    \lim_{x \to 0} \int_x^\infty\frac{\dd u}{u}\,e^{2iu} = \alpha + \frac{i \pi}{2} - 2 - 2i x - \ln x + {\cal{O}}(x^2) \,,
\end{equation}
where $\alpha = 2-\gamma-\ln 2$. Analogously, using the exponential integral properties, 
we can obtain
\begin{equation}
    \lim_{x \to 0} \int_x^\infty\frac{\dd u}{u}\,e^{2iu}\ln u
    = 2 - i\pi - \frac{\pi^2}{24} + 2i x - 2\alpha + \frac{i\pi\alpha}{2} - 2ix\ln x - \frac{\ln^2 x}{2} + {\cal{O}}(x^2) \,,
\end{equation}
\begin{align}
    \lim_{x \to 0} \int_x^\infty\,\frac{\dd u}{u}\,e^{2iu}\ln^2u 
    =\,& -\frac{8}{3} + 2i\pi + \frac{\pi ^2}{6} + \frac{i \pi ^3}{24}  -4 i x +4 \alpha -2 i \pi  \alpha 
    -\frac{\pi ^2 \alpha }{12} -2 \alpha ^2  \notag\\
    &+ \frac{i \pi  \alpha ^2}{2} + \frac{\alpha ^3}{3} -\frac{2 \zeta (3)}{3} +4 i x \ln x -2 i x \ln^2 x  -\frac{\ln^3 x}{3} + {\cal{O}}(x^2) \,,
\end{align}
see for details Refs.~\cite{1969NISTJ..73B.191G,1972hmfw.book.....A}.

The double integral in~\cref{eqn:y_21} has a known solution, see Ref.~\cite{1969NISTJ..73B.191G}. 
We take again the leading terms in $x$, corresponding to
\begin{align}\label{eqn:F00}
    \lim_{x \to 0} \int_x^\infty\frac{\dd u}{u}\,e^{-2iu}\int_u^\infty\frac{\dd t}{t}\,e^{2it} 
    =\,& 2 - i\pi + \frac{\pi^2}{8} - \pi x - 2\alpha + \frac{i\pi\alpha}{2} + 2ix\alpha + \frac{\alpha^2}{2} \notag\\
    &+ \left(2 - \frac{i\pi}{2} - 2ix - \alpha\right)\ln x + \frac{\ln^2 x}{2} + {\cal O}(x^2) \,.
\end{align}
For the double integral in~\cref{eqn:F01}, we expand the integrand in the limit $u \to 0$ and then integrate the expansion 
between $x$ and $1$ to ensure convergence of the leading terms. This is possible because for large arguments the 
exponential function is suppressed by $\ln x / x$. We obtain 
\begin{align} \label{eqn:int_2}
\lim_{x \to 0} \int_x^\infty\,\frac{\dd u}{u}\,e^{2iu}\ln u\int_u^\infty\frac{\dd t}{t}\,e^{-2it} 
    \simeq\,& \lim_{x \to 0} \int_x^1\, \dd u\, \left[-2 i \ln^2 u+2 i \ln u \right.\notag\\
    &\left.-\frac{1}{2} i \ln (16) \ln u - \frac{\ln (2 u) \ln u}{u} - \gamma \frac{\ln u}{u} \right.\notag\\
    &\left.-\frac{i \pi}{2}  \frac{\ln u}{u} + \pi  \ln u -2 i \gamma  \ln u\right] \notag\\
    =\,& -2i - \pi +2ix + \pi x - 2i\alpha + 2ix\alpha - \left(2ix + \pi x + 2ix\alpha \right)\ln x \notag\\
    &+ \left(1 + \frac{i\pi}{4} +2ix - \frac{\alpha}{2}\right)\ln^2 x + \frac{\ln^3 x}{x} + {\cal O}(x^2) \,.    
\end{align}
This allow us to calculate the double integral entering~\cref{eqn:y_31}, whose asymptotic form is
\begin{align}
    \lim_{x \to 0} \int_x^\infty\,\frac{\dd u}{u}\,e^{-2iu}\int_u^\infty\frac{\dd t}{t}\,e^{2it} \ln t  
    =\,& (-4+2i) +(1+i)\pi - \frac{5\pi^2}{12} +\frac{i\pi^3}{48} -2ix+2\pi x -\frac{i\pi^2 x}{12} \notag\\
    &+ (6+2i)\alpha -i\pi\alpha +\frac{5\pi^2\alpha}{24} -4ix\alpha -\pi x\alpha-3\alpha^2 +\frac{i\pi\alpha^2}{4} \notag\\ 
    &+ i\alpha^2x +\frac{\alpha^3}{2} +\left(-2+i\pi +\frac{\pi^2}{24}+4ix +2\alpha - \frac{i\pi\alpha}{2}-\frac{\alpha^2}{2}\right)\ln x \notag\\ 
    &-ix\ln^2 x +\frac{\ln^3 x}{6} + {\cal O}(x^2) \,.
\end{align}
In~\cref{eqn:y_32}, we also have the complex conjugate of~\cref{eqn:int_2}
\begin{align}
    \lim_{x \to 0} \int_x^\infty\,\frac{\dd u}{u}\,e^{-2iu}\ln u\int_u^\infty\frac{\dd t}{t}\,e^{2it} 
    =\,& 2i - \pi -2ix + \pi x + 2i\alpha - 2ix\alpha + \left(2ix - \pi x + 2ix\alpha \right)\ln x \notag\\
    &+ \left(1 - \frac{i\pi}{4} -2ix - \frac{\alpha}{2}\right)\ln^2 x + \frac{\ln^3 x}{x} + {\cal O}(x^2) \,.
\end{align}

Finally, to solve the triple integral entering~\cref{eqn:y_31}, we take the limit for $u \to 0$ of the double integral in the 
integrand, which is~\cref{eqn:F00}, and then we integrate the leading contributions of order ${\cal O}(x^0)$, multiplied 
by $e^{2iu}/u$, between $x$ and $\infty$. 
This approximation is justified since the integrand rapidly decays to zero for $u > 1$, making contributions from this region negligible. Then we also expand in the super-Hubble limit
\begin{align} \label{eqn:int_3}
    \lim_{x \to 0} &\int_x^\infty\,\frac{\dd u}{u}\,e^{2iu}\int_u^\infty\,\frac{\dd t}{t}\,e^{-2it}\int_t^\infty\,\frac{\dd s}{s}\,e^{2is} 
    = 
    -\frac{4}{3} + i \pi - \frac{\pi^2}{4} + \frac{5i \pi^3}{48} + 2 \alpha - i \pi \alpha  \notag \\ 
    &+ \frac{\pi^2 \alpha}{8} - \alpha^2 + \frac{i \pi \alpha^2}{4} + \frac{\alpha^3}{6}- \frac{\zeta(3)}{3} + \left( - 2  + i \pi  - \frac{\pi^2 }{8} + 2 \alpha  
    - \frac{i \pi \alpha }{2} - \frac{\alpha^2 }{2} \right)\ln x \notag\\
    &+\left(- 1 + \frac{i \pi }{4} + \frac{\alpha }{2} \right)\ln^2 x- \frac{\ln^3 x}{6} + {\cal O}(x) \,,
\end{align}
where $\zeta$ is the Riemann zeta function.

An alternative and more justified strategy is the following. 
We have divided the integral into two parts: one from $x$ to $1$ and one from $1$ to $+\infty$. 
The contribution of the integral from $1$ to $+\infty$ is negligible because the integrand $f(u)$ decreases rapidly for $u \gg 1$. 
In particular, the asymptotic behaviour of $f(u)$ is such that $f(u) \to 0$ quickly for large $u$, so the contribution of the 
interval $[1,+\infty)$ is insignificant compared to the interval $[x,1]$. This observation has been verified numerically, 
confirming that the contribution of the interval $[1,+\infty)$ is negligible.
Analogous to before, we can expand the double integral in the integrand, now keeping contributions of order ${\cal O}(x^2)$ and 
restricting the integral to the interval $[x,1]$ on which the expansion is justified. We obtain
\begin{align} \label{eqn:int_3_alternative}
    \lim_{x \to 0} &\int_x^\infty\,\frac{\dd u}{u}\,e^{2iu}\int_u^\infty\,\frac{\dd t}{t}\,e^{-2it}\int_t^\infty\,\frac{\dd s}{s}\,e^{2is} 
    = 
    -\frac{25}{4} +\left(-1+\pi +\frac{5i\pi}{2} + 5\alpha -2i\alpha\right) \frac{e^{2i}}{4}\notag\\
    & - \frac{i\pi}{4}  - \frac{3\pi^2}{4} - \frac{i\pi^3}{16} + 6\alpha  - i\pi \alpha + \frac{3\pi^2 \alpha}{8} - 3\alpha^2 + \frac{i\pi \alpha^2}{4} + \frac{\alpha^3}{2}\notag\\
    & 
+ \left( \frac{13}{4} - i\pi + \frac{\pi^2}{8} - 2\alpha + \frac{i\pi \alpha}{2} + \frac{\alpha^2}{2} \right) {\rm Ei}(2i) \notag\\
    &
+ \left( -4i - \pi + 2i\alpha \right) {}_3F_3(1,1,1;2,2,2; 2i)
+ 2i \, {}_4F_4(1,1,1,1;2,2,2,2; 2i) \notag\\
    &
 + \left( - 2  + i \pi  - \frac{\pi^2 }{8} + 2 \alpha  
    - \frac{i \pi \alpha }{2} - \frac{\alpha^2 }{2}\right)\ln x + \left(- 1 + \frac{i \pi }{4} + \frac{\alpha }{2}\right)\ln^2 x - \frac{\ln^3 x}{6} + {\cal O}(x) \, ,
\end{align}
where ${}_pF_q$ is the generalized hypergeometric function.

We note that~\cref{eqn:int_3,eqn:int_3_alternative} differ only in the constant terms. Moreover, the solution presented in 
Ref.~\cite{Auclair:2022yxs} also present a different constant term. For all these reasons, we decide to present the super-Hubble 
solution of~\cref{eqn:int_3} as
\begin{align} \label{eqn:int_3_Z}
    =\,&-\frac{4}{3} + i \pi - \frac{\pi^2}{4} + \frac{5i \pi^3}{48} + 2 \alpha - i \pi \alpha + \frac{\pi^2 \alpha}{8} - \alpha^2 + \frac{i \pi \alpha^2}{4} + \frac{\alpha^3}{6} - {\cal Z} \notag \\ 
    &+ \left( - 2  + i \pi  - \frac{\pi^2 }{8} + 2 \alpha  
    - \frac{i \pi \alpha }{2} - \frac{\alpha^2 }{2}\right)\ln x \notag\\
    &+ \left(- 1 + \frac{i \pi }{4} + \frac{\alpha }{2}\right)\ln^2 x - \frac{\ln^3 x}{6} + {\cal O}(x) 
\end{align}
where ${\cal Z} = \zeta(3)/3$ from~\cref{eqn:int_3}, ${\cal Z} \sim 2.97353 - 0.0273557 i$ from~\cref{eqn:int_3_alternative}, 
and ${\cal Z} = 7 \zeta(3)/3$ in Ref.~\cite{Auclair:2022yxs}. Different choices of ${\cal Z}$ do not lead to numerically 
significant differences in the final PPS.

Summarising, all the asymptotic solutions entering the PPS equations agree with the results previously calculated in 
Ref.~\cite{Auclair:2022yxs}, except for one term in~\cref{eqn:int_3}.

\section{Parameterisation of the Power Spectra} \label{app:an_bn}
The power spectra of scalar and tensor perturbations can be estimated through analytical methods. Typically, this involves expanding 
the power spectra around a specific wavenumber, denoted as $k_*$, and then determining the coefficients through a slow-roll expansion 
or another suitable approximation technique. Because the analysis must span multiple orders of magnitude in $k$, the most effective 
expansion variable is $\ln k$. In this regard, two expansions have been proposed in the literature \cite{Leach:2002ar}. The first one, 
already presented up to third-order terms in Ref.~\cite{Auclair:2022yxs}, consists in expanding directly the power spectrum in $\ln k$, 
leading to
\begin{equation}\label{eqn:PSina}
    \frac{\mathcal{P}_X(k)}{\mathcal{P}_{X0}(k_*)} = a_{X0} + a_{X1} \ln\left(\frac{k}{k_*}\right) + \frac{a_{X2}}{2} \ln^2\left(\frac{k}{k_*}\right)+\frac{a_{X3}}{3!}\ln^3\left(\frac{k}{k_*}\right) \,,
\end{equation}
where $X = [{\rm \zeta,\, t}]$ and 
\begin{equation}
    \mathcal{P}_{\zeta 0}(k_*) =  \frac{H^2_*}{8\pi^2 M_{\rm Pl}^2 \epsilon_{1*}} \,,\qquad
    \mathcal{P}_{\rm t 0}(k_*)= \frac{2H^2_*}{\pi^2 M_{\rm Pl}^2} \,.
\end{equation}
Therefore, the coefficients $a_{\rm \zeta i}$ and $a_{\rm t i}$ at third order are respectively
\begin{align}\label{eqn:a0s}
    a_{\zeta 0} =  & 1 - 2 (1 - \alpha) \epsilon_{1*} + \left( -3 - 2\alpha + 2\alpha^2 + \frac{\pi^2}{2} \right) \epsilon_{1*}^2 + \alpha\epsilon_{2*} \notag \\
& + \left( -6 +\alpha + \alpha^2 + \frac{7 \pi^2}{12} \right) \epsilon_{1*} \epsilon_{2*} + \frac{1}{8} \left( -8 + 4\alpha^2 + \pi^2 \right) \epsilon_{2*}^2 \notag\\
& + \frac{1}{24} \left( -12\alpha^2 + \pi^2 \right) \epsilon_{2*} \epsilon_{3*} - \frac{1}{24} \left( -16 + 24\alpha - 4\alpha^3 - 3\alpha\pi^2 + 6{\cal Z} \right) (8 \epsilon_{1*}^3 + \epsilon_{2*}^3)\notag \\
& + \frac{1}{12} \left( -72\alpha + 36\alpha^2 + 13 \pi^2 + 8\alpha\pi^2 - 36{\cal Z} \right) \epsilon_{1*}^2 \epsilon_{2*} \notag\\
& - \frac{1}{24} \left( 16 + 24\alpha - 12\alpha^2 - 8\alpha^3 - 15 \pi^2 - 6\alpha\pi^2 + 84{\cal Z} \right) \epsilon_{1*} \epsilon_{2*}^2 \notag\\
& + \frac{1}{24} \left( 16 + 4\alpha^3 - \alpha\pi^2 - 24{\cal Z} \right) ( \epsilon_{2*} \epsilon_{3*}^2 + \epsilon_{2*} \epsilon_{3*} \epsilon_{4*}) \notag\\
& + \frac{1}{24} \left( 48\alpha - 12\alpha^3 - 5\alpha\pi^2 \right) \epsilon_{2*}^2 \epsilon_{3*} \notag\\
& + \frac{1}{12} \left( -8 + 72\alpha - 12\alpha^2 - 8\alpha^3 + \pi^2 - 6\alpha\pi^2 - 24{\cal Z} \right) \epsilon_{1*} \epsilon_{2*} \epsilon_{3*} \,,
\end{align}
\begin{align}\label{eqn:a1s}
    a_{\zeta 1} = & -2 \epsilon_{1*} + 2 (-2\alpha + 1) \epsilon_{1*}^2 - \epsilon_{2*} - (2\alpha + 1) \epsilon_{1*} \epsilon_{2*} - \alpha\epsilon_{2*}^2 + \alpha\epsilon_{2*} \epsilon_{3*} \notag\\
& - \frac{1}{8} \left( -8 + 4\alpha^2 + \pi^2 \right) (8 \epsilon_{1*}^3 + \epsilon_{2*}^3) - 6\left( -1 + \alpha + \frac{\pi^2}{9} \right) \epsilon_{1*}^2 \epsilon_{2*} \notag\\
& - \frac{1}{4} \left( -4 + 4\alpha + 4\alpha^2 + \pi^2 \right) \epsilon_{1*} \epsilon_{2*}^2 + \left( -6 + 2\alpha + 2\alpha^2 + \frac{\pi^2}{2} \right) \epsilon_{1*} \epsilon_{2*} \epsilon_{3*} \notag\\
& + \frac{1}{24} \left( -12\alpha^2 + \pi^2 \right) ( \epsilon_{2*} \epsilon_{3*}^2 + \epsilon_{2*} \epsilon_{3*} \epsilon_{4*}) + \frac{1}{24} \left( -48 + 36\alpha^2 + 5 \pi^2 \right) \epsilon_{2*}^2 \epsilon_{3*} \,,
\end{align}
\begin{align}\label{eqn:a2s}
    a_{\zeta 2} = & 4 \epsilon_{1*}^2 + 2 \epsilon_{1*} \epsilon_{2*} + 6 \epsilon_{1*}^2 \epsilon_{2*} + \epsilon_{2*}^2 - \epsilon_{2*} \epsilon_{3*}  + (1 + 2\alpha) (\epsilon_{1*} \epsilon_{2*}^2 - 2 \epsilon_{1*} \epsilon_{2*} \epsilon_{3*})\notag \\
& + \alpha (8 \epsilon_{1*}^3 + \epsilon_{2*}^3 - 3 \epsilon_{2*}^2 \epsilon_{3*} + \epsilon_{2*} \epsilon_{3*}^2 + \epsilon_{2*} \epsilon_{3*} \epsilon_{4*}) \,,
\end{align}
\begin{align}\label{eqn:a3s}
    a_{\zeta 3} = &-8 \epsilon_{1*}^3 - 2 \epsilon_{1*} \epsilon_{2*}^2 - \epsilon_{2*}^3 + 4 \epsilon_{1*} \epsilon_{2*} \epsilon_{3*}  + 3 \epsilon_{2*}^2 \epsilon_{3*} - \epsilon_{2*} \epsilon_{3*}^2 - \epsilon_{2*} \epsilon_{3*} \epsilon_{4*} \,,
\end{align}
\begin{align}\label{eqn:a0t}
   a_{\rm t0} = & 1 + 2(-1 + \alpha) \epsilon_{1*} 
+ \left(-3 - 2\alpha + 2\alpha^2 + \frac{\pi^2}{2}\right) \epsilon_{1*}^2  \notag \\ 
& + \left(-2 + 2\alpha - \alpha^2 + \frac{\pi^2}{12}\right) \epsilon_{1*} \epsilon_{2*} 
- \frac{1}{3} \left(-16 + 24\alpha - 4\alpha^3 - 3\alpha \pi^2 + 6{\cal Z}\right) \epsilon_{1*}^3 \notag \\ 
& + \frac{1}{12} (-96 + 72\alpha + 36\alpha^2 - 24\alpha^3 + 13 \pi^2 - 10\alpha \pi^2) \epsilon_{1*}^2 \epsilon_{2*}  \notag \\ 
&  - \frac{1}{12} \left(8 - 24\alpha + 12\alpha^2 - 4\alpha^3 - \pi^2 + \alpha \pi^2 + 24{\cal Z}\right) (\epsilon_{1*} \epsilon_{2*}^2 + \epsilon_{1*} \epsilon_{2*} \epsilon_{3*}) \,,
\end{align}
\begin{align}\label{eqn:a1t}
    a_{\rm t1} = &-2 \epsilon_{1*} + 2(-2\alpha+1) \epsilon_{1*}^2 + (-2 + 2\alpha) \epsilon_{1*} \epsilon_{2*} - (-8 + 4\alpha^2 + \pi^2) \epsilon_{1*}^3 \notag \\ 
    &    + 6\left(- 1 - \alpha + \alpha^2 + \frac{5 \pi^2}{36}\right) \epsilon_{1*}^2 \epsilon_{2*} \notag \\
    &+ 
   \left(-2 + 2\alpha - \alpha^2 + \frac{\pi^2}{12}\right) (\epsilon_{1*} \epsilon_{2*}^2 + \epsilon_{1*} \epsilon_{2*} \epsilon_{3*}) \,,
\end{align}
\begin{align}\label{eqn:a2t}
    a_{\rm t2} = &  4 \epsilon_{1*}^2 + 8 \alpha \epsilon_{1*}^3 -2 
   \epsilon_{1*} \epsilon_{2*} + 2(3 - 6\alpha) \epsilon_{1*}^2 \epsilon_{2*} - 2(1 - \alpha) (\epsilon_{1*} \epsilon_{2*}^2 + \epsilon_{1*} \epsilon_{2*} \epsilon_{3*}) \,,
\end{align}
\begin{align}\label{eqn:a3t}
    a_{\rm t3} = & -8 \epsilon_{1*}^3 + 12 \epsilon_{1*}^2 \epsilon_{2*} -2 \epsilon_{1*} \epsilon_{2*}^2 - 2\epsilon_{1*} \epsilon_{2*} \epsilon_{3*} \,.
\end{align}
Note that ${\cal Z}$ enters only the third-order corrections of the LO parts, that are~\cref{eqn:a0s,eqn:a0t}.

Alternatively, the second kind of expansion consists of expanding the logarithm of the power spectrum in $\ln k$ as
\begin{equation}\label{eqn:PSinb}
    \ln \left[\frac{\mathcal{P}_X(k)}{\mathcal{P}_{X0}(k_*)}\right] = b_{X0} + b_{X1} \ln\left(\frac{k}{k_*}\right) + \frac{b_{X2}}{2} \ln^2\left(\frac{k}{k_*}\right)+\frac{b_{X3}}{3!}\ln^3\left(\frac{k}{k_*}\right) \,.
\end{equation}
Here we present the coefficients $b_{\rm \zeta i}$ and $b_{\rm t i}$ up to third-order corrections, they are respectively
\begin{align}\label{eqn:b0s}
b_{\zeta 0} =  & -2 (1 - \alpha) \epsilon_{1*} + \left(-5 + 2\alpha + \frac{\pi^2}{2}\right) \epsilon_{1*}^2 + \alpha\epsilon_{2*} + \left(-6 + 3\alpha -\alpha^2 + \frac{7 \pi^2}{12}\right) \epsilon_{1*} \epsilon_{2*} \notag \\
& + \frac{1}{24} \left(-12\alpha^2 + \pi^2\right) \epsilon_{2*} \epsilon_{3*} - \frac{1}{24} \left(80 - 48\alpha - 24 \pi^2 + 48{\cal Z}\right) \epsilon_{1*}^3 + \left(-1 + \frac{\pi^2}{8}\right) \epsilon_{2*}^2  \notag \\
& - \frac{1}{12} \left(-8 +  3{\cal Z}\right) \epsilon_{2*}^3 + \left(-12 + 15\alpha - 3\alpha^2 + \frac{27 \pi^2}{12} - \alpha\pi^2 - 3{\cal Z}\right) \epsilon_{1*}^2 \epsilon_{2*} \notag \\
& - \frac{1}{24} \left(64 - 168\alpha + 36\alpha^2 - 8\alpha^3 - 21 \pi^2 + 14\alpha\pi^2 + 84{\cal Z}\right) \epsilon_{1*} \epsilon_{2*}^2 \notag \\
& + \frac{1}{24} \left(16 + 4\alpha^3 - \alpha\pi^2 - 24{\cal Z}\right) (\epsilon_{2*} \epsilon_{3*}^2 + \epsilon_{2*} \epsilon_{3*} \epsilon_{4*})  + \left(2\alpha - \frac{\alpha \pi^2}{4}\right) \epsilon_{2*}^2 \epsilon_{3*} \notag \\
& + \frac{1}{12} \left(-8 + 72\alpha - 24\alpha^2 + 4\alpha^3 + 2 \pi^2 - 7\alpha \pi^2 - 24{\cal Z}\right) \epsilon_{1*} \epsilon_{2*} \epsilon_{3*} \,,
\end{align}

\begin{align}\label{eqn:b1s}
    b_{\zeta 1} = & -2 \epsilon_{1*} - 2 \epsilon_{1*}^2 - 2 \epsilon_{1*}^3 - \epsilon_{2*} - (3 - 2\alpha) \epsilon_{1*} \epsilon_{2*} + \alpha \epsilon_{2*} \epsilon_{3*} - \frac{2}{3} \left( \frac{45}{2} - 9\alpha - \frac{3\pi^2}{2} \right) \epsilon_{1*}^2 \epsilon_{2*} \notag \\
&  - \left( 7 - 3\alpha + \alpha^2 - \frac{7 \pi^2}{12} \right) \epsilon_{1*} \epsilon_{2*}^2 + \left( -6 + 4\alpha - \alpha^2 + \frac{7 \pi^2}{12} \right) \epsilon_{1*} \epsilon_{2*} \epsilon_{3*} \notag \\
&  + \frac{1}{24} \left( -12\alpha^2 + \pi^2 \right) \left( \epsilon_{2*} \epsilon_{3*}^2 + \epsilon_{2*} \epsilon_{3*} \epsilon_{4*} \right) + \left( -2 + \frac{\pi^2}{4} \right) \epsilon_{2*}^2 \epsilon_{3*} \,,
\end{align}
\begin{align}\label{eqn:b2s}
    b_{\zeta 2} = & -2 \epsilon_{1*} \epsilon_{2*} - 6 \epsilon_{1*}^2 \epsilon_{2*} - (3 - 2\alpha) \epsilon_{1*} \epsilon_{2*}^2 - \epsilon_{2*} \epsilon_{3*} \notag \\
&  - 2 (2 - \alpha) \epsilon_{1*} \epsilon_{2*} \epsilon_{3*} + \alpha \epsilon_{2*} \epsilon_{3*}^2 + \alpha \epsilon_{2*} \epsilon_{3*} \epsilon_{4*} \,,
\end{align}
\begin{equation}\label{eqn:b3s}
    b_{\zeta 3} =- 2 \epsilon_{1*} \epsilon_{2*}^2 - 2 \epsilon_{1*} \epsilon_{2*} \epsilon_{3*} - \epsilon_{2*} \epsilon_{3*}^2 - \epsilon_{2*} \epsilon_{3*} \epsilon_{4*} \,,
\end{equation}
\begin{align}\label{eqn:b0t}
    b_{\rm t0} = & -2 (1 - \alpha) \epsilon_{1*} + \frac{1}{2} (-10 + 4\alpha + \pi^2) \epsilon_{1*}^2  + \left(-2 + 2\alpha - \alpha^2 + \frac{\pi^2}{12}\right) \epsilon_{1*} \epsilon_{2*} \notag \\
& - \frac{1}{3} \left(10 - 6\alpha - 3 \pi^2 + 6{\cal Z}\right) \epsilon_{1*}^3  + \left(-12 + 14\alpha - 3\alpha^2 + \frac{15 \pi^2}{12} - \alpha \pi^2\right) \epsilon_{1*}^2 \epsilon_{2*} \notag \\
& - \frac{1}{12} \left(8 - 24\alpha + 12\alpha^2 - 4\alpha^3 - \pi^2 + \alpha\pi^2 + 24{\cal Z}\right) (\epsilon_{1*} \epsilon_{2*}^2 + \epsilon_{1*} \epsilon_{2*} \epsilon_{3*}) \,,
\end{align}
\begin{align}\label{eqn:b1t}
b_{\rm t1} = & -2 \epsilon_{1*} - 2 \epsilon_{1*}^2 + 2(-1 + \alpha) \epsilon_{1*} \epsilon_{2*} - 2 \epsilon_{1*}^3  + (-14 + 6\alpha +  \pi^2) \epsilon_{1*}^2 \epsilon_{2*} \notag \\
& + \left(-2 + 2\alpha - \alpha^2 + \frac{\pi^2}{12}\right) (\epsilon_{1*} \epsilon_{2*}^2 + \epsilon_{1*} \epsilon_{2*} \epsilon_{3*}) \,,
\end{align}
\begin{equation}\label{eqn:b2t}
    b_{\rm t2} =  -2 \epsilon_{1*} \epsilon_{2*} - 6 \epsilon_{1*}^2 \epsilon_{2*} - 2 (1 - \alpha) \epsilon_{1*} \epsilon_{2*}^2 - 2 (1 - \alpha) \epsilon_{1*} \epsilon_{2*} \epsilon_{3*} \,,
\end{equation}
\begin{equation}\label{eqn:b3t}
b_{\rm t3} =  -2 \epsilon_{1*} \epsilon_{2*}^2 - 2 \epsilon_{1*} \epsilon_{2*} \epsilon_{3*} \,.
\end{equation}

Using the latter parameterisation, it is possible to write directly both the scalar and tensor spectral indices, 
runnings, and runnings of the running with respect to the coefficients $b_{\rm X i}$ as follows
\begin{equation} \label{eqn:ns}
    n_{\rm s} = 1 + b_{\zeta 1} + b_{\zeta 2} \ln\left(\frac{k}{k_*}\right) + \frac{b_{\zeta 3}}{2}\ln^2\left(\frac{k}{k_*}\right) \,,
\end{equation}
\begin{equation} \label{eqn:as}
    \alpha_{\rm s} = b_{\zeta 2} + b_{\zeta 3}\ln\left(\frac{k}{k_*}\right) \,,
\end{equation}
\begin{equation} \label{eqn:bs}
    \beta_{\rm s} = b_{\zeta 3} \,,
\end{equation}
\begin{equation}
    n_{\rm t} = b_{\rm t1} + b_{\rm t2} \ln\left(\frac{k}{k_*}\right) + \frac{b_{\rm t3}}{2}\ln^2\left(\frac{k}{k_*}\right) \,,
\end{equation}
\begin{equation}
    \alpha_{\rm t} = b_{\rm t2} + b_{\rm t3}\ln\left(\frac{k}{k_*}\right) \,,
\end{equation}
\begin{equation}
    \beta_{\rm t} = b_{\rm t3} \,.
\end{equation}

\section{Impact of the Prior Range} \label{app:an_prior}
In this section, we investigate the sensitivity of the cosmological constraints on the HFF parameters to the choice of prior width, 
as previously done in Ref.~\cite{Martin:2024nlo} for third-order results, but using a different combination of datasets and 
different priors. To assess the effect of the prior width on the results, we repeated our analysis by varying the sampled 
range of the HFF parameters as $\epsilon_{i \geq 2} \in [-1,\,1]$. By expanding and contracting the prior ranges, we aim to quantify 
how these choices affect the posterior distributions, especially with respect to the validity of the analytical equations. This 
analysis allows us to test the robustness of our results and ensure that the numbers presented in~\cref{sec:results} are weakly 
dependent on the range of priors. In particular, the prior effects arise from the correlation between $\epsilon_3$, which is not 
well constrained by current cosmological data, and $\epsilon_2$; as shown in Ref.~\cite{Martin:2024nlo}.

\begin{figure}[h!]
\centering
\includegraphics[width=\textwidth]{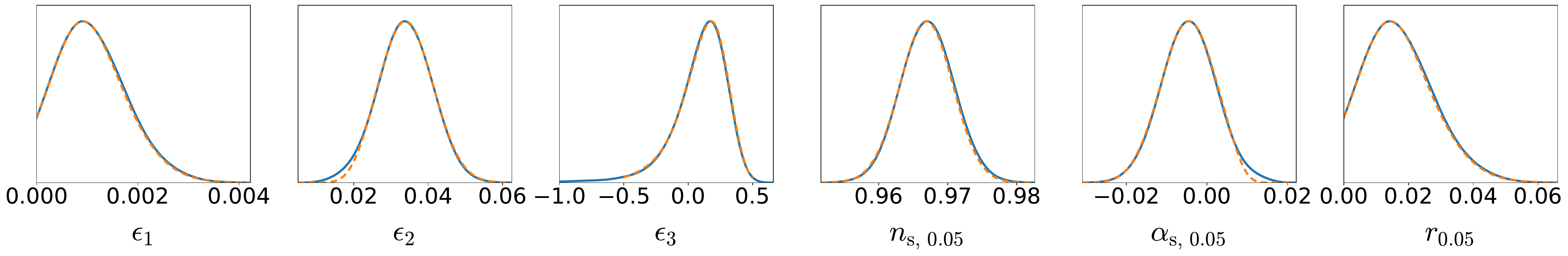}
\includegraphics[width=\textwidth]{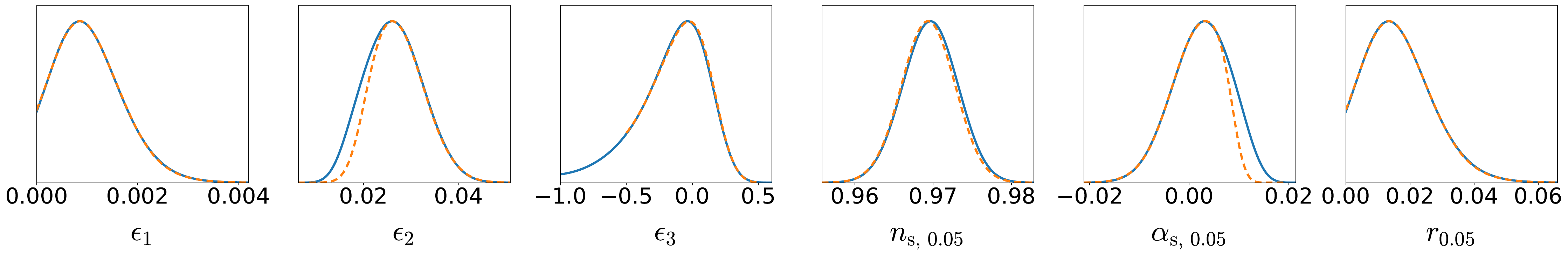}
\includegraphics[width=\textwidth]{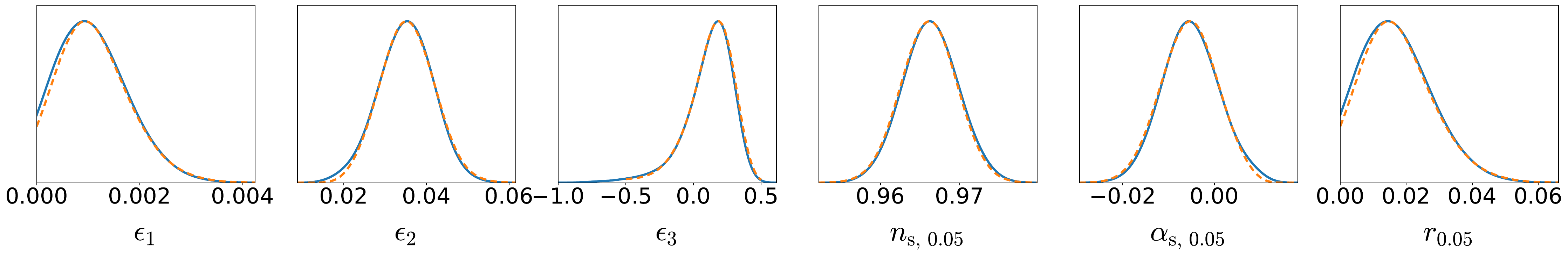}
\caption{One-dimensional marginalised posterior distributions for the first three HFF parameters $\epsilon_1$, 
$\epsilon_2$, and $\epsilon_3$ and for the scalar spectral index $\ns$, its running $\as$ assuming, and the tensor-to-scalar ratio $r$ 
derived using second-order slow-roll equations. Solid blue lines correspond to larger prior range $\epsilon_{2,3} \in [-1,\,1]$ while 
dashed orange lines correspond to tighter priors used in the main text, that are $\epsilon_{2,3} \in [-0.5,\,0.5]$. 
Different rows correspond to different datasets: P18+BK18 (upper row), P18+ACT+BK18 with ACT data truncated (central row), and 
P18+SPT+BK18 (lower row), all in combination with the external datasets.}\label{fig:eps123_prior1}
\end{figure}

\begin{figure}[h!]
\centering
\includegraphics[width=\textwidth]{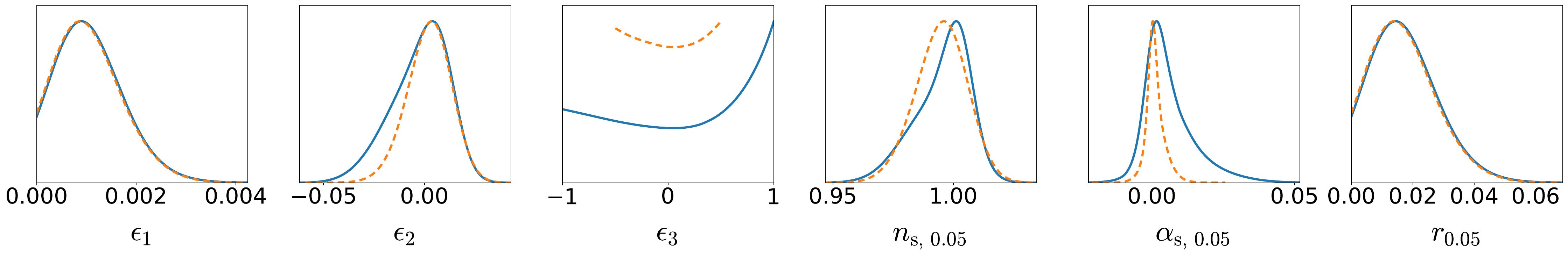}
\includegraphics[width=\textwidth]{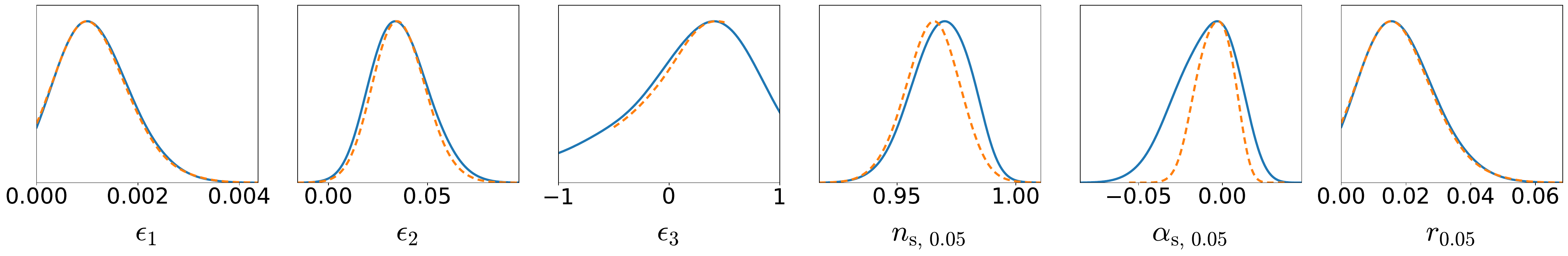}
\includegraphics[width=\textwidth]{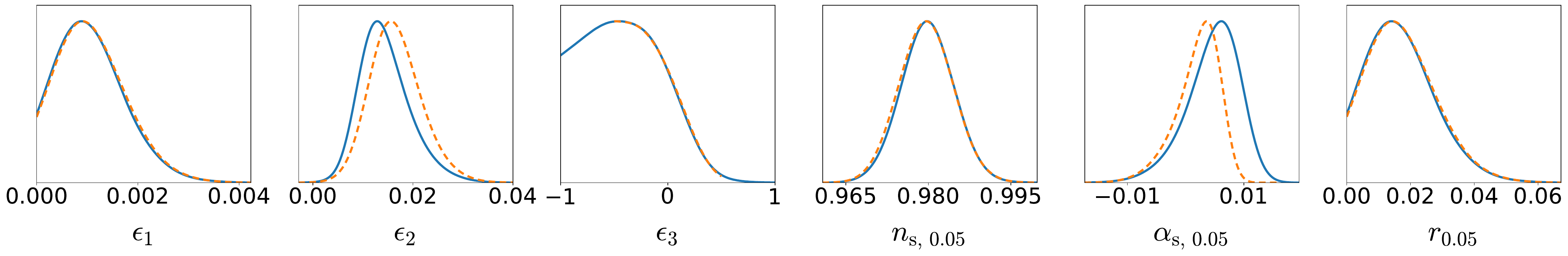}
\caption{Same as \cref{fig:eps123_prior1} for ACT+BK18 (upper row), and SPT+BK18 (central row), and for the 
combination P18+ACT+BK18 with \Planck\ data truncated, all in combination with the external datasets.}\label{fig:eps123_prior2}
\end{figure}

When the full \Planck\ data are included (meaning the combinations P18+BK18, P18+ACT+BK18 with ACT temperature data truncated, and 
P18+SPT+BK18), no significant effect on the mean and width of the one-dimensional posterior distributions is observed, as shown 
in~\cref{fig:eps123_prior1}. However, in cases where \Planck\ data are excluded or the temperature data are truncated when combined 
with ACT data, $\epsilon_2$ and $\epsilon_3$ are less constrained, leading to a small impact on the uncertainties on $\epsilon_2$ and 
$\as$, with a shift of less than $0.5\sigma$ in the mean value of $\as$ for the combination P18+ACT+BK18 with the \Planck\ temperature 
data truncated. This is shown in~\cref{fig:eps123_prior2}.
For the cases including SPT data, where the posterior distributions of the HFF parameters are centred around zero, these effects are 
smaller compared to the analysis performed with the full ACT dataset.

We conclude that while the narrower prior ranges may appear safer in terms of perturbative expressions and to minimise the spread of 
the tails of the posterior distributions of the poorly constrained parameters, there is no strong indication against using 
$\epsilon_{i \geq 2} \in [-1,\,1]$ which should ensure the validity of the perturbative regime under which we derived the equations.

\bibliographystyle{elsarticle-num} 
\bibliography{cas-refs}

\end{document}